\documentclass{scrartcl}
\usepackage[utf8]{inputenc}
\usepackage{xcolor}
\usepackage{amsmath}
\usepackage{amssymb}
\usepackage{mathtools}
\usepackage{bbold}
\usepackage[left=2cm, right=2cm, top=2cm, bottom=3cm]{geometry}
\usepackage{authblk}
\usepackage{lipsum}
\usepackage{microtype}
\usepackage[numbers,sort&compress]{natbib}
\usepackage{hyperref}

\newcommand{\ie}{i.\,e.~}

\newcommand{\eg}{e.\,g.~}

\newcommand{\wrt}{w.\,r.\,t.~}

\newcommand{\mi}{\mathrm{i}}

\newcommand{\md}{\mathrm{d}}

\newcommand{\e}{\mathrm{e}}

\newcommand{\LArrow}[1]{\parbox{#1}{\tikz{\draw[<-](0,0)--(#1,0);}}}
\newcommand{\ShortLArrow}{\LArrow{0.2cm}}

\newcommand{\ini}[1]{#1^{(\mathrm{i})}}
\newcommand{\vini}[1]{\vec{#1}^{\,(\mathrm{i})}}
\newcommand{\tini}[1]{\boldsymbol{#1}^{(\mathrm{i})}}

\newcommand{\tens}[1]{\boldsymbol{#1}}

\newcommand{\fin}[1]{#1^{(\mathrm{f})}}

\newcommand{\tfin}[1]{\boldsymbol{#1}^{(\mathrm{f})}}

\newcommand{\mean}[1]{\langle #1 \rangle}

\usepackage{tikz}
\usepackage{pgfplots}
\pgfplotsset{compat=1.11}
\usepackage{xparse}

\usetikzlibrary{
arrows.meta,
bending,
positioning,
decorations.pathmorphing,
decorations.pathreplacing,
decorations.markings,
shapes.geometric,
calc
}

\newlength{\gs}
\newlength{\feyn}
\newlength{\pdotdiam}
\NewDocumentCommand{\mytikz}{O{-0.6ex}}{\tikz[baseline=#1,radius=\gs]}

\tikzset{
fdot/.style={draw, black, circle, fill, minimum size=\pdotdiam, inner sep=0},
bdot/.style={draw, black, circle, minimum size=\pdotdiam, inner sep=0},
intline/.style={draw, black, line cap=round, thick, postaction={decorate, decoration={markings, mark=at position 0.6 with {\arrow{stealth}}} } },
propline/.style={draw, black, line cap=round, thick, postaction={decorate, decoration={markings, mark=at position 0.6 with {\arrow{Stealth}} } }},
vertex/.style={draw, black, circle, fill, minimum size=2\pdotdiam, inner sep=0},
}

\NewDocumentCommand \vertex{O{}}{
\coordinate[vertex] (v1) at (0,0);
\node[below=2pt] at (v1) {$#1$};
}

\NewDocumentCommand \vertexnew{O{}}{
\coordinate[vertex] (v1) at (2\gs,0);
\node[below=2pt] at (v1) {$#1$};
}

\NewDocumentCommand \fdot{O{}}{
\coordinate[fdot] (f1) at (0,0);
\node[below=2pt] at (f1) {$#1$};
}

\NewDocumentCommand \bdot{O{}}{
\coordinate[bdot] (b1) at (0,0);
\node[below=0.5pt] at (b1) {$#1$};
}

\NewDocumentCommand \ellipsis{O{}}{
\node[draw=none] (ellipsis1) at (0,0) {$\cdots$};
}

\NewDocumentCommand \dotpattern{O{}O{}O{}O{}}{
\coordinate[fdot] (f1) at (0,0);
\node [below=2pt] at (f1) {$#1$};
\node[draw=none] at (\gs,0) {$\cdots$};
\coordinate[fdot] (f2) at (2\gs,0);
\node [below=2pt] at (f2) {$#2$};
\coordinate[bdot] (b1) at (3\gs,0);
\node[below=0.5pt] at (b1) {$#3$};
\node[draw=none] at (4\gs,0) {$\cdots$};
\coordinate[bdot] (b2) at (5\gs,0);
\node[below=0.5pt] at (b2) {$#4$};
}

\NewDocumentCommand \ffpattern{O{}O{}}{
\coordinate[fdot] (f1) at (0,0);
\node [below] at (f1) {$#1$};
\coordinate[fdot] (f2) at (2\gs,0);
\node [below] at (f2) {$#2$};
}

\NewDocumentCommand \bbpattern{O{}O{}}{
\coordinate[bdot] (b1) at (0,0);
\node [below] at (b1) {$#1$};
\coordinate[bdot] (b2) at (\gs,0);
\node [below] at (b2) {$#2$};
}

\NewDocumentCommand \fbpattern{O{}O{}}{
\coordinate[fdot] (f1) at (0,0);
\node [below] at (f1) {$#1$};
\coordinate[bdot] (b1) at (\gs,0);
\node [below] at (b1) {$#2$};
}

\NewDocumentCommand \bffpattern{O{}O{}O{}}{
\coordinate[bdot] (b1) at (0,0);
\node [below] at (b1) {$#1$};
\coordinate[fdot] (f1) at (\gs,0);
\node [below] at (f1) {$#2$};
\coordinate[fdot] (f2) at (2\gs,0);
\node [below] at (f2) {$#3$};
}

\NewDocumentCommand \ffbpattern{O{}O{}O{}}{
\coordinate[fdot] (f1) at (0,0);
\node [below] at (f1) {$#1$};
\coordinate[fdot] (f2) at (\gs,0);
\node [below] at (f2) {$#2$};
\coordinate[bdot] (b1) at (2\gs,0);
\node [below] at (b1) {$#3$};
}

\NewDocumentCommand \fbbpattern{O{}O{}O{}}{
\coordinate[fdot] (f1) at (0,0);
\node [below] at (f1) {$#1$};
\coordinate[bdot] (b1) at (\gs,0);
\node [below] at (b1) {$#2$};
\coordinate[bdot] (b2) at (2\gs,0);
\node [below] at (b2) {$#3$};
}

\NewDocumentCommand \bbffpattern{O{}O{}O{}O{}}{
\coordinate[bdot] (b1) at (0,0);
\node [below] at (b1) {$#1$};
\coordinate[bdot] (b2) at (\gs,0);
\node [below] at (b2) {$#2$};
\coordinate[fdot] (f1) at (2\gs,0);
\node [below] at (f1) {$#3$};
\coordinate[fdot] (f2) at (3\gs,0);
\node [below] at (f2) {$#4$};
}

\NewDocumentCommand \ffbbpattern{O{}O{}O{}O{}}{
\coordinate[fdot] (f1) at (0,0);
\node [below=2pt] at (f1) {$#1$};
\coordinate[fdot] (f2) at (\gs,0);
\node [below=2pt] at (f2) {$#2$};
\coordinate[bdot] (b1) at (2\gs,0);
\node [below=0.5pt] at (b1) {$#3$};
\coordinate[bdot] (b2) at (3\gs,0);
\node [below=0.5pt] at (b2) {$#4$};
}

\NewDocumentCommand \fbbbpattern{O{}O{}O{}O{}}{
\coordinate[fdot] (f1) at (0,0);
\node [below=2pt] at (f1) {$#1$};
\coordinate[bdot] (b1) at (\gs,0);
\node [below=0.5pt] at (b1) {$#2$};
\coordinate[bdot] (b2) at (2\gs,0);
\node [below=0.5pt] at (b2) {$#3$};
\coordinate[bdot] (b3) at (3\gs,0);
\node [below=0.5pt] at (b3) {$#4$};
}

\NewDocumentCommand \fffbbbpattern{O{}O{}O{}O{}O{}O{}}{
\coordinate[fdot] (f1) at (0,0);
\node [below=2pt] at (f1) {$#1$};
\coordinate[fdot] (f2) at (\gs,0);
\node [below=2pt] at (f2) {$#2$};
\coordinate[fdot] (f3) at (2\gs,0);
\node [below=2pt] at (f3) {$#3$};
\coordinate[bdot] (b1) at (3\gs,0);
\node [below=0.5pt] at (b1) {$#4$};
\coordinate[bdot] (b2) at (4\gs,0);
\node [below=0.5pt] at (b2) {$#5$};
\coordinate[bdot] (b3) at (5\gs,0);
\node [below=0.5pt] at (b3) {$#6$};
}

\NewDocumentCommand \fvvfpattern{O{}O{}O{}O{}}{
\coordinate[fdot] (f1) at (0,0);
\node [below] at (f1) {$#1$};
\coordinate[vertex] (v1) at (1.5\gs,0);
\node [below] at (v1) {$#2$};
\coordinate[vertex] (v2) at (3\gs,0);
\node [below] at (v2) {$#3$};
\coordinate[fdot] (f2) at (4.5\gs,0);
\node [below] at (f2) {$#4$};
}

\NewDocumentCommand \straightline{mm}{
    {\draw[intline] (#1) -- (#2);}
}

\NewDocumentCommand \arcline{mm}{
    {\draw[intline] (#1) to [out=135,in=45] node [sloped,above] {} (#2)}
}

\NewDocumentCommand \bfprop{mm}{
    { \draw[densely dashed, -Stealth, thick] (#1) -- ($(#1)!0.55!(#2)$) ;
    \draw[solid, thick]  ($(#1)!0.54!(#2)$) -- (#2);
    }
}

\NewDocumentCommand \fbprop{mm}{
    { \draw[solid, thick] (#1) -- ($(#1)!0.46!(#2)$) ;
    \draw[densely dashed, Stealth-, thick] ($(#1)!0.45!(#2)$) -- (#2) ;
    }
}

\NewDocumentCommand \fbpropCircle{m}{
    {\coordinate[draw=none] (c1) at (3.5\gs,0); 
    \coordinate[draw=none] (cu) at (2.75\gs,0.75\gs);
    \coordinate[draw=none] (cd) at (2.75\gs,-0.75\gs);
    \draw[solid, thick] (#1) to [out=90,in=180] node [sloped,above] {} (cu) ;
    \draw[solid, thick] (cu) to [out=0,in=90] node [sloped,above] {} (c1) ;
    \draw[densely dashed, Stealth-, thick]  (c1) to [out=-90,in=0] node [sloped,above] {} (cd);
    \draw[densely dashed, thick]  (cd) to [out=180,in=-90] node [sloped,above] {} (#1);
    }
}

\NewDocumentCommand \ffpropCircle{m}{
    {\coordinate[draw=none] (c1) at (3.5\gs,0); 
    \coordinate[draw=none] (cu) at (2.75\gs,0.75\gs);
    \coordinate[draw=none] (cd) at (2.75\gs,-0.75\gs);
    \draw[solid, thick] (#1) to [out=90,in=180] node [sloped,above] {} (cu) ;
    \draw[solid, -Stealth,thick] (c1) to [out=90,in=0] node [sloped,above] {} (cu) ;
    \draw[solid, -Stealth,thick]  (c1) to [out=-90,in=0] node [sloped, above] {} (cd);
    \draw[solid, thick]  (cd) to [out=180,in=-90] node [sloped,above] {} (#1);
    }
}

\NewDocumentCommand \ffprop{mm}{
    { \draw[solid, thick] (#1) -- ($(#1)!0.26!(#2)$) ;
    \draw[Stealth-, thick] ($(#1)!0.25!(#2)$) -- ($(#1)!0.5!(#2)$);
    \draw[-Stealth, thick] ($(#1)!0.5!(#2)$) -- ($(#1)!0.75!(#2)$) ;
    \draw[solid, thick] ($(#1)!0.72!(#2)$) -- (#2);
    }
}

\NewDocumentCommand \ffpropcomb{mm}{
    {     
    \draw[solid, thick] (#1) -- ($(#1)!0.21!(#2)$) ;
    \draw[densely dashed, Stealth-, thick] ($(#1)!0.2!(#2)$) -- ($(#1)!0.5!(#2)$) ;
    \coordinate[vertex] (v1) at ($(#1)!0.5!(#2)$);
    \draw[densely dashed, -Stealth, thick] (v1) -- ($(v1)!0.55!(#2)$) ;
    \draw[solid, thick] ($(v1)!0.54!(#2)$) -- (#2) ;
    }
}

\NewDocumentCommand \nvertex{}{
    {
    \draw[densely dashed, gray!60, thick] (0,0) -- ( 30:0.5cm) ;
    \draw[densely dashed, gray!60, thick] (0,0) -- ( 150:0.5cm) ;
    \draw[loosely dotted, thick] ( 70:0.5cm) to [out=135,in=45] node [sloped,above] {} ( 110:0.5cm);
    \draw ( 130:0.5cm) node [left] {$X_1$};
    \draw ( 40:0.5cm) node [right] {$X_r$};
    \draw[solid, gray!60, thick] (0,0) -- ( 210:0.5cm) ;
    \draw[solid, gray!60, thick] (0,0) -- ( 330:0.5cm) ;
    \draw[loosely dotted, thick] ( 250:0.5cm) to [out=-45,in=-135] node [sloped,above] {} ( 290:0.5cm);
    \draw ( 230:0.5cm) node [sloped, left] {$X_{1}^\prime$};
    \draw ( 310:0.5cm) node [sloped,right] {$X_{s}^\prime$};
    \coordinate[vertex] (v1) at (0,0);
    }
}

\NewDocumentCommand \ffdisconnected{mm}{
    \coordinate[fdot] (f1) at (0,0.7\gs);
    \node [below=2pt] at (f1) {$#1$};
    \coordinate[vertex] (v1) at (2\gs,0.7\gs);
    \fbprop{f1}{v1}; 
    \coordinate[fdot] (f2) at (0, -0.7\gs);
    \node [below=2pt] at (f2) {$#2$};
    \coordinate[vertex] (v2) at (2\gs,-0.7\gs);
    \fbprop{f2}{v2};
}

\title{Field Theory Approach to Classical $N$-Particle Systems In and Out of Equilibrium}
\author[1]{Tristan Daus\thanks{t.daus@thphys.uni-heidelberg.de}}
\author[1]{Elena Kozlikin\thanks{elena.kozlikin@uni-heidelberg.de}}
\affil[1]{Institute for Theoretical Physics, Heidelberg University, Philosophenweg 12, 69120 Heidelberg, Germany}
\date{}
\begin{document}
\maketitle
\begin{abstract}
    We present an approach to solving the evolution of a classical $N$-particle ensemble based on the path integral approach to classical mechanics. This formulation provides a perturbative solution to the Liouville equation in terms of a propagator which can be expanded in a Dyson series. We show that this perturbative expansion exactly corresponds to an iterative solution of the BBGKY-hierarchy in orders of the interaction potential. Using the path integral formulation, we perform a Hubbard-Stratonovich transformation (HST) to obtain an effective field theoretic description in terms of macroscopic fields, which contains the full microscopic dynamics of the system in its vertices. Naturally, the HST leads to a new perturbative expansion scheme which contains an infinite order of microscopic interactions already at the lowest order of the perturbative expansion. Our approach can be applied to in and out of equilibrium systems with arbitrary interaction potentials and initial conditions. We show how (unequal-time) cumulants of the Klimontovich phase space densities can be computed within this framework and derive results for density and momentum correlations for a spatially homogeneous system. Under the explicit assumptions for the interaction potential and initial conditions, we show that well-known results related to plasma oscillations and the Jeans instability criterion for gravitational collapse can be recovered in the lowest order perturbative expansion and that both are the effect of the same collective behaviour of the many-body system.  
\end{abstract}
\tableofcontents

\section{Introduction}
The overwhelming success of the path integral formulation of quantum field theory had sparked a keen interest in an analogous formulation for classical mechanics more than half a century ago. The pioneering work of \cite{MSR1973}, - now known as the MSR formalism -, laid the groundwork for a path integral formulation for stochastic systems \cite{Jensen1981} and for classical Hamiltonian systems \cite{Gozzi1988,Gozzi1989}. The idea behind the introduction of the path integral formalism to classical mechanics was to apply the wealth of already developed functional methods to classical systems \cite{ZINNJUSTIN_1986}. Perhaps more importantly, the goal was to give to the theory of classical statistical mechanics the same generality and rigour as quantum field theory\cite{MSR1973}. The path integral formulation was, thus, intended to encompass the properties of a system, be it conservative or dissipative, or in or out of equilibrium. It would also provide a graphical representation in terms of Feynman diagrams and it would 
allow to apply non-perturbative methods from quantum field theory to classical systems. While the works of \cite{MSR1973,Jensen1981,ZINNJUSTIN_1986} have given rise to many applications of the path integral formalism for stochastic systems \cite{kleinert2009path, Peliti_1985, Harsh_2023}, not much attention -- beside a purely academic interest \cite{Cattaruzza_2013, Abrikosov:2004cf, Penco_2006, Das_2012} -- has been paid to the path integral formulation for classical Hamiltonian systems introduced by \cite{Gozzi1988,Gozzi1989}. However, there are many applications for the latter, ranging from plasma systems \cite{Kozlikin_2023} to cosmic Dark Matter large-scale structures \cite{Lilow_2019, Pixius_2022}. The common denominator for such systems is that typical mean-field approximations or standard fluid descriptions become insufficient for their description when they are dominated by effects that go beyond collective behaviour. To study such systems, numerical $N$-body simulations have become the standard method. These are, however, computationally expensive, especially for systems with long-ranged interactions, and heavily tailored to each particular application. Although, they can successfully reproduce results for those systems they were designed to simulate, they are not suitable as a general framework to gain a fundamental understanding of many-body physics. 

To establish such a framework, we thus, return to the original idea of the MSR formalism in order to establish a general and rigorous analytical 
approach for the study of classical $N$-particle systems based on the path integral formulation of classical mechanics proposed by \cite{Gozzi1988,Gozzi1989}. 
Our main interest is the study of structure formation and evolution in terms of $n$-point correlation functions of the Klimontovich phase space densities. We obtain a perturbative solution to the Liouville equation in terms of a propagator which can be expanded in a Dyson series, which corresponds to an iterative solution of the BBGKY-hierarchy in orders of the interaction potential. This microscopic perturbative expansion is not new. However, the path integral formulation allows us to establish a macroscopic field theory by applying a Hubbard-Stratonovich transformation (HST) \cite{Stratonovich1957,Hubbard1959} to the generating functional. The HST leads to an effective field theoretic description which contains the full statistics of the microscopic system in its vertices. A similar approach was explored in \cite{Lilow_2019}. 

We set up a new macroscopic perturbation theory in terms of macroscopic propagators and vertices, which corresponds to a re-ordering of the microscopic perturbation theory. The macroscopic propagator contains a re-summation of microscopic terms of a certain class to infinite order in the interaction potential. Furthermore, the macroscopic field theory allows us to apply functional (non-perturbative) methods from (quantum) field theory which we shall discuss in future work. The macroscopic theory is therefore especially interesting for long-ranged interaction potentials where perturbative approaches do not suffice. The approach presented here is completely general, it can be applied to in- and out-of-equilibrium systems for any (two-particle) interaction potential and arbitrary initial conditions.\\ 

The contents of this paper are structured in the following way: In Sec.\ \ref{sec:Basics} we introduce our notation and refresh the readers memory on a few well-known quantities from statistical mechanics that will be used later on. We then derive the path integral for classical mechanics in Sec.\ \ref{sec:Construction} and discuss the microscopic perturbation theory extensively in Sec.\ \ref{sec:MicroscopicTheory}. We provide calculations of Klimontovich phase space densities up to second order in perturbation theory and draw the connection to the BBGKY-hierarchy, the Boltzmann and the Vlasov equation. In addition, we provide an intuitive picture of the microscopic perturbation theory, as it will play an important role in the macroscopic field theory, which we derive in Sec.\ \ref{sec:MacroTheory}. First, we construct a generating functional and perform a (modified) HST in Sec.\ \ref{sec:MacroscopicGeneral} in order to obtain the macroscopic field theory. We then derive Feynman rules in Sec.\ \ref{sec:Feynman} and discuss the free or tree-level theory and its connection to the microscopic theory in Sec.\ \ref{sec:MicroMacro} and Sec.\ \ref{sec:TreeLevel}. Finally, we provide an instructive toy-model application in Sec.\ \ref{sec:HomogeneousSystem} where we compute the tree-level density and momentum correlations for a spatially homogeneous system. At the end of Sec.\ \ref{sec:HomogeneousSystem} we recover two well-known effects -- one from plasma physics and one from cosmology -- in our tree-level theory and show that both are, in fact, the result of the same collective behaviour of a many-body system.

\section{Time Evolution of the Classical Ensemble}
\label{sec:PathIntegralFormulation}

\subsection{Basics and Notation}
\label{sec:Basics}
We denote the microstate of an $N$-particle ensemble in three-dimensional Euclidean space by
the set of phase space coordinates $x_j = (\vec{q}_j, \vec{p}_j)^\top\,,\,\,j=1\cdots N$ of the individual particles, with respective canonical positions and momenta $\vec{q}_j$, $\vec{p}_j$. Furthermore, we assume the Hamilton function of the system to be of the form 
\begin{equation}\label{eq:Hamiltonfunction}
    H(x_1\,,\cdots \,,x_N, t) =  H_0(x_1\,,\cdots \,,x_N, t)+V(x_1\,,\cdots \,,x_N, t)\,,
\end{equation}
with
\begin{equation}\label{eq:TwoParticlePotential}
    H_0(x_1\,,\cdots \,,x_N, t) = \sum_{i=1}^N\frac{{\vec{p}_i}^{\,2}}{2m}\,,\quad
    V(x_1\,,\cdots \,,x_N, t) = \frac{1}{2}\sum_{i\neq j = 1}^Nv(|\vec{q}_i-\vec{q}_j|, t)\,,
\end{equation}
where $v(|\vec{q}_i-\vec{q}_j|, t)$ is a possibly time dependent two-particle interaction potential depending only on the absolute value of the distance of the two particles. Moreover, all particles are assumed to have the same mass $m$. The temporal evolution of the system is now described by the set of phase space trajectories $\{x_j(t)\}$ of the particles which are subject to Hamilton's equations of motion
\begin{align}
    \dot{\vec{q}}_j=&\nabla_{\vec{p}_j}H(x_1\,,\cdots \,,x_N, t)\\
    \dot{\vec{p}}_j=&-\nabla_{\vec{q}_j}H(x_1\,,\cdots \,,x_N, t)\,,\,\,\,j=1\cdots N\,
\end{align}
which can be brought into the symplectic form 
\begin{equation}\label{eq:eom}
    \dot{x}_j=\omega\cdot\nabla_jH(x_1\,,\cdots \,,x_N, t)\,,
\end{equation}
where $\omega=\begin{pmatrix}
    0 & \mathbb{1}_{3\times3} \\
    -\mathbb{1}_{3\times3} & 0
\end{pmatrix}$ and $\nabla_j=(\nabla_{\vec{q}_j}, \nabla_{\vec{p}_j})^\top$ are the symplectic matrix and the phase space gradient, respectively. Due to the indistinguishable nature of the particles almost all of the quantities we are interested in will depend on the whole particle ensemble. Thus, in order to increase readability, we introduce a tensorial notation, bundling all objects which carry a particle index, \eg $q_j$, into multiparticle objects, 
\begin{equation}
    \tens{A}=\sum_{j=1}^N A_j\otimes e_j\,,
\end{equation}
where $e_j$ is the $N$-dimensional canonical unit-vector. The set of phase space coordinates $\{x_j\}$ is now, for instance, represented by the tensor $\tens{x}$. Note that from the definition of the tensor product, the scalar product between two quantities $\tens{A}$ and $\tens{B}$ decomposes as
\begin{equation}
    \tens{A\cdot B}=\sum_{j,i=1}^N(A_i\otimes e_i)\cdot(B_j\otimes e_j)=\sum_{j,i=1}^NA_iB_j\underbrace{e_i\cdot e_j}_{=\delta_{ij}}=\sum_{i=1}^NA_iB_i\,.
\end{equation}
Functions depending on the whole set of particle positions will be denoted by 
\begin{equation}
    F(\{x_j\}, t)=F(x_1\,,\cdots \,,x_N,t)\equiv F(\tens{x},t)\,.
\end{equation}
Furthermore, the integration measure of the $N$-particle phase space is written as 
\begin{equation}
    \md^{6N}x=\md^{3N}q\,\md^{3N}p \equiv \md\tens{q}\,\md\tens{p}=\md \tens{x}\,.
\end{equation}
Using this notation, the equations of motion for the whole bundle of particles can be brought into the convenient form 
\begin{equation}\label{eq:tensoreom}
    0 = \tens{\mathcal{E}}(\tens{x})=\dot{\tens{x}} - \tens{\omega}\tens{\cdot}\tens{\nabla}H(\tens{x}, t)\,,
\end{equation}
where $\tens{\omega}=\omega\otimes\mathbb{1}_{N\times N}$ is the tensorial generalization of the symplectic structure. \\

Note, that by the above equation of motion the phase space trajectories travelled by particles are uniquely defined by the Hamilton function once an appropriate set of initial conditions $\tini{x}=\tens{x}(t=\ini{t})$ at initial time $\ini{t}$ is given. This makes the determinism of classical mechanics manifest.\\

In general we do not know the initial phase space coordinates explicitly but only on a statistical level, \ie through an initial phase space probability density distribution 
\begin{equation}
    \varrho_N(\tini{x}, \ini{t})=\varrho_N(x_1,\cdots, x_N, \ini{t})\,.
\end{equation}
Consequently, the specific final positions of the particles are only known on a probabilistic level, even though the trajectories themselves are deterministic. Drawing different realizations of initial coordinates from $\varrho_N(\tini{x}, \ini{t})$ and evolving them according to \eqref{eq:eom} will eventually result in a final phase space density distribution $\varrho_N(\tens{x}, t)$. Thus at any given time $t>\ini{t}$, the probability density of the system is given by 
 \begin{align}
     \varrho_N(\tens{x}, t)=&\int\md\tini{x}K(\tens{x}, t|\tini{x}, \ini{t})\varrho_N(\tini{x}, \ini{t})\,,\label{eq:evolution}\\
     K(\tens{x}, t|\tini{x}, \ini{t})=&\,\,\delta_D(\tens{x}-\tens{x}_{cl}(t;\tini{x}))\,,\label{eq:transition}
 \end{align}
where $\tens{x}_{cl}(t;\tini{x})$ denotes the classical solution to \eqref{eq:tensoreom} evaluated at time $t$, with parametric dependence on the initial conditions. In the above equation, $K(\tens{x}, t|\tini{x}, \ini{t})$ can be seen as the \textit{transition probability} from the initial phase space density to the final one. Classical determinism, as discussed above, implies that $K(\tens{x}, t|\tini{x}, \ini{t})$ is exactly given by the Dirac-Delta distribution in \eqref{eq:transition}, ensuring that the particles are found at $\tens{x}_{cl}(t;\tini{x})$. \\

A classical, time-independent observable is defined as a phase space function $\mathcal{O}(\tens{x})$ of the $6N$ variables. Its value at time $t$, given by $\mathcal{O}(t)=\mathcal{O}(\tens{x}(t))$, is, by the above discussion, also a random variable. Consequently, its statistical average at time $t$ can be computed as  
\begin{equation}\label{eq:expectationstatphys}
   \mean{\mathcal{O}(t)}=\int\md \tens{x}\mathcal{O}(\tens{x})\varrho_N(\tens{x}, t)\,.
\end{equation}
Inserting equation \eqref{eq:evolution} we find 
\begin{equation}\label{eq:PropagateForExpectation}
    \mean{\mathcal{O}(t)}=\int\md \tens{x}\,\md \tini{x}\mathcal{O}(\tens{x})K(\tens{x}, t|\tini{x}, \ini{t})\varrho_N(\tini{x}, \ini{t})\,.
\end{equation}
Integrating the above equation over the phase space coordinates $\md\tens{x}$ we eventually arrive at an expression for the classical expectation value of $\mathcal{O}$ in terms of the initial phase space density
\begin{equation}\label{eq:expectationsingleoperator}
     \mean{\mathcal{O}(t)}=\int\md\tini{x}\mathcal{O}(\tens{x}_{cl}(t;\tini{x}))\varrho_N(\tini{x}, \ini{t})\,.
\end{equation}
This result can be generalized to correlations between operators $\mathcal{O}_1,\cdots\mathcal{O}_k$ at respective times $t_1,\cdots t_k$,
\begin{equation}
\label{eq:expectation1}
     \mean{\mathcal{O}_1(t_1)\cdots\mathcal{O}_k(t_k)}=\int\md\tini{x}\mathcal{O}_1(\tens{x}_{cl}(t_1;\tini{x}))\cdots\mathcal{O}_k(\tens{x}_{cl}(t_k;\tini{x}))\varrho_N(\tini{x}\ini{t})\,.
\end{equation}

For simple systems with analytic solutions $\tens{x}_{cl}(t;\tini{x})$ equation \eqref{eq:expectation1} gives the prescription on how to compute expectation values. In general, however, the analytical solution to \eqref{eq:eom} is not known, such that we need a different approach. Our main objective in this work will, therefore, be to construct a path integral expression for the transition amplitude \eqref{eq:transition}, thereby facilitating a more quantum-like treatment of the $N$-particle statistical system.

\subsubsection{Macroscopic Observables}\label{sec:MacroscopicObservables}

Let us now briefly define the primary observables and statistical objects that will be of interest to us. Given the probability distribution at some time $t$, we are typically interested in the computation of collective properties of the whole $N$-particle ensemble. Let us therefore introduce the (local) Klimontovich phase space density,
\begin{equation}\label{eq:Klimontovich}
    f(\vec{  {q}}_1, \vec{  {p}}_1, t_1)=\sum_{i=1}^N\delta_D(\vec{  {q}}_1-\vec{q}_i(t_1))\delta_D(\vec{  {p}}_1-\vec{p}_i(t_1))\equiv \sum_{i=1}^N\delta_D(x_1-x_i(t_1))\,,
\end{equation}
which counts the number of microscopic particles of the ensemble, living in $6N$-dimensional $\Gamma$-space, that occupy a phase space point  $  {x}_1=(\vec{  {q}}_1, \vec{  {p}}_1)$ in $6$-dimensional $\mu$-space at time $t_1$.
Each particle thus contributes a term $\delta_D(\vec{  {q}}_1-\vec{q}_i(t_1))\delta_D(\vec{  {p}}_1-\vec{p}_i(t_1))$ to the local macroscopic phase space density. In the following, we will refer to $  {x}=(\vec{  {q}}, \vec{  {p}})$ and $x_i=(\vec{q}_i, \vec{p}_i)$ as the macroscopic and microscopic variables respectively. The phase space density is consequently a function of external macroscopic variables $(\vec{q}_1, \vec{p}_1)$ and the microscopic variables $\tens{x}(t_1)$ describing the state of the system at time $t_1$. For better readability however, we generally keep the latter dependence implicit as the state of the system is already characterized by the time $t_1$. It should also be noted that the index $i$ in equation \eqref{eq:Klimontovich} is a particle label, while the index $1$ attached to the $\mu$-space coordinates of $f(\vec{  {q}}_1, \vec{  {p}}_1, t_1)$ refers to the phase space point at which we read off the Klimontovich phase space density. We trust that the appropriate meaning of the indices should become apparent from context. Further observables such as the physical density $\rho(\vec{  {q}}_1, t_1)$ or the momentum density $\vec{\Pi}(\vec{  {q}}_1, t_1)$ can be obtained from \eqref{eq:Klimontovich} according to 
\begin{align}
    \rho(\vec{  {q}}_1, t_1)=&\sum_{i=1}^N\delta_D(\vec{  {q}}_1-\vec{q}_i(t_1))=\int\md^3  {p}_1f(\vec{  {q}}_1, \vec{  {p}}_1, t_1)\,,\label{eq:particledensity}\\
    \vec{\Pi}(\vec{  {q}}_1, t_1)=&\sum_{i=1}^N\vec{p}_i(t_1)\,\delta_D(\vec{  {q}}_1-\vec{q}_i(t_1))=\int\md^3  {p}_1\,\vec{  {p}}_1\,f(\vec{  {q}}_1, \vec{  {p}}_1, t_1)\label{eq:momentumdensity}\,.
\end{align}
In general we will also be interested in higher order correlations of \eqref{eq:particledensity} and \eqref{eq:momentumdensity}, connecting different points $\vec{  {q}}_1\cdots\vec{  {q}}_k$ and times $t_1\cdots t_k$ in configuration space, such as unequal-time two-point density correlations. 
We will thus be looking at observables of the form 
\begin{equation}
    \mathcal{O}(\vec{  {q}}_1, t_1, \cdots\vec{  {q}}_k, t_k)=\int\md^3  {p}_1\cdots\md^3  {p}_k F(\vec{  {p}}_1,\cdots,\vec{  {p}}_k)f(\vec{  {q}}_1, \vec{  {p}}_1, t_1)\cdots f(\vec{  {q}}_k, \vec{  {p}}_k, t_k)\,,\label{eq:generalDensityExpectation}
\end{equation}
where $F(\vec{  {p}}_1,\cdots,\vec{  {p}}_k)$ is some function of the momenta.
The collective properties of the system are then captured by taking the average over the microscopic degrees of freedom at the respective times, giving  
\begin{equation}
    \mean{\mathcal{O}(\vec{  {q}}_1, t_1,\cdots\vec{  {q}}_k,t_k)}=\int\md^3  {p}_1\cdots\md^3  {p}_k F(\vec{  {p}}_1,\cdots,\vec{  {p}}_k)\mean{f(\vec{  {q}}_1, \vec{  {p}}_1, t_1)\cdots f(\vec{  {q}}_k, \vec{  {p}}_k, t_k)}\,.
    \label{eq:expectation_obs}
\end{equation}
Consequently, our main task reduces to the computation of averages of products of the Klimontovich phase space density, which can be done within the path integral framework.

\subsubsection{Reduced phase space densities}\label{sec:reducedphasespacedensities}

For a given phase space density $\varrho_N(\tens{x}, t)$ we define the $s$-particle reduced phase space density according to 
\begin{equation}\label{eq:reduceddensity}
    f_s(x_1,\cdots,x_s,t)\equiv\frac{N!}{(N-s)!}\int\md^3x_{s+1}\cdots\md^3x_N\varrho(x_1,\cdots,x_s,x_{s+1},\cdots,x_N, t)\,.
\end{equation}
It captures the statistical information of an $s$-particle subsystem at time $t$. Depending on the concrete observable to compute we may thus describe the system using the reduced phase space densities $f_s$ instead of the full phase space statistics $\varrho_N(\tens{x},t)$ which reduces the amount of complexity considerably. Due to the indistinguishable nature of the particles and the resulting symmetry of $\varrho_N(\tens{x},t)$ under the exchange of any two particles, the definition of the reduced densities in \eqref{eq:reduceddensity} is equivalent to 
\begin{equation}\label{eq:reduceddensityalternative}
    f_s(  {x}_1,\cdots,  {x}_s,t)=\int\md\tens{x}\sum_{i_1\neq \cdots\neq i_s=1}^N\delta_D(  {x}_1-x_{i_1})\cdots\delta_D(  {x}_s-x_{i_s})\varrho_N(\tens{x},t)\,.
\end{equation}
By equation \eqref{eq:expectationstatphys} the equal time expectation value of a product of $k$-phase space densities at time $t$ is given by 
\begin{equation}\label{eq:standardexpectationvalue}
    \mean{f(\vec{q}_1, \vec{p}_1, t)\cdots f(\vec{q}_k, \vec{p}_k, t)}=\int\md\tens{x}f(\vec{q}_1, \vec{p}_1; \tens{x})\cdots f(\vec{q}_k, \vec{p}_k;\tens{x})\varrho_N(\tens{x}, t)\,,
\end{equation}
where we made the dependence on the microscopic degrees of freedom explicit. Using \eqref{eq:reduceddensityalternative}, this integral decomposes into reduced densities, such that 
\begin{equation}\label{eq:decomposition}
    \mean{f(\vec{q}_1, \vec{p}_1, t)\cdots f(\vec{q}_k, \vec{p}_k, t)}=\cdots+f_k(  {x}_1,\cdots,  {x}_k,t)\,.
\end{equation}
The ellipsis in the above equation stands for all terms which involve reduced phase space densities of degree lower than $k$. They arise from terms in the product of sums in \eqref{eq:standardexpectationvalue} in which at least two indices are identified, \eg for $k=2$ we find 
\begin{align}
    \mean{f(\vec{  {q}}_1, \vec{  {p}}_1,t)f(\vec{  {q}}_2, \vec{  {p}}_2,t)}=&\,\delta_D(  {x}_1-  {x}_2)f_1(  {x}_1,t)+f_2(  {x}_1,  {x}_2, t)\label{eq:k=2}\,.
\end{align}
We call the respective terms in the decomposition in \eqref{eq:decomposition} the $r$-particle contribution with $1\leq r\leq k$ to the $k$-point correlation function, as they depend on the statistics of $r$-particle subsets. On an intuitive level, they correspond to the possibility of randomly picking the same particle multiple times in order to measure the $k$-point correlation function. Thus, in order to study the evolution of correlation functions \eqref{eq:decomposition}, it is enough to understand the time-evolution of the $s$-particle densities. In the case of equal-time correlation functions this corresponds to the standard approach leading to the BBGKY hierachy. For comparison to our approach we summarize the key points in appendix \ref{sec:BBGKY}.\\

If our system is in an uncorrelated state, \ie the particles are distributed in a statistically independent way sucht that the probability of finding a particle at some point is independent of any of the other particles, the $s$-particle reduced distribution function takes the form 
\begin{equation}
    f_s(x_1, \cdots, x_s, t)=\prod_{j=1}^sf_1(x_j, t)\,.
\end{equation}
In order to measure how much the system differs from such an uncorrelated state, we decompose a set of $s$-particles into all possible disjoint subsets which contain at least one particle. Each such cluster represents a group of mutually correlated particles. The clusters themselves are statistically independent from each other such that we can write, for instance,
\begin{align}
    f_2(x_1, x_2, t)=&f_1(x_1, t)f_1(x_2, t)+g_2(x_1, x_2, t)\,,\label{eq:twopointconnected}\\
    f_3(x_1, x_2, x_3, t)=&f_1(x_1, t)f_1(x_2, t)f_1(x_3, t)+f_1(x_1, t)g_2(x_2, x_3, t)\nonumber\\&+f_1(x_2, t)g_2(x_1, x_3, t)+f_1(x_3, t)g_2(x_1, x_2, t)\label{eq:threepointconnected}\\&+g_3(x_1, x_2, x_3, t)\,,\nonumber\\
    \vdots\nonumber
\end{align}
where $g_r(x_1, \cdots,x_r, t)$ denotes the irreducible $r$-particle distribution function containing the subset of correlated particles within $f_s$. Importantly, this decomposition of the total phase space density $\varrho_N(\tens{x}, t)$ into reducible and irreducible $s$-particle distributions can be done in general at any time. Our approach described in the following will make use of it in order to decompose the initial phase space density into initial correlations, which are then propagated in time.

\subsection{Time Evolution via Classical Path Integrals}\label{sec:Construction}

In quantum mechanics, the intuitive picture behind the path integral representation of the transition probability is to allow the particles to travel all possible trajectories connecting two fixed end points in configuration space. Each trajectory is then functionally integrated over, weighted by the exponential of the classical action. This idea can be transferred to the classical situation as described in the following. To keep things short we mention the main ideas here and postpone a more detailed derivation and further discussion to appendix \ref{sec:discretization} and especially the literature \cite{Abrikosov:2004cf, Gozzi:1993tm, kleinert2009path}. \\

The classical transition probability \eqref{eq:transition} between an initial time $\ini{t}$ and a final time $\fin{t}$ satisfies the important relation
\begin{equation}
    K(\tfin{x}, \fin{t}|\tini{x}, \ini{t})=\int \md x  K(\tfin{x}, \fin{t}|\tens{x}, t)\,K(\tens{x}, t|\tini{x}, \ini{t})\,,
\end{equation}
which can be used to decompose the evolution into $N$ steps. The Dirac-Deltas then ensure that at every time step the particles can be found on the respective classical trajectories, which in the continuum limit $N\rightarrow\infty$ converges to the classical trajectory and thus leads to the path integral representation of the classical transition amplitude,
\begin{equation}\label{eq:PathPropagator}
    K(\tfin{x}, \fin{t}|\tini{x}, \ini{t})=\int\limits_{\tini{x}}^{\tfin{x}} \mathcal{D}\tens{x}(t)\delta_D[\tens{x}(t)-\tens{x}_{cl}(t;\tini{x})]\,,
\end{equation}
where the functional integration measure $\mathcal{D}\tens{x}(t)\delta_D[\tens{x}(t)-\tens{x}_{cl}(t;\tini{x})]$ gives weight $0$ to all trajectories but the classical one, with fixed end points at $\tini{x}$ and $\tfin{x}$. This ensures classical determinism in the path integral approach. 
We can now perform a transformation $\tens{x}\rightarrow \tens{\mathcal{E}}[\tens{x}]$ inside the functional Dirac-Delta distribution in order to explicitly use the equations of motion instead of the classical solution, which is not explicitly known in general. To this end we make use of the functional relation 
\begin{equation} \label{eq:variabletransform}
    \delta_D[\tens{x}(t)-\tens{x}_{cl}(t;\tini{x})]=\delta_D[\tens{\mathcal{E}}[\tens{x}(t)]]\cdot \mathcal{J}\,,
\end{equation}
where $\mathcal{J}=\left|\frac{\delta \tens{\mathcal{E}}[\tens{x}]}{\delta \tens{x}}\right|$ is the Jacobian emerging from the change of variables. In appendix \ref{sec:discretization} we compute the Jacobian explicitly during the rigorous discretization procedure and discuss how its explicit form depends on the convention for the discretization. Furthermore, we show that by using the pre-point prescription in the discretized theory which corresponds to defining $\Theta(0)=0$, the determinant is constant and can be absorbed in the functional integration measure. This phenomenon is also known from the theory of stochastic dynamics and has been discussed in great detail in the literature \cite{Nakazato:1990kk, Ezewa}. Keeping this in mind, we plug \eqref{eq:variabletransform} into \eqref{eq:PathPropagator}, absorb the Jacobian and introduce a doublet $\tens{\chi}=(\tens{\chi_q}, \tens{\chi_p})^\top$ in order to express the remaining Dirac-Delta distribution in terms of its Fourier transform. We then arrive at the expression 
\begin{equation}\label{eq:PropagatorPathIntegral}
     K(\tfin{x}, \fin{t}|\tini{x}, \ini{t})=\int\limits_{\tini{x}}^{\tfin{x}} \mathcal{D}\tens{x}(t)\mathcal{D}\tens{\chi}(t)\,\exp\Bigg[\mi\int\limits_{\ini{t}}^{\fin{t}}\md t\tens{\chi}(t)\tens{\cdot}\left(\dot{\tens{x}}(t)-\tens{\omega\cdot\nabla}H(\tens{x}(t))\right)\Bigg]\,,
\end{equation}
where we explicitly inserted the equations of motion \eqref{eq:eom}. Note that the functional integration is performed over all $\tens{x}(t)$ trajectories with fixed endpoints and over all $\tens{\chi}(t)$ trajectories with no restrictions on the end points. The representation \eqref{eq:PropagatorPathIntegral} for the classical transition probability together with \eqref{eq:evolution} opens the possibility for a functional approach to the time evolution of classical $N$-particle systems in and out of equilibrium, depending on the explicit form of the initial phase space density $\varrho_N(\tini{x}, \ini{t})$. The out of equilibrium case will be the main goal of our application. \\

As is well known, the time evolution of a classical statistical system is governed by the \textit{Liouville} equation, 
\begin{equation}\label{eq:Liouville}
    \partial_t\,\varrho_N(\tens{x}, t)= \,-\mi \hat{L}\,\varrho_N(\tens{x}, t)\,,
\end{equation}
where the Liouville operator is defined by 
\begin{align}
    \hat{L}=&\mi\tens{\nabla}H(\tens{x})\tens{\cdot}\tens{\omega}\tens{\cdot}\tens{\nabla}=\mi\tens{\nabla_q}H(\tens{x})\cdot\tens{\nabla_p} - \mi\tens{\nabla_p}H(\tens{x})\cdot\tens{\nabla_q}\,.
\end{align}
The formal solution to \eqref{eq:Liouville} is then given by 
\begin{equation}
    \varrho_N(\tens{x}, t) = \hat{U}(t, \ini{t}) \, \varrho_N(\tini{x}, \ini{t})
\end{equation}
with the classical time evolution operator\footnote{Here we assume a time independent Hamilton function for simplicity.} 
\begin{equation}\label{eq:classicalTimeEvolution}
     \hat{U}(t, \ini{t})= \, \e^{-\mi(t-\ini{t})\hat{L}}\,.
\end{equation}
In fact, as has been shown in \cite{Gozzi1989, Abrikosov:2004cf}, \eqref{eq:PropagatorPathIntegral} corresponds to the path integral \textit{representation} of the time evolution operator \eqref{eq:classicalTimeEvolution}. Indeed, expanding \eqref{eq:PropagatorPathIntegral} for small time steps, one recovers the Liouville equation \eqref{eq:Liouville}. Thus, it is the functional formulation of the associated operator formalism famously known as the Koopman-von Neumann formalism \cite{Mauro:2001rm, vonNeumann, Koopman}, which describes classical mechanics as an operatorial theory on the Hilbert space of complex square-integrable classical wavefunctions, in complete analogy to quantum mechanics. As is clear form the exponent in \eqref{eq:PropagatorPathIntegral}, $\tens{\chi}$ plays the role of the conjugate momenta to $\tens{x}$ in an extended phase space \cite{Abrikosov:2004cf}. With that intuition in mind we finally note that the whole path integral construction is very similar to the one used in stochastic dynamics describing the Brownian motion of a particle subject to a Langevin equation \cite{Das:2014jia}. However, in contrast to the latter case, we do not have a noise term spoiling the determinism of the time evolution. The only probabilistic element in our system enters through the initial phase space density distribution. It is thus the initial state which describes the statistical properties of the system, and thus determines whether it is in an equilibrium state or not.\\

From the path integral representation of the transition probability we can now construct analogous expressions for time ordered expectation values \eqref{eq:expectation1} in the following way: We insert the observables inside the path integration which enforces the microscopic classical trajectories. Furthermore, we integrate out the final microscopic degrees of freedom $\tfin{x}$ and formally let $t_f\rightarrow\infty$, which leaves us with 
\begin{align}\label{eq:PathIntegralExpectation}
    \mean{\hat{\mathcal{T}}\mathcal{O}_1(t_1)\cdots\mathcal{O}_k(t_k)}=\int\md\tfin{x}&\int\md\tini{x}\varrho_N(\tini{x}, \ini{t})\times\\
    &\times\int\limits_{\tini{x}}^{\tfin{x}} \mathcal{D}\tens{x}(t)\mathcal{D}\tens{\chi}(t)\,\mathcal{O}_1(\tens{x}(t_1))\cdots\mathcal{O}_k(\tens{x}(t_k))\,\exp\Bigg[\mi\mathcal{S}[\tens{x}(t),\tens{\chi}(t)]\Bigg]\,.\nonumber
\end{align}
The effective classical action is given by 
\begin{equation}\label{eq:effectiveaction}
    \mathcal{S}[\tens{x}(t),\tens{\chi}(t)]=\int\limits_{\ini{t}}^{\infty}\md t\,\tens{\chi}(t)\tens{\cdot}\Big(\dot{\tens{x}}(t)-\tens{\omega\cdot\nabla}H(\tens{x}(t))\Big)\,,
\end{equation}
which, inserting \eqref{eq:Hamiltonfunction}, can be split into a free and an interacting part 
\begin{align}\label{eq:interactionaction}
    \mathcal{S}[\tens{x}(t),\tens{\chi}(t)]=&\mathcal{S}_0[\tens{x}(t),\tens{\chi}(t)] + \mathcal{S}_\mathrm{I}[\tens{x}(t),\tens{\chi}(t)]\,,
\end{align}
with 
\begin{equation}
    \mathcal{S}_0[\tens{x}(t),\tens{\chi}(t)]=\int\limits_{\ini{t}}^{\infty}\md t\,\Big\{\tens{\chi}(t)\tens{\cdot}\Big(\dot{\tens{x}}(t)-\tens{\omega\cdot\nabla}H_0(\tens{x}(t))\Big)\Big\}\,,\,\,\,\,\,\,\,\mathcal{S}_\mathrm{I}[\tens{x}(t),\tens{\chi}(t)]=\int\limits_{\ini{t}}^{\infty}\md t\,\tens{\chi_p}(t)\tens{\cdot}\tens{\nabla_q}V(\tens{q}(t), t)\,.
\end{equation}
There are several ways to proceed with the computation of \eqref{eq:PathIntegralExpectation}. The most straight forward way is to introduce sources for $\tens{q}$ and $\tens{\chi_p}$ and to pull the interactions and the operators out of the path integral by replacing the variables with appropriate functional derivatives. This has been done in \cite{Bartelmann_2014,Pixius_2022, Penco_2006} and results in a perturbation theory in terms of the interaction potential, leading to particle trajectories which deviate from their inertial free motion governed by $H_0(\tens{x})$. If the interactions are weak, the full trajectory can be approximated very accurately by low orders of the perturbative expansion. If, however, the operators one is interested in are non-polynomial functions of the underlying degrees of freedom $\tens{x}(t)$, this approach requires Taylor expanding these functions, introducing a further approximation and consequently a loss of accuracy. In the next section we therefore take a different path which, as we will see, is more closely related to standard kinetic theory and exhibits similarities to many-body quantum mechanics, such that it eventually results in a better approximation of the time evolution of the density. \\

\subsection{Microscopic Perturbation Theory}\label{sec:MicroscopicTheory}
In this section we describe the perturbative approach to the computation of expectation values. Without loss of generality we assume that the times of the operators are ordered, $\fin{t}> t_1\geq\cdots\geq t_k\geq\ini{t}$. As described in appendix \ref{sec:OperatorInPathint}, the path integral may then be split at the respective points $\tens{x}(t_k)\equiv\tens{x}_k$, \ie 
\begin{align}\label{eq:ManyOperatorsinPathintegral}
    \int\limits_{\tini{x}}^{\tfin{x}} &\mathcal{D}^\prime\tens{x}(t)\mathcal{D}\tens{\chi}(t)\,\mathcal{O}_1(\tens{x}(t_1))\cdots\mathcal{O}_k(\tens{x}(t_k))\,\exp\Bigg[\mi\mathcal{S}[\tens{x}(t),\tens{\chi}(t)]\Bigg]=\\
    &\int\md\tens{x}_1\cdots\md\tens{x}_k\,K(\tfin{x}, \fin{t}|\tens{x}_1, t_1)\mathcal{O}_1(\tens{x}_1)K(\tens{x}_1, t_1|\tens{x}_2, t_2)\cdots\mathcal{O}_k(\tens{x}_1)K(\tens{x}_k, t_k|\tini{x}, \ini{t})\,.
\end{align}
Note that the integration $\md \tens{x}_k$ signifies integration over the whole ensemble at time $t_k$. Furthermore, due to the full propagator being a Dirac-Delta distribution of the classical trajectories \eqref{eq:transition}, we find 
\begin{equation}
    1=\int\md\tfin{x}K(\tfin{x}, \fin{t}|\tens{x}_1, t_1)\,,
\end{equation}
such that we can integrate out $\tfin{x}$ in \eqref{eq:PathIntegralExpectation} which thus leaves us with
\begin{align}\label{eq:UnequalExpectationPropagator}
    \mean{\hat{\mathcal{T}}\mathcal{O}_1(t_1)\cdots\mathcal{O}_k(t_k)}=
    \int\md\tens{x}_1\cdots\md\tens{x}_k&\md\tini{x}\mathcal{O}_1(\tens{x}_1)K(\tens{x}_1, t_1|\tens{x}_2, t_2)\cdots\\&\cdots K(\tens{x}_{k-1}, t_{k-1}|\tens{x}_k, t_k)\mathcal{O}_k(\tens{x}_k)K(\tens{x}_k, t_k|\tini{x}, \ini{t})\varrho_N(\tini{x}, \ini{t})\,.\nonumber
\end{align}
Equation \eqref{eq:UnequalExpectationPropagator} is a direct generalization of \eqref{eq:PropagateForExpectation}. We evolve the initial density with the full propagator until we take the first average at time $t_k$ by integrating over the ensemble $\tens{x}_k$. The density is then further evolved until we take the next average over $\tens{x}_{k-1}$ at $t_{k-1}$. This procedure is repeated up to $t_{1}$ where we take the last average over $\tens{x}_{1}$ giving the full unequal-time correlation between the observables $\mathcal{O}_1\,,\cdots\,,\mathcal{O}_k$. The microscopic perturbation theory now consists 
 of expanding the full propagator \eqref{eq:PropagatorPathIntegral} in a Dyson series and sorting the resulting terms in \eqref{eq:UnequalExpectationPropagator} by powers of the potential. Let us therefore write 
\begin{align}\label{eq:DysonSeries}
    K(\tfin{x}, \fin{t}|\tini{x}, \ini{t})=&\int\limits_{\tini{x}}^{\tfin{x}} \mathcal{D}^\prime\tens{x}(t)\mathcal{D}\tens{\chi}(t)\,\exp\Bigg[\mi\mathcal{S}_0[\tens{x}(t),\tens{\chi}(t)] + \mi\mathcal{S}_\mathrm{I}[\tens{x}(t),\tens{\chi}(t)]\Bigg]\\
    =&\int\limits_{\tini{x}}^{\tfin{x}} \mathcal{D}^\prime\tens{x}(t)\mathcal{D}\tens{\chi}(t)\,\exp\Bigg[\mi\mathcal{S}_0[\tens{x}(t),\tens{\chi}(t)]\Bigg]\sum_{n=0}^\infty\frac{\mi^n}{n!}\Big(\mathcal{S}_\mathrm{I}[\tens{x}(t),\tens{\chi}(t)]\Big)^n\\
    =&\sum_{n=0}^\infty K_n(\tfin{x}, \fin{t}|\tini{x}, \ini{t})\,.
\end{align}
In the following we occasionally abbreviate 
$$K_n(k|l)\equiv K_n(\tens{x}_k, t_k| \tens{x}_l, t_l)\,,\,\,\,\,\, K_n(f|i)\equiv K_n(\tfin{x}, \fin{t}|\tini{x}, \ini{t})\,,$$
in order to condense the notation. We easily find the zeroth order propagator,
\begin{align}
    K_0(f|i)=&\int\limits_{\tini{x}}^{\tfin{x}} \mathcal{D}^\prime\tens{x}(t)\mathcal{D}\tens{\chi}(t)\,\exp\Bigg[i\mathcal{S}_0[\tens{x}(t),\tens{\chi}(t)]\Bigg]\\
    =&\delta_D\left(\tfin{x}-\tens{x}_{cl,0}(\fin{t};\tini{x})\right)\\
    =&\delta_D\left(\tfin{q}-\tini{q}-\frac{\tini{p}}{m}(\fin{t}-\ini{t})\right)\delta_D\left(\tfin{p}-\tini{p}\right)\,,
\end{align}
describing the free evolution of the particles. In order to restrict ourselves to causal propagation, we multiply the free propagator with $\Theta(\fin{t}-\ini{t})$ such that
\begin{align}\label{eq:FreePropagator}
    K_0(f|i)=\delta_D\left(\tfin{q}-\tini{q}-\frac{\tini{p}}{m}(\fin{t}-\ini{t})\right)\delta_D\left(\tfin{p}-\tini{p}\right)\,\Theta(\fin{t}-\ini{t})\,,
\end{align}
\ie the density evolves from earlier to later times. Inserting the explicit form of the two particle potential \eqref{eq:TwoParticlePotential} into $\mathcal{S}_\mathrm{I}[\tens{x}(t),\tens{\chi}(t)]$ we find for $K_1$
\begin{align}
    K_1(f|i)=\mi \int\limits_{\tini{x}}^{\tfin{x}}\mathcal{D}^\prime\tens{x}(t)\mathcal{D}\tens{\chi}(t)\,\exp\Bigg[\mi \mathcal{S}_0[\tens{x}(t),\tens{\chi}(t)]\Bigg]\int_{\ini{t}}^{\fin{t}}\md \Bar{t}\sum_{i\neq j=1}^N\vec{\chi}_{p_i}(\Bar{t})\cdot\nabla_{q_i}v(|\vec{q}_i(\Bar{t})-\vec{q}_j(\Bar{t})|,\Bar{t})\,.
\end{align}
In appendix \ref{sec:OperatorInPathint} we show that this expression can be brought into the form 
\begin{align}\label{eq:FirstOrderPropagator}
    K_1(f|i)=\int_{\ini{t}}^{\fin{t}}\md \Bar{t}_1\,\int\md\Bar{\tens{x}}_1\,K_0(f|\Bar{1})\,\hat{\mathcal{V}}(\Bar{1})\,K_0(\Bar{1}|i)\,,
\end{align}
where the potential operator $\hat{V}(1)$ defined as 
\begin{equation}
    \hat{\mathcal{V}}(1)\equiv\hat{\mathcal{V}}(\tens{x}_1, t_1)\equiv\sum_{i\neq j=1}^N\nabla_{\vec{q}_{1_i}}v(|\vec{q}_{1_i}-\vec{q}_{1_j}|,t_1)\cdot\nabla_{\vec{p}_{1_i}}
\end{equation}
acts on the free propagator on its right and describes the reaction of the system to inter particle interactions. Again, we recall that $x_{1_i}$ refers to the $i$-th particle at time $t_1$. To second order we analogously find 
\begin{align}\label{eq:SecondOrderPropagator}
    K_2(f|i)=\int_{\ini{t}}^{\fin{t}}\md \Bar{t}_1\int_{\ini{t}}^{\Bar{t}_1}\md \Bar{t}_2\,\int\md\Bar{\tens{x}}_1\md\Bar{\tens{x}}_2\,K_0(f|\Bar{1})\,\hat{\mathcal{V}}(\Bar{1})\,K_0(\Bar{1}|\Bar{2})\,\hat{\mathcal{V}}(\Bar{2})K_0(\Bar{2}|i)\,.
\end{align}
 Attention has to be paid to the time ordering, which we have made explicit in the boundaries of the time integration. In general there are $n!$ possibilities of ordering $n$ interactions, which exactly cancels the $\frac{1}{n!}$ appearing in the expansion of the Dyson Series \eqref{eq:DysonSeries}. Iterating the above procedure, we eventually arrive at the familiar integral equation for the full propagator
 \begin{align}\label{eq:LippmanSchwinger}
    K(f|i)=K_0(f|i) + \int_{\ini{t}}^{\fin{t}}\md \Bar{t}_1\,\int\md\Bar{\tens{x}}_1K_0(f|\Bar{1})\,\hat{\mathcal{V}}(\Bar{1})\,K(\Bar{1}|i)\,.
\end{align}
If we multiply \eqref{eq:LippmanSchwinger} with the initial phase space density distribution and integrate over the initial state, we find the following \textit{Lippmann-Schwinger} like equation for the full phase space density
\begin{align}\label{eq:LippmanSchwingerDensity}
    \varrho_N(\tfin{x}, \fin{t})=\varrho_N^{(0)}(\tfin{x}, \fin{t}) + \int_{\ini{t}}^{\fin{t}}\md \Bar{t}_1\,\int\md\Bar{\tens{x}}_1K_0(\tfin{x}, \fin{t}|\Bar{\tens{x}}_1, \Bar{t}_1)\,\hat{\mathcal{V}}(\Bar{\tens{x}}_1, \Bar{t}_1)\,\varrho_N(\Bar{\tens{x}}_1, \Bar{t}_1)\,,
\end{align}
where $\varrho_N^{(0)}(\tfin{x}, \fin{t})$ is the freely evolving density. Equation \eqref{eq:LippmanSchwingerDensity} describes how the full final density arises as a scattering process\footnote{The analogy with the scattering theory of many-body quantum mechanics is not a coincidence. In quantum mechanics the Lippmann-Schwinger equation describes a scattering process of a wave function with a potential \wrt a free evolution. Famously, classical mechanics ca be formulated within the same framework as quantum mechanics, know as Koopman-von Neumann formalism \cite{Mauro:2001rm}. Within this framework it can be shown that a postulated classical wavefunction obeys the same evolution equation as the classical phase space density. Thus the scattering process described by the wave function translates to a scattering process of the density.} in which the incoming full phase space density is deformed by the interaction operator $\hat{\mathcal{V}}(\tens{x}, t)$ due to the presence of inter particle interactions. The first order approximation of this equation thus corresponds to the analogy of the Born approximation in scattering theory. Importantly, integrating \eqref{eq:LippmanSchwingerDensity} over the full final phase space, and using the normalization of the free and the full phase space density, we find the condition
\begin{equation}
   \int_{\ini{t}}^{\fin{t}}\md \Bar{t}_1\,\int\md\Bar{\tens{x}}_1\,\hat{\mathcal{V}}(\Bar{\tens{x}}_1, \Bar{t}_1)\,\varrho_N(\Bar{\tens{x}}_1, \Bar{t}_1)=0\,,
\end{equation}
from which we obtain the normalization condition
\begin{equation}\label{eq:NormalizationCondition}
  \int\md x_{i}\,\nabla_{\vec{q}_{i}}v(|\vec{q}_i-\vec{q}_j|,t)\cdot\nabla_{\vec{p}_i}\,\varrho_N(\tens{x}, t)=0\,,
\end{equation}
which also holds order by order in perturbation theory. It can also be verified by partially integrating over the momenta $\vec{p}_i$ and using the fact, that the density vanishes at the boundaries in momentum space. Thus, all terms vanish in which the momentum of particle $i$, affected by the force induced by particle $j$, is integrated over. This normalization argument will be used in the following in order to find all contributing terms in perturbation theory. \\

Keeping the above in mind, we can further simplify the integrals over $\tens{x}_1$ and $\tens{x}_2$ in \eqref{eq:FirstOrderPropagator} and \eqref{eq:SecondOrderPropagator}, respectively. Performing the integrals over the Dirac-Delta distributions, we find 
\begin{align}\label{eq:FirstOrderIntegrated}
    K_1(1|2)=&\int_{t_2}^{t_1}\md \Bar{t}_1\,\sum_{i\neq j=1}^N\nabla_{\vec{q}_{1_i}}v\left(\left|\vec{q}_{1_i}(\Bar{t}_1;t_1)-\vec{q}_{1_j}(\Bar{t}_1;t_1)\right|,\Bar{t}_1\right)\cdot \nabla_{\vec{p}_{1_i}}\,K_0(1|2)\\
    =&\int_{t_2}^{t_1}\md \Bar{t}_1\,\sum_{i\neq j=1}^N\,\hat{\mathcal{L}}_{i\ShortLArrow j}(t_1, t_1; \Bar{t}_1)\,K_0(1|2)\,,
\end{align}
where we introduced the short-hand notation 
\begin{equation}
       \vec{  {q}}_1(t^\prime;t_1)=\vec{  {q}}_1-\frac{\vec{  {p}}_1}{m}(t_1-t^\prime)\,,
 \end{equation}
describing the history of a freely evolving particle as seen from $  {x}_1=(\vec{  {q}}_1, \vec{  {p}}_1)$ at time $t_1$ and also defined the interaction operator 
\begin{equation}\label{eq:InteractionOperator}
    \hat{\mathcal{L}}_{i\ShortLArrow j}(t_1, t_1; \Bar{t}_1)\equiv\nabla_{\vec{q}_{1_i}}v\left(\left|\vec{q}_{1_i}(\Bar{t}_1;t_1)-\vec{q}_{1_j}(\Bar{t}_1;t_1)\right|,\Bar{t}_1\right)\cdot \nabla_{\vec{p}_{1_i}}\,.
\end{equation}
The latter describes how particle $j$ exhibits a force on particle $i$ at time $\Bar{t}_1$. Both particles positions are described by their backwards evolution as seen from $t_1$. As expected, $K_1$ describes the deviation from the free density propagation due to the interaction between $i$ and $j$ integrated over all possible times $\Bar{t}_1$. The corresponding expression for $K_2(1|2)$ reads
\begin{align}\label{eq:SecondOrderIntegrated}
    K_2(1|2)=\int_{t_2}^{t_1}\md \Bar{t}_1\,\int_{t_2}^{\Bar{t}_1}\md \Bar{t}_2\sum_{i\neq j=1}^N\,\hat{\mathcal{L}}_{i\ShortLArrow j}(t_1, t_1; \Bar{t}_1)\,\left[\sum_{k\in\{i,j\}}\sum_{\substack{l=1 \\ l\neq k}}^N\,\hat{\mathcal{L}}_{k\ShortLArrow l}(t_1, t_1; \Bar{t}_2)\right]\,K_0(1|2)\,,
\end{align}
and describes how a second interaction operator deforms the propagation of the initial density which has already been deformed by an interaction at earlier times. Importantly, the particle with index $k$, which is deflected in the earlier interaction has to be either the deflected particle $i$ or the deflecting particle $j$ in the later interaction. This is again due to the normalization condition \eqref{eq:NormalizationCondition}. The same logic applies to all higher orders. \\

As an illustrative example we shall present the computations of the Klimontovich phase space density correlation functions
$\mean{f(\vec{  {q}}_1, \vec{  {p}}_1, t_1)}$ and $\mean{f(\vec{  {q}}_1, \vec{  {p}}_1, t_1)f(\vec{  {q}}_2, \vec{  {p}}_2, t_2)}$ up to second order in the perturbation theory.

\subsubsection{Free Theory}\label{sec:FreeTheory}

In the simplest case we only have free propagators appearing. The expectation value of the Klimontovich phase space density at time $t_1$ is obtained by averaging over the ensemble $\tens{x}(t_1)\equiv\tens{x}_1$ at the respective time, \ie 
\begin{align}
    \mean{f(\vec{{q}}_1, \vec{{p}}_1, t_1)}^{(0)}=&\int\md \tens{x}_1\md \tini{x}f(\vec{  {q}}_1, \vec{{p}}_1, t_1)K_0(\tens{x}_1, t_1|\tini{x},\ini{t})\varrho_N(\tini{x}, \ini{t})\,.
\end{align}
Inserting the definitions \eqref{eq:Klimontovich} and \eqref{eq:FreePropagator} of the Klimontovich density and the free propagator, we easily compute 
\begin{align}
    \mean{f(\vec{{q}}_1, \vec{  {p}}_1, t_1)}^{(0)}=&\sum_{i=1}^N\int\md \tini{x}_i\delta_D\left(\vec{  {q}}_1-\ini{\vec{q}\,}_i-\frac{\ini{\vec{p}\,}_i}{m}(t_1-\ini{t})\right)\delta_D(\vec{  {p}}_1-\ini{\vec{p}\,}_i)\varrho_N(\tini{x}, \ini{t})\\
    =&f_1\left(\vec{  {q}}_1-\frac{\vec{  {p}}_1}{m}(t_1-\ini{t}), \vec{  {p}}_1, \ini{t}\right)\\\equiv &f_1^{(0)}(  x_1, t_1)\label{eq:freeOnePoint}\,,
\end{align}
where in the first step we inserted the definition of the one-particle reduced initial phase space density \eqref{eq:reduceddensityalternative} and in the last line we defined the freely evolved one-particle reduced distribution function. The final result is therefore, as expected, simply the shifted initial phase space density. A similar computation for $\mean{\hat{\mathcal{T}}f(\vec{  {q}}_1, \vec{  {p}}_1, t_1)f(\vec{  {q}}_2, \vec{  {p}}_2, t_2)}$ reveals 
\begin{align}
    \mean{\hat{\mathcal{T}}f(\vec{  {q}}_1, \vec{  {p}}_1, t_1)f(\vec{  {q}}_2, \vec{  {p}}_2, t_2)}^{(0)}=&\int\md \tens{x}_1\,\md \tens{x}_2\,\md \tini{x}f(\vec{  {q}}_1, \vec{  {p}}_1, t_1)\,K_0(1|2)\,f(\vec{  {q}}_2, \vec{  {p}}_2, t_2)\,K_0(2|i)\,\varrho_N(\tini{x}, \ini{t})\\
    =&\delta_D(  {x}_2-  {x}_1(t_2;t_1))f_1^{(0)}(  {x}_1, t_1)+f_2^{(0)}( x_1, t_1, x_2, t_2)\label{eq:freeTwoPoint}\,,
\end{align}
which corresponds to \eqref{eq:k=2} in the case of the free theory case with unequal times where 
\begin{equation}\label{eq:freelyevolvedf2}
    f_2^{(0)}( x_1, t_1, x_2, t_2)= f_2\left( \vec{  {q}}_1-\frac{\vec{  {p}}_1}{m}(t_1-\ini{t}), \vec{  {p}}_1, \vec{  {q}}_2-\frac{\vec{  {p}}_2}{m}(t_2-\ini{t}), \vec{  {p}}_2, \ini{t}\right)\,
\end{equation}
is the freely evolved unequal-time generalization of $f_2$. Only the initial reduced phase space densities $f_1$ and $f_2$ appear in \eqref{eq:freeOnePoint} and \eqref{eq:freeTwoPoint}, respectively, since the free evolution does not add any further correlations to the ones already present at initial time. Thus, the result is, as expected, simply the initial correlation function shifted by the free trajectories. For the free theory we can generalize the result to $n$-th order correlation functions and find 
\begin{align}
    \mean{\hat{\mathcal{T}}f(\vec{  {q}}_1, \vec{  {p}}_1, t_1)\cdots f(\vec{  {q}}_n, \vec{  {p}}_n, t_n)}^{(0)}
    =&\cdots + f_n^{(0)}( x_1, t_1, \cdots, x_n, t_n)\label{eq:freenPoint}\,,
\end{align}
where again the ellipsis stands for all lower-order particle contributions. Furthermore, the freely evolved $k$-particle reduced distribution function is, in analogy to \eqref{eq:freelyevolvedf2}, given by
\begin{equation}\label{eq:freelyevolvedfk}
    f_k^{(0)}(x_1, t_1, \cdots, x_k, t_k)=f_2\left( \vec{  {q}}_1-\frac{\vec{  {p}}_1}{m}(t_1-\ini{t}), \vec{  {p}}_1, \cdots, \vec{  {q}}_k-\frac{\vec{  {p}}_k}{m}(t_k-\ini{t}), \vec{  {p}}_k, \ini{t}\right)\,,
\end{equation}
where each particle is shifted individually.

\subsubsection{First and Second Order Perturbation Theory}

The first order correction to $\mean{f(\vec{  {q}}_1, \vec{  {p}}_1, t_1)}$ now includes an interaction operator \eqref{eq:InteractionOperator} and can be computed using \eqref{eq:FirstOrderIntegrated} as
\begin{align}
    \mean{f(\vec{  {q}}_1, \vec{  {p}}_1, t_1)}^{(1)}=&\int\md \tens{x}_1\,\md \tini{x}\,f(\vec{  {q}}_1, \vec{  {p}}_1, t_1)K_1(\tens{x}_1, t_1|\tini{x},\ini{t})\varrho_N(\tini{x}, \ini{t})\,\\
    =&\int_{\ini{t}}^{t_1}\md \Bar{t}_1\int\md \tens{x}_1\,\md \tini{x}\sum_{i=1}^N\delta_D(x_1-x_{1_i})\sum_{j\neq l=1}^N\hat{\mathcal{L}}_{j\ShortLArrow l}(t_1, t_1; \Bar{t}_1)K_0(1|i)\varrho_N(\tini{x}, \ini{t})\\
    =&\int_{\ini{t}}^{t_1}\md \Bar{t}_1\,\int\md   {x}_2\,\hat{\mathcal{L}}_{x_1\ShortLArrow x_2}(t_1, \Bar{t}_1; \Bar{t}_1)\,\left[f_2^{(0)}(  x_1, t_1, x_2, \Bar{t}_1)\right]\,.\label{eq:OnePointFirstOrder}
\end{align}
In the second equality we inserted the definitions of the Klimontovich phase space density. From the second to the third line we made use of the normalization condition \eqref{eq:NormalizationCondition} to identify the particle indices $j$ with $i$ and performed the remaining integrals over the Dirac-Delta distributions. Furthermore, we substituted $x_{1,l}(\Bar{t}_1;t_1)$ by $x_2$, such that the interaction operator now has the indices $x_1$ and $x_2$. The interpretation of the result is clear: A freely evolved two-particle distribution connecting the phase space points $x_1$ and $x_2$ is deformed by the force acting from $x_2$ on $x_1$. The result is then averaged over $x_2$ to yield the first-order correction to the mean density at $x_1$. Note, that the two-particle distribution $f_2$ evolves $x_2$ until it interacts at time $\Bar{t}_1$, while the density at $x_1$ is evolved to the final time $t_1$.\\

\begin{figure}
	\centering
	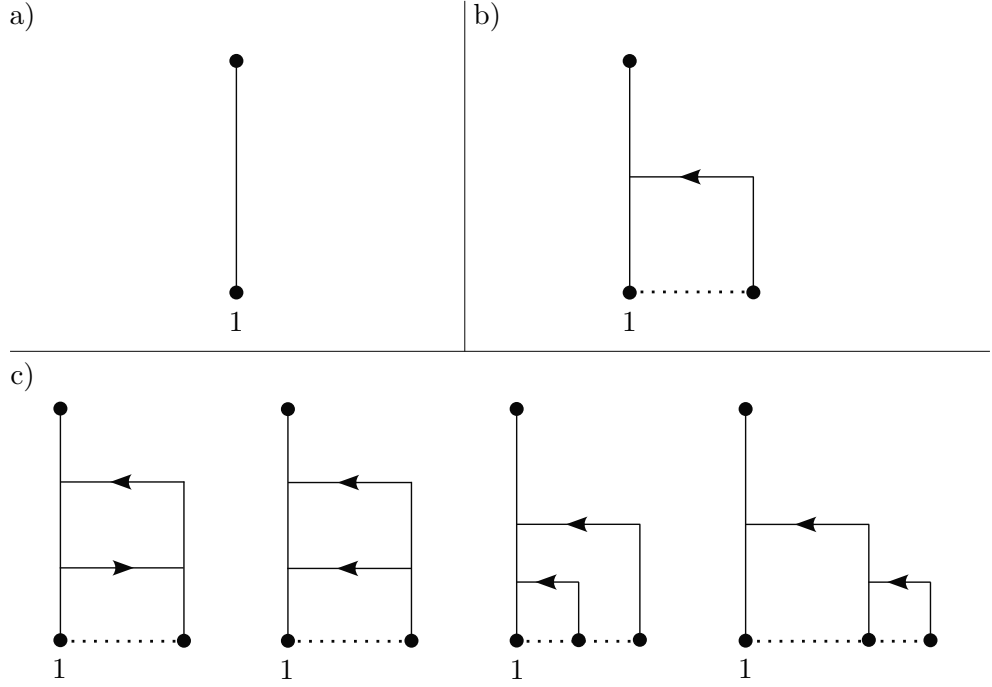
	\caption{A schematic diagram representation of the terms up to second order in perturbation theory for the mean Klimontovich phase space density $ \mean{f(\vec{{q}}_1, \vec{{p}}_1, t_1)}$ is given. In a) we present the free theory \eqref{eq:freeOnePoint}, in b) the first-order perturbation theory \eqref{eq:OnePointFirstOrder} and in c) the second-order perturbation theory \eqref{eq:OnePointSecondOrder}. The filled circles represent positions in phase space. Interactions are depicted as arrows. The direction of the arrows indicates that, for instance, in b) a force is acting on the one-particle distribution at phase space position $x_1$ labeled 1. The dotted lines represent initial correlations. All phase space positions are connected by a dotted line, representing the respective s-particle phase space density. We present all possible topologies of diagrams. The last two terms for the second order contributions in c) introduce a new type of diagram. It represents an indirect interaction between the rightmost phase space position and 1 via an intermediate interaction.}
	\label{fig:PT_onePoint}
\end{figure}

\begin{figure}
  \centering
  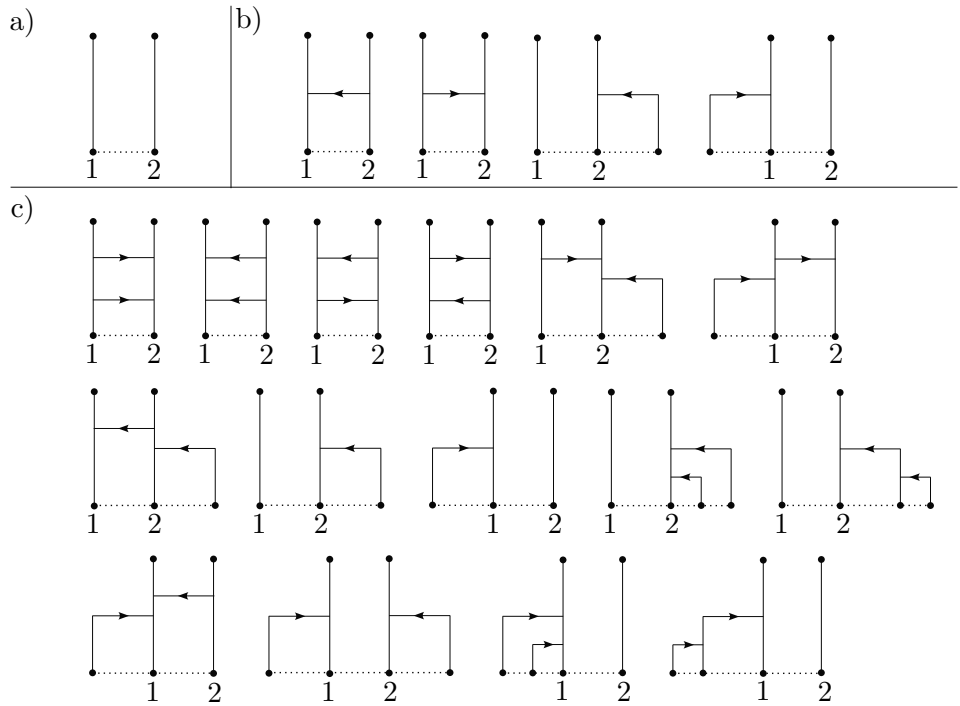
  \caption{All diagrams up to the second order in perturbation theory for the Klimontovich two-point correlation $\mean{\hat{\mathcal{T}}f(  {x}_1, t_1)f(  {x}_2, t_2)}$ are represented. We list in a) the free theory \eqref{eq:freeTwoPoint}, in b) the first-order perturbation theory \eqref{eq:ff_firstOrder} and in c) the second order perturbation theory. }
  \label{fig:PT_twoPoint}
\end{figure}

The first order correction to $\mean{\hat{\mathcal{T}}f(\vec{  {q}}_1, \vec{  {p}}_1, t_1)f(\vec{  {q}}_2, \vec{  {p}}_2, t_2)}$ comes from the following two terms,
\begin{align}
    \mean{\hat{\mathcal{T}}f(\vec{  {q}}_1, \vec{  {p}}_1, t_1)f(\vec{  {q}}_2, \vec{  {p}}_2, t_2)}^{(1)}=&\int\md \tens{x}_1\,\md \tens{x}_2\,\md \tini{x}f(\vec{  {q}}_1, \vec{  {p}}_1, t_1)\,K_1(1|2)\,f(\vec{  {q}}_2, \vec{  {p}}_2, t_2)\,K_0(2|i)\,\varrho_N(\tini{x}, \ini{t})\\
    &+\int\md \tens{x}_1\,\md \tens{x}_2\,\md \tini{x}f(\vec{  {q}}_1, \vec{  {p}}_1, t_1)\,K_0(1|2)\,f(\vec{  {q}}_2, \vec{  {p}}_2, t_2)\,K_1(2|i)\,\varrho_N(\tini{x}, \ini{t})\,,\nonumber
\end{align}
since both terms come with a single power of the interaction potential. The first term describes an interaction taking place after the first average over $\tens{x}_2$ has been taken, while the second term describes the interaction even before the first averaging. Note that in the first term, the interaction operator again enforces, by the normalization argument \eqref{eq:NormalizationCondition}, that the particle deflected by the potential corresponds to the particle contributing to the density $f(\vec{  {q}}_1, \vec{  {p}}_1, t_1)$, while in the second term, the deflected particle can either be the particle contributing to $f(\vec{  {q}}_1, \vec{  {p}}_1, t_1)$ or to $f(\vec{  {q}}_2, \vec{  {p}}_2, t_2)$. Nicely, both contributions to the deflection of the particle belonging to $f(\vec{  {q}}_1, \vec{  {p}}_1, t_1)$ add up to a single time integration over the whole time interval. Thus, upon performing all of the remaining integrals over the free propagators as above, we get the result 
\begin{align}\label{eq:ff_firstOrder}
    \mean{\hat{\mathcal{T}}f(  {x}_1, t_1)f(  {x}_2, t_2)}^{(1)}=&\,\nonumber\delta_D(  {x}_2-  {x}_1(t_2;t_1))\int_{\ini{t}}^{t_1}\md \Bar{t}_1\,\int\md   {x}_3\,\hat{\mathcal{L}}_{x_1\ShortLArrow x_3}(t_1, \Bar{t}_1; \Bar{t}_1)\,f_2^{(0)}( x_1, t_1, x_3, \Bar{t}_1)\nonumber\\
    &+\int_{\ini{t}}^{t_1}\md \Bar{t}_1\,\hat{\mathcal{L}}_{x_1\ShortLArrow x_2}(t_1, t_2; \Bar{t}_1)\,f_2^{(0)}(  x_1, t_1, x_2, t_2)\nonumber\\
    &+\int_{\ini{t}}^{t_2}\md \Bar{t}_1\,\hat{\mathcal{L}}_{x_2\ShortLArrow x_1}(t_2, t_1; \Bar{t}_1)\,f_2^{(0)}(  x_1, t_1, x_2, t_2)\\
    &+\int_{\ini{t}}^{t_1}\md \Bar{t}_1\,\int\md   {x}_3\,\hat{\mathcal{L}}_{x_1\ShortLArrow x_3}(t_1, \Bar{t}_1; \Bar{t}_1)\,f_3^{(0)}(  x_1, t_1, x_2, t_2, x_3, \Bar{t}_1)\nonumber\\
    &+\int_{\ini{t}}^{t_2}\md \Bar{t}_1\,\int\md   {x}_3\,\hat{\mathcal{L}}_{x_2\ShortLArrow x_3}(t_2, \Bar{t}_1; \Bar{t}_1)\,f_3^{(0)}(  x_1, t_1, x_2, t_2, x_3, \Bar{t}_1)\nonumber\,.  
\end{align}
The first term is the first order correction to the one-particle contribution to the two point density correlation. The second and third terms are contributions in which densities at $\vec{  {q}}_1$ and $\vec{  {q}}_2$ influence each other respectively, while the last two terms represent contributions in which an external densities at $\vec{  {q}}_3$ influences the densities at $\vec{  {q}}_1$ and $\vec{  {q}}_2$. See Fig.\ \ref{fig:PT_onePoint} and Fig.\ \ref{fig:PT_twoPoint} for a diagrammatic representation of the first order corrections to $\mean{f(  {x}_1, t_1)}^{(1)}$ and $\mean{f(  {x}_1, t_1)f(  {x}_2, t_2)}^{(1)}$.\\

Last but not least, we present the second order corrections to the mean density. Using \eqref{eq:SecondOrderIntegrated} we find by the same computations as above
\begin{align}
    \mean{f(\vec{  {q}}_1, \vec{  {p}}_1, t_1)}^{(2)}=&\int\md \tens{x}_1\,\md \tini{x}\,f(\vec{  {q}}_1, \vec{  {p}}_1, t_1)K_2(\tens{x}_1, t_1|\tini{x},\ini{t})\varrho_N(\tini{x}, \ini{t})\\=&\int_{\ini{t}}^{t_1}\md \Bar{t}_1\int_{\ini{t}}^{\Bar{t}_1}\md \Bar{t}_2\int\md   {x}_2\hat{\mathcal{L}}_{x_1\ShortLArrow x_2}(t_1, \Bar{t}_1; \Bar{t}_1)\,\Big[\hat{\mathcal{L}}_{x_1\ShortLArrow x_2}(t_1, \Bar{t}_1; \Bar{t}_2)f_2^{(0)}( x_1, t_1, x_2, \Bar{t}_1)\Big] \label{eq:OnePointSecondOrder} \\ 
    +&\int_{\ini{t}}^{t_1}\md \Bar{t}_1\int_{\ini{t}}^{\Bar{t}_1}\md \Bar{t}_2\int\md   {x}_2\hat{\mathcal{L}}_{x_1\ShortLArrow x_2}(t_1, \Bar{t}_1; \Bar{t}_1)\,\Big[\hat{\mathcal{L}}_{x_2\ShortLArrow x_1}(\Bar{t}_1, t_1; \Bar{t}_2)f_2^{(0)}(  x_1, t_1, x_2, \Bar{t}_1)\Big]\nonumber\\
    +&\int_{\ini{t}}^{t_1}\md \Bar{t}_1\int_{\ini{t}}^{\Bar{t}_1}\md \Bar{t}_2\int\md   {x}_2\md   {x}_3\hat{\mathcal{L}}_{x_1\ShortLArrow x_2}(t_1, \Bar{t}_1; \Bar{t}_1)\,\Big[\hat{\mathcal{L}}_{x_1\ShortLArrow x_3}(t_1, \Bar{t}_2; \Bar{t}_2)f_3^{(0)}( x_1, t_1, x_2, \Bar{t}_1, x_3, \Bar{t}_2)\Big]\nonumber\\
    +&\int_{\ini{t}}^{t_1}\md \Bar{t}_1\int_{\ini{t}}^{\Bar{t}_1}\md \Bar{t}_2\int\md   {x}_2\md   {x}_3\hat{\mathcal{L}}_{x_1\ShortLArrow x_2}(t_1, \Bar{t}_1; \Bar{t}_1)\,\Big[\hat{\mathcal{L}}_{x_2\ShortLArrow x_3}(\Bar{t}_1, \Bar{t}_2; \Bar{t}_2)f_3^{(0)}( x_1, t_1, x_2, \Bar{t}_1, x_3, \Bar{t}_2)\Big]\nonumber
\end{align}

A diagrammatic representation of $\mean{f(\vec{  {q}}_1, \vec{  {p}}_1, t_1)}^{(2)}$ can be found in Fig.\ \ref{fig:PT_onePoint}. The second order correction emerges from all possibilities of how a single particle can be influenced by two interactions: Particle 1 is either deflected twice by the same particle, or it first deflects another particle which then acts again on particle 1. The last two terms involve three particles. Here, two different particles can interact with particle 1 or, particle 1 is deflected by a particle which has been previously deflected itself. The free reduced density distribution evolves the particles individually to their latest respective appearance in the contribution. The same pattern emerges for the second order correction of the two-point density $\mean{f(\vec{  {q}}_1, \vec{  {p}}_1, t_1)f(\vec{  {q}}_2, \vec{  {p}}_2, t_2)}^{(2)}$ whose diagrammatic representation is also included in Fig.\ \ref{fig:PT_twoPoint}. This scheme is continued with increasing order in the perturbative expansion. Note, that the derivatives of interactions at later times act on the potentials of all interactions at earlier times and on the initial density distribution. Thus by the product rule the number of terms increases rapidly in higher orders of the perturbation theory. \\

The microscopic perturbation theory thus generates additional correlations in the system to the ones already present at initial time by connecting all possible clusters of particles at a given order of the two particle potential. It becomes clear that each increasing order in the perturbative expansion requires the knowledge of higher freely evolved reduced $s$-particle densities, which can easily be computed, given an initial phase space distribution as shown above. It can be directly seen from the examples we have presented, that each order in the perturbative expansion introduces higher derivatives which act on the interaction potential as well as the initial density distribution, thus causing further deformation of the initial phase space density. \\

We note that the microscopic perturbative expansion corresponds to the formal solution of Liouville's equation \eqref{eq:Liouville} in terms of an iterative solution of its Greens \cite{vonRoos}. In fact, this solution scheme corresponds to a truncation of the BBGKY-hierarchy where the reduced $s$-particle distributions are computed iteratively, meaning that the one-point distribution to first order in interactions contains the free solution of the two-particle distribution and the second order contains the solution for the two-particle distribution to first order in the interactions and so on. The order of the interaction potential, thus, introduces a truncation criterion. However, the perturbation scheme we have presented here is not a typical truncation scheme for the BBGKY-hierarchy. In appendix \ref{sec:BBGKY} we discuss the typical procedures used to truncate the hierarchy which lead to the Boltzmann and the Vlasov equation and how they differ from the approach presented here. It is important to note that, unlike the Vlasov equation, our evolution equations do contain collision terms. These appear, for instance, in the second and third line of equation \eqref{eq:ff_firstOrder}. Each order in our perturbation series will bring in additional collision terms. Hence, we do not make additional assumptions on the collision terms, as is typically done for the Boltzmann equation, nor do we introduce additional truncation criteria for higher-order correlations that appear in our evolution equations. These collision terms are especially important, if we start out with initial conditions that contain no correlations, or if we wish to work with finite ranged interaction potentials.\\

Just as an iterative solution of the BBGKY-hierarchy, our approach directly provides evolution equations for the correlation functions. Although the information content is the same, our path integral formulation offers two major advantages: first of all, it is easily possible to compute unequal-time correlators which can be seen as a generalisation of the BBGKY-based approach. More importantly, due to the path integral formulation it is now possible to apply field theory methods which allow us to construct a macroscopic field theory, implicitly restructuring the perturbation theory in such a way that we can resum microscopic interactions to infinite order, as we will show in the following.

\section{Macroscopic Field Theory}
\label{sec:MacroTheory}
We construct a non-perturbative description of the system in terms of macroscopic fields
using the path integral formulation introduced in Sec. \ref{sec:Construction}. Our approach is based on the Hubbard-Stratonovich transformation (HST) known from equilibrium statistical mechanics. The underlying idea is to describe the system of $N$ particles interacting via a two-body potential in terms of a fluctuating macroscopic field whose correlations can directly be related to the macroscopic observables discussed in Sec. \ref{sec:MacroscopicObservables}. In order to apply the HST to out-of-equilibrium systems with arbitrary potentials, we introduce a slight modification of the HST. We, thus, obtain an effective field theoretic description of the underlying, microscopic physics in terms of macroscopic quantities, which contains the full statistics of the microscopic system in its vertices. One important advantage of this approach is that already the free (or tree-level) theory contains an infinite order of microscopic interactions in terms of the macroscopic propagator. 
A similar approach was pursued in \cite{Lilow_2019}, where a macroscopic field theory was constructed based on the microscopic theory of \cite{Bartelmann_2014}.\footnote{As described at the end of Sec.\ \ref{sec:PathIntegralFormulation}, however, our microscopic perturbation theory differs from that of \cite{Bartelmann_2014}, which consequently also leads to a different macroscopic theory.}

We apply our macroscopic field theory to a homogeneous many-body particle system and provide the expressions for density and momentum correlations of the tree-level theory. These equations are exactly solvable by means of the Laplace transformation if the system is time-translation invariant. 
In a last step, we specify the initial conditions to a Maxwell-Boltzmann distribution and show that in the low-temperature limit we recover two important results which show the power of this formalism: (a) For a Coulomb potential, we find the that the system exhibits a collective oscillating behaviour at the Langmuir frequency. (b) For the Newtonian gravitational potential, we find collective behaviour leading to a gravitational instability with the characteristic time-scale known from Jeans theory. We show that our approach describes both effects fully, already at lowest order in the macroscopic perturbation theory, and that both phenomena share the same origin.

\subsection{Field Theory Construction for Classical $N$-particle Systems}\label{sec:MacroscopicGeneral}

In this section we present the derivation of the macroscopic field theory approach, based on a generalization of the HST to arbitrary potentials. The main ideas are based on the equilibrium description of self-gravitating gases by \cite{deVega1996}. Our goal will be to ultimately obtain a field theory from which correlation functions of the Klimontovich phase space density $f(\vec{  {q}}, \vec{  {p}}, t)$ can be obtained. \\

Let us therefore observe that the interacting part of the action \eqref{eq:interactionaction} can be written as 
\begin{align}
    \mi\mathcal{S}_I[\tens{x}(t),\tens{\chi}(t)]=&\,\mi\int\limits_{\ini{t}}^{\infty}\md t\sum_{i\neq j=1}^N\vec{\chi}_{p_i}(t)\cdot \nabla_{\vec{q}_i}v(|\vec{q}_i(t)-\vec{q}_j(t)|,t)\\
    =&\int\md X_1\int\md X_2 \,\mathcal{B}(X_1)\,\delta_D(X_1-X_2)\,f(X_2)\,,\label{eq:ActionBeforeHST}
\end{align}
where we defined the phase space and time coordinates $X=(\vec{q}, \vec{p}, t)$, with 
\begin{equation}
    \md X_1 = \md^3 q_1\md^3 p_1\md t_1\,,\,\,\,\,\,A(X_1)=A(\vec{q}_1, \vec{p}_1, t_1)\,, 
\end{equation}
and the Dirac delta distribution 
\begin{equation}
    \delta_D(X_1-X_2)=\delta_D(\vec{  {q}}_1-\vec{  {q}}_2)\delta_D(\vec{  {p}}_1-\vec{  {p}}_2)\delta_D(t_1-t_2)\equiv\mathbb{1}(X_1, X_2)\,.
\end{equation}
Beside the already known Klimontovich phase space density $f(X)$ we introduced the response field $\mathcal{B}(X)$, defined as 
\begin{align}\label{eq:ResponseField}
    \mathcal{B}(\vec{  {q}}_1, \vec{  {p}}_1, t_1)&=\mi\sum_{i=1}^{N}\vec{\chi}_{p_i}(t_1)\cdot\nabla_{\vec{q}_i}v(|\vec{q}_i-\vec{  {q}}_1|, t_1)\,.
\end{align}
It is, just like $f(X)$, a function of the microscopic and macroscopic variables and, as its name and form suggests, describes the reaction of the system to the presence of a density $f(X)$. Note, that although the momentum integrals are trivial and can be performed immediately, we included them explicitly in order to keep the momentum information in $f(X)$. The above action can be brought into a quadratic form upon introducing a macroscopic field doublet, 
\begin{equation}
    \Phi(X_1)=(\mathcal{B}(X_1),\,\,  f(X_1))^\top\,,
\end{equation}
and the matrix 
\begin{equation}
    \sigma(X_1, X_2)=\begin{pmatrix}
    0 & \mathbb{1}(X_1, X_2) \\ 
    \mathbb{1}(X_1, X_2) & 0
    \end{pmatrix}\,.
\end{equation}
We then find 
\begin{align}\label{eq:interactionactionmacroscopic}
    \mi\mathcal{S}_I[\tens{x}(t),\tens{\chi}(t)]=\frac{1}{2}\int\md X_1\int\md X_2\Phi(X_1)^\top\cdot\sigma(X_1,X_2)\cdot\Phi(X_2)\,.
\end{align}
We now include a source doublet, 
\begin{equation}
    \mathcal{J}(X)=(J_f(X),\,\, J_{\mathcal{B}}(X))^\top\,,
\end{equation}
into \eqref{eq:interactionactionmacroscopic} such that 
\begin{equation}
    \mi\mathcal{S}_I[\tens{x}(t),\tens{\chi}(t);\mathcal{J}]=\frac{1}{2}\int\md X_1\int\md X_2\Big(\Phi(X_1)+\mathcal{J}(X_1)\Big)^\top\cdot\sigma(X_1,X_2)\cdot\Big(\Phi(X_2)+\mathcal{J}(X_2)\Big)\,.
\end{equation}
Our key object, the macroscopic generating functional, is then defined as 
\begin{align}\label{eq:macroscopicGenFuncBeforeTrafo}
    \mathcal{Z}[J_f, \,J_{\mathcal{B}}]=\int\md \tfin{x}\int\md\tini{x}\varrho_N(\tini{x}, \ini{t})\int\limits_{\tini{x}}^{\tfin{x}} \mathcal{D}\tens{x}(t)\mathcal{D}\tens{\chi}(t)\,\exp\Bigg[\mi\mathcal{S}_0[\tens{x}(t),\tens{\chi}(t)] + \mi\mathcal{S}_\mathrm{I}\left[\tens{x}(t),\tens{\chi}(t);\mathcal{J}\right]\Bigg]\,.
\end{align}
It is immediately evident from the discussion around \eqref{eq:UnequalExpectationPropagator} that applying functional derivatives on $\mathcal{Z}[J_f, \,J_{\mathcal{B}}]$ \wrt $J_f$ and $J_{\mathcal{B}}$ results in correlation functions of the Klimontovich phase space density and the response field, \ie
\begin{align}
    \frac{\delta\mathcal{Z}[J_f, \,J_{\mathcal{B}}]}{\delta J_f(X_1)}\Bigg\rvert_{\mathcal{J}=0}=\mean{f(X_1)}\,,\,\,\,\,\,\frac{\delta\mathcal{Z}[J_f, \,J_{\mathcal{B}}]}{\delta J_\mathcal{B}(X_1)}\Bigg\rvert_{\mathcal{J}=0}=\mean{\mathcal{B}(X_1)}\,,
\end{align}
which extends to all higher-order and mixed correlation functions. We now apply a HST to the exponential of $\mathcal{S}_I$ by expressing it as a Gaussian functional integral,
\begin{equation}
    \e^{\frac{1}{2}\left[\Phi+\mathcal{J}\right]^\top\cdot\sigma\cdot\left[\Phi+\mathcal{J}\right]}=\mathcal{N}\int\mathcal{D}\Psi \e^{-\frac{1}{2}\Psi^\top\cdot\sigma^{-1}\cdot\Psi+\Psi^\top\cdot\left[\Phi+\mathcal{J}\right]}\,,
\end{equation}
with the doublet $\Psi(X_1)=(\Psi_f(X_1),\,\,\Psi_\mathcal{B}(X_1))$, conjugate to $\Phi(X_1)$, and the matrix
\begin{equation}
    \sigma^{-1}(X_1, X_2)=\begin{pmatrix}
    0 & \mathbb{1}(X_1, X_2) \\ 
    \mathbb{1}(X_1, X_2) & 0
    \end{pmatrix}\,,
\end{equation}
inverse to $\sigma(X_1, X_2)$. The normalization factor $\mathcal{N}$ will later be chosen such that $\mathcal{Z}[0,0]=1$. Importantly, in contrast to $f(X_1)$ and $\mathcal{B}(X_1)$ the fields $\Psi_{f}(X_1)$ and $\Psi_{\mathcal{B}}(X_1)$ no longer depend on the microscopic degrees of freedom. For that reason we will refer to them as macroscopic conjugate fields whose physical interpretation will be clarified in a moment. Thus, the generating function \eqref{eq:macroscopicGenFuncBeforeTrafo} assumes the form
\begin{equation}\label{eq:macroscopicGenFuncAfterTrafo}
    \mathcal{Z}[J_f, \,J_{\mathcal{B}}]=\mathcal{N}\int\mathcal{D}\Psi\,\e^{-\frac{1}{2}\Psi^\top\cdot\sigma^{-1}\cdot\Psi+\mathcal{J}^\top\cdot\Psi}\cdot\mathcal{I}_{f\mathcal{B}}^{(0)}[\Psi_f,\,\Psi_{\mathcal{B}}] \,,
\end{equation}
where we absorbed all the remaining microscopic degrees of freedom into the functional $\mathcal{I}_{f\mathcal{B}}^{(0)}[\Psi_f,\,\Psi_{\mathcal{B}}]$, defined as 
\begin{equation}\label{eq:MomentGenFunc}
\begin{aligned}
    \mathcal{I}_{f\mathcal{B}}^{(0)}[\Psi_f,\,\Psi_{\mathcal{B}}]=\int\md \tfin{x}\int\md\tini{x}\varrho_N(\tini{x}, \ini{t})&\int\limits_{\tini{x}}^{\tfin{x}} \mathcal{D}\tens{x}(t)\mathcal{D}\tens{\chi}(t)\,\times\\&\times\exp\Bigg[\mi\mathcal{S}_0[\tens{x}(t),\tens{\chi}(t)] + \int\md X_1\Psi^\top(X_1)\cdot\Phi(X_1)\Bigg]\,.
\end{aligned}
\end{equation}
Clearly, $\mathcal{I}_{f\mathcal{B}}^{(0)}[\Psi_f,\,\Psi_{\mathcal{B}}]$ is the moment generating functional of $f$-$\mathcal{B}$ correlation functions for the free microscopic theory. This can be seen by functionally differentiating $\mathcal{I}_{f\mathcal{B}}^{(0)}[\Psi_f,\,\Psi_{\mathcal{B}}]$ \wrt its arguments which yields the respective $f$-$\mathcal{B}$ correlator (see \eqref{eq:UnequalExpectationPropagator}). Consequently, the respective cumulant-generating functional is given by 
\begin{align}\label{eq:CumulantGenFunc}
    \mathcal{W}^{(0)}_{f\mathcal{B}}[\Psi_f,\,\Psi_{\mathcal{B}}]\equiv& \ln\Big[\mathcal{I}_{f\mathcal{B}}^{(0)}[\Psi_f,\,\Psi_{\mathcal{B}}]\Big]\,,
\end{align}
and has the series expansion
\begin{equation}
\begin{aligned}\label{eq:MacroscopicTheoryGeneralInteractions}
    \mathcal{W}_{f\mathcal{B}}^{(0)}[\Psi_{f}, \Psi_\mathcal{B}] = \sum_{r,s=0}^{\infty}\frac{1}{r!s!}\Bigg[\prod_{m=1}^r\int\md X_m\Psi_\mathcal{B}(X_m)\Bigg]\Bigg[\prod_{n=1}^{s}&\int\md X_n^\prime\Psi_f(X_n^\prime)\Bigg]\, \times\\&\times G_{f\cdots f \mathcal{B}\cdots \mathcal{B}}^{(0)}(X_1,\cdots X_r,X_1^\prime,\cdots X_s^\prime)\,,
\end{aligned}
\end{equation}
where the coefficients $G_{f\cdots f\mathcal{B}\cdots\mathcal{B}}^{(0)}$ denote the free $f$-$\mathcal{B}$ cumulants. They are defined as the connected part of the respective correlation functions containing $r$ phase space densities and $s$ response fields, \ie
\begin{equation}\label{eq:timeorderedfB}
    G_{f\cdots f \mathcal{B}\cdots \mathcal{B}}^{(0)}(X_1,\cdots X_r,X_1^\prime,\cdots X_s^\prime)=\mean{\hat{\mathcal{T}}f(X_1)\cdots f(X_r)\mathcal{B}(X_1^\prime)\cdots \mathcal{B}(X_s^\prime)}^{(0)}_c\,.
\end{equation}
As earlier, the combinatorial prefactor in \eqref{eq:MacroscopicTheoryGeneralInteractions} accounts for all permutations of fields of the same kind in the above time ordered product. The above correlation functions are computed as described in Sec. \ref{sec:FreeTheory}. In appendix \ref{sec:Cumulants} we derive general expressions for the free cumulants. Note that from the coupling between $\Psi$ and $\Phi$ in \eqref{eq:MomentGenFunc} the $f$ fields inside the cumulant couple to the macroscopic $\Psi_{\mathcal{B}}$ field and vice-versa. Plugging \eqref{eq:CumulantGenFunc} into \eqref{eq:macroscopicGenFuncAfterTrafo} we get the final result for the macroscopic generating functional,
\begin{align}\label{eq:MacroscopicTheoryGeneral}
    \mathcal{Z}[J_f, J_\mathcal{B}]=&\mathcal{N}\int\mathcal{D}\Psi_f\mathcal{D}\Psi_\mathcal{B}\exp\Bigg[-\mathcal{S}[\Psi_f, \Psi_\mathcal{B}]+ \int\md X_1J_f(X_1)\Psi_f(X_1)+\int\md X_1 J_{\mathcal{B}}(X_1)\Psi_\mathcal{B}(X_1)\Bigg]\,,
\end{align}
where the macroscopic action is given by 
\begin{equation}
    \mathcal{S}[\Psi_f, \Psi_\mathcal{B}]=\int\md X_1\Psi_f(X_1)\Psi_{\mathcal{B}}(X_1)-\mathcal{W}_{f\mathcal{B}}^{(0)}[\Psi_{f}, \Psi_\mathcal{B}]\,,
    \label{eq:macroAction}
\end{equation}
with $\mathcal{W}_{f\mathcal{B}}^{(0)}$ given by its series representation \eqref{eq:MacroscopicTheoryGeneralInteractions}. By construction we find 
\begin{equation}
    \mean{f(X_1)}=\frac{\delta\mathcal{Z}[J_f, J_\mathcal{B}]}{\delta J_f(X_1)}\Bigg\rvert_{J_f, J_\mathcal{B}=0}=\mean{\Psi_{f}(X_1)}\,,
\end{equation}
which extends to all higher correlators, \ie
\begin{equation}
    \mean{\hat{\mathcal{T}}f(X_1)\cdots f(X_n)}=\mean{\hat{\mathcal{T}}\Psi_{f}(X_1)\cdots\Psi_{f}(X_n)}\;.
\end{equation}
We have thus constructed a non-local effective field theoretic description of a general systems with two-particle interactions from which phase space density correlation functions can be obtained, by applying appropriate field theoretic methods. The vertices contain the full information on the free microscopic statistics. At this point we still have, in principle, the full knowledge of the $N$-particle ensemble. The microscopic degrees of freedom are encoded in the vertices of our new field theory. The term `effective' therefore refers to the fact that our field theory contains the full hierarchy of possible vertices (or couplings) and it is therefore inevitable that a truncation scheme must be chosen in order to obtain physical quantities. In the following sections we study the structure of the field theory and put it in relation to the microscopic perturbation theory.

\subsection{Feynman Rules}
\label{sec:Feynman}
As is usual in field theory, we split the macroscopic action into a part that contains all terms that are at most quadratic in the fields $\Psi_{f}$ and $\Psi_\mathcal{B}$ and a part containing higher orders. We then define the free macroscopic action\footnote{As derived in appendix \ref{sec:Cumulants} all pure $G^{(0)}_{\mathcal{B}\cdots\mathcal{B}}(1,\cdots n)$ cumulants vanish identically.},
\begin{align}
    \mathcal{S}_0[\Psi_{f},\Psi_\mathcal{B}]=&\int\md X_1\md X_2\Psi_\mathcal{B}(X_1)\left[\mathbb{1}(X_1, X_2) - G^{(0)}_{f\mathcal{B}}(X_1,X_2)\right]\Psi_f(X_2)\\&-\int\md X_1\Psi_\mathcal{B}(X_1)G^{(0)}_f(X_1)-\frac{1}{2!}\int\md X_1\md X_2\Psi_\mathcal{B}(X_1)\Psi_\mathcal{B}(X_2)G^{(0)}_{ff}(X_1,X_2)\;,
\end{align}
and the interaction part which contains all higher order vertices,
\begin{align}\label{eq:Vertexpart}
    \mathcal{S}_\mathrm{I}[\Psi_{f},\Psi_\mathcal{B}]=&\sum_{r+s>2}^{\infty}\frac{1}{r!s!}\Bigg[\prod_{m=1}^r\int\md X_m\Psi_\mathcal{B}(X_m)\Bigg]\Bigg[\prod_{n=1^\prime}^s\int\md X_n^\prime\Psi_f(X_n^\prime)\Bigg]\, G_{f\cdots f \mathcal{B}\cdots \mathcal{B}}^{(0)}(X_1,\cdots X_r,X_1^\prime,\cdots X_s^\prime)\,.
\end{align}
Note that in this context `free' refers to all such terms that can be computed exactly within the path integration. In order to avoid confusion with the free microscopic theory, we will talk about the tree-level theory in this context. We can then pull the vertex part out of the generating functional, by replacing the fields $\Psi_{f}$ and $\Psi_\mathcal{B}$ by a functional derivative \wrt to the respective source. The full generating functional than assumes the form 
\begin{equation}
     \mathcal{Z}[J_f, J_\mathcal{B}]=\mathcal{N}^\prime\exp\bigg[\mathcal{S}_\mathrm{I}\bigg[\frac{\delta}{\delta J_f},\frac{\delta}{\delta J_\mathcal{B}}\bigg]\bigg]\cdot\mathcal{Z}_0[J_f, J_\mathcal{B}]\;,
\end{equation}
with the free generating functional given by
\begin{equation}\label{eq:FreeMacrosGeneratingFunctionalPathInt}
    \mathcal{Z}_0[J_f, J_\mathcal{B}]=\mathcal{N}^{\prime\prime}\int\mathcal{D}\Psi_f\mathcal{D}\Psi_\mathcal{B}\exp\bigg[-\mathcal{S}_0[\Psi_\mathcal{B},\Psi_{f}]+\int\md X_1 J_f(X_1)\Psi_f(X_1)+\int\md X_1 J_\mathcal{B}(X_1)\Psi_\mathcal{B}(X_1)\bigg]\,.
\end{equation}
The normalization $\mathcal{N}=\mathcal{N}^\prime\mathcal{N}^{\prime\prime}$ is chosen such that $\mathcal{Z}[0,0]=1=\mathcal{Z}_0[0,0]$. In appendix \ref{sec:MacroscopicZ0} we show that the free generating functional is given by
\begin{equation}\label{eq:FreeMacroGeneratingFunctional}
    \mathcal{Z}_0[J_f, J_\mathcal{B}]=\exp\Bigg[\frac{1}{2}\begin{pmatrix}
        J_f \\ J_\mathcal{B}
    \end{pmatrix}^\top\cdot\begin{pmatrix}
        \Delta_{ff} & \Delta_{f\mathcal{B}}\\ \Delta_{\mathcal{B}f} & 0
    \end{pmatrix}\cdot\begin{pmatrix}
        J_f \\ J_\mathcal{B}
    \end{pmatrix}+J_f\cdot\Delta_{f\mathcal{B}}\cdot G_{f}^{(0)}\Bigg]\,,
\end{equation}
where the integration is implied in the products. Furthermore, we introduced two different types of propagators, defined by 
\begin{align}
    \Delta_{f\mathcal{B}}(X_1,X_2)=&\Big[\mathbb{1} - G^{(0)}_{f\mathcal{B}}\Big]^{-1}(X_1,X_2)\label{eq:RetardedPropagator}\;,\\
     \Delta_{\mathcal{B}f}(X_1,X_2)\equiv&\Delta_{f\mathcal{B}}(X_2,X_1)\label{eq:AdvancedPropagator}\;,\\
    \Delta_{ff}(X_1,X_2)=&\int\md\Bar{X}_1\md\bar{X}_2\,\Delta_{f\mathcal{B}}(X_1,\bar{X}_1)\,G^{(0)}_{ff}(\bar{X}_1,\bar{X}_2)\,\Delta_{\mathcal{B}f}(\bar{X}_2,X_2)\label{eq:CorrelationPropagator}\,.
\end{align}
Since $G^{(0)}_{f\mathcal{B}}(X_1,X_2)\propto\Theta(t_1-t_2)$ as shown in appendix \ref{sec:Cumulants}, we will refer to $\Delta_{f\mathcal{B}}(X_1,X_2)$ and $\Delta_{\mathcal{B}f}(X_1,X_2)$ as the retarded- and advanced causal propagators, respectively. They describe the propagation of interactions through the system as we will discuss shortly. The statistical propagator $\Delta_{ff}(X_1,X_2)$, on the other hand, carries information about fluctuations of the macroscopic density field, which are imprinted in the two-point cumulant $G^{(0)}_{ff}$. Note, that the inverse in \eqref{eq:RetardedPropagator} has to be taken in the functional sense, \ie as the solution to the integral equation 
\begin{equation}\label{eq:Delta_R}
    \int\md\bar{X}_1\Big[\mathbb{1}(X_1,\bar{X}_1) - G^{(0)}_{f\mathcal{B}}(X_1,\bar{X}_1)\Big]\cdot\Delta_{f\mathcal{B}}(\Bar{X}_1,X_2)=\mathbb{1}(X_1,X_2)\,.
\end{equation}
We define a diagrammatic representation for the propagators and vertices so that we may represent terms of the perturbative expansion in terms of Feynman diagrams. For the causal propagators $\Delta_{f\mathcal{B}}$ and $\Delta_{\mathcal{B}f}$ we define the following lines:
\begin{align}\label{eq:fB_prop}
        \Delta_{f\mathcal{B}}(X_1,X_2) &= \mytikz{
            \ffpattern[X_1][X_2]\fbprop{f1}{f2}
            } \,,\\
        \Delta_{\mathcal{B}f}(X_1,X_2) &= \mytikz{
            \ffpattern[X_1][X_2]\bfprop{f1}{f2}
            }        \,.
\end{align}
The time flow inherited from the Heaviside function in $G^{(0)}_{f\mathcal{B}}$ is always from a dashed line to a solid line, indicated by the arrows.
The vertices are represented by 
\begin{equation}\label{eq:vertex}
    \int \md X_1 \dots \md X_r\,\md X_{1}^\prime \dots \md X_{s}^\prime\, G^{(0)}_{f\dots f\mathcal{B}\dots \mathcal{B}}(X_1,\dots,X_r,X_{1}^\prime,\dots,X_{s}^\prime) = \mytikz{
    \nvertex{}{}
    }\;,
\end{equation}
where each $f$ field corresponds to a dashed line while each $\mathcal{B}$ field corresponds to a solid line. The statistical propagator $\Delta_{ff}$ itself is a composite diagram of the two-point vertex $G^{(0)}_{ff}$ and the retarded and advanced propagators $ \Delta_{f\mathcal{B}} $ and $ \Delta_{\mathcal{B}f}$, respectively. We introduce a separate rule, as $\Delta_{ff}$ plays an important role in carrying initial two-point correlations,
\begin{equation}\label{eq:ff_prop}
    \Delta_{ff}(X_1,X_2) = \mytikz{
            \ffpattern[X_1][X_2]\ffpropcomb{f1}{f2}
            }\equiv \mytikz{
            \ffpattern[X_1][X_2]\ffprop{f1}{f2}
            }
\end{equation}
With those rules, all diagrams can be constructed by attaching propagators with their appropriate endpoint to the vertices and integrating over every vertex argument. 

\subsection{Connection to the Microscopic Theory}
\label{sec:MicroMacro}
Let us now discuss the physical interpretation of the propagators, the vertices and the Feynman rules and compare it to the physical picture of the microscopic perturbation theory. In appendix \ref{sec:Cumulants} we derive general expressions for the free $f$-$\mathcal{B}$-cumulants with the help of a diagrammatic representation. As is easily seen, the pure $f$-cumulant $G^{(0)}_{f\cdots f}$ simply corresponds to the connected part of the free phase space density correlator \eqref{eq:freenPoint}. Hence, it only consists of freely evolved irreducible phase space densities $g^{(0)}_r(x_1, t_1, \cdots, x_r, t_r)$, which contain the information on the initial correlation between the respective points. On the other hand, the mixed cumulants $G^{(0)}_{f\cdots f\mathcal{B}\cdots\mathcal{B}}$ describe how the respective pure density cumulant responds to forces exhibited by external densities that are not connected to the given cluster. These forces are represented by the $\mathcal{B}$ fields. For instance, we find
\begin{align}
G_{f\mathcal{B}}^{(0)}(X_1, X_2) = & \left[\hat{\mathcal{L}}_{x_1\ShortLArrow x_2}(t_1,t_2;t_2)\right]\cdot G_{f}^{(0)}(X_1)\,,\\
    G_{ff\mathcal{B}}^{(0)}(X_1,X_2,X_3) =  &\left[\hat{\mathcal{L}}_{x_1\ShortLArrow x_3}(t_1,t_3;t_3) + \hat{\mathcal{L}}_{x_2\ShortLArrow x_3}(t_2,t_3;t_3)\right]\cdot G_{ff}^{(0)}(X_1,X_2)\,,\\
    G_{f\mathcal{B}\mathcal{B}}^{(0)}(X_1,X_2,X_3)  = &\left[\hat{\mathcal{L}}_{x_1\ShortLArrow x_2}(t_1,t_2;t_2)\hat{\mathcal{L}}_{x_1\ShortLArrow x_3}(t_1,t_3;t_3)\right]\cdot G_{f}^{(0)}(X_1)\,,
\end{align}
where the interaction operator $\hat{\mathcal{L}}_{x_1\ShortLArrow x_2}(t_1,t_2;t_2)$ is defined as\footnote{The reader might have noticed, that in contrast to \eqref{eq:InteractionOperator}, we absorbed the Heaviside function into the definition of the interaction operator in order to keep the boundaries of the time integration independent from each other. The causality is then ensured by $\Theta(t_1-t_2)$.} 
\begin{equation}
    \hat{\mathcal{L}}_{x_1\ShortLArrow x_2}(t_1,t_2;t_2)\equiv\Theta(t_1-t_2)\nabla_{\vec{q}_1}v(|\vec{q}_1-\frac{\vec{p}_1}{m}(t_1-t_2)-\vec{q}_2\,|, t_2)\cdot\nabla_{\vec{p}_1}\,.
\end{equation}
 The expression for $G^{(0)}_{f\mathcal{B}}$ is clearly reminiscent of the first-order correction to the one-point correlation function of the Klimontovich phase space density \eqref{eq:OnePointFirstOrder}. It thus describes how the single particle distribution at $x_1$ at time $t_1$ responds to a force originating from $x_2$ at an earlier time $t_2$. Next, $G_{ff\mathcal{B}}^{(0)}$ describes the same process in which $G_{ff}^{(0)}$ is deformed by an interaction with an external density, while $G_{f\mathcal{B}\mathcal{B}}^{(0)}$ represents the deviation of the single particle distribution function at $x_1$ from its free evolution due to two consecutive interactions with two external densities. Importantly, the Heaviside functions inside the interaction operator ensure that the interactions act from later times to earlier times. In general, a $G_{f\cdots f\mathcal{B}\cdots \mathcal{B}}^{(0)}$ thus describes the deflection of the corresponding freely evolved $G_{f\cdots f}^{(0)}$ cumulant due to forces exhibited by external particles located at the positions and times of the respective $\mathcal{B}$ field. These different types of cumulants thus represent the building blocks of the macroscopic field theory. They appear as vertices of our macroscopic field theory and comprise all possible ways to correlate different points in space and time through either initial correlations or interactions. Since our macroscopic field theory contains an infinite number of vertices, we must choose some truncation criterion. This leads us to an effective field theory. By determining which vertices are taken into account, we implicitly control the type of microscopic effects contained in the macroscopic description. We could, for instance, neglect all pure $f$ vertices beyond the three-point vertex $G_{fff}^{(0)}$ which would mean that we neglect all higher-order \textit{initial} $n$-point correlations while higher order $n$-point correlations induced by the \textit{interactions} enter through the mixed $f-\mathcal{B}$-vertices. \\

\begin{figure}
  \centering
\begingroup%
  \makeatletter%
  \providecommand\color[2][]{%
    \errmessage{(Inkscape) Color is used for the text in Inkscape, but the package 'color.sty' is not loaded}%
    \renewcommand\color[2][]{}%
  }%
  \providecommand\transparent[1]{%
    \errmessage{(Inkscape) Transparency is used (non-zero) for the text in Inkscape, but the package 'transparent.sty' is not loaded}%
    \renewcommand\transparent[1]{}%
  }%
  \providecommand\rotatebox[2]{#2}%
  \newcommand*\fsize{\dimexpr\f@size pt\relax}%
  \newcommand*\lineheight[1]{\fontsize{\fsize}{#1\fsize}\selectfont}%
  \ifx\svgwidth\undefined%
    \setlength{\unitlength}{156.89310269bp}%
    \ifx\svgscale\undefined%
      \relax%
    \else%
      \setlength{\unitlength}{\unitlength * \real{\svgscale}}%
    \fi%
  \else%
    \setlength{\unitlength}{\svgwidth}%
  \fi%
  \global\let\svgwidth\undefined%
  \global\let\svgscale\undefined%
  \makeatother%
  \begin{picture}(1,1.02058962)%
    \lineheight{1}%
    \setlength\tabcolsep{0pt}%
    \put(0,0){\includegraphics[width=\unitlength,page=1]{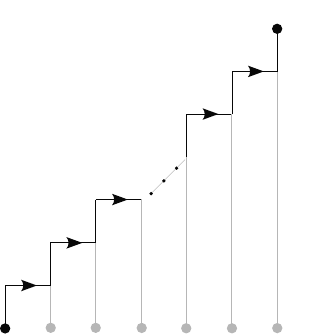}}%
    \put(0.83137847,0.97699595){\color[rgb]{0,0,0}\makebox(0,0)[lt]{\lineheight{1.25}\smash{\begin{tabular}[t]{l}$X_1$\end{tabular}}}}%
  \end{picture}%
\endgroup%

  \caption{A diagrammatic representation of an interaction chain contributing to the tree-level evolution of the phase space density $\mean{\Psi_f(X_1)}^{(\mathrm{tree})}$ in \eqref{eq:Psi_f_diagram}. The tree-level correlator $\mean{\Psi_f(X_1)}^{(\mathrm{tree})}$ is given by an infinite sum of such chains with increasing number of links, starting at zero. phase space points which take part in the interactions are shown in gray.}
  \label{fig:tree}
\end{figure}

With that intuition in mind, we can now take a closer look at $\Delta_{f\mathcal{B}}(X_1,X_2) $ defined by the functional inversion \eqref{eq:RetardedPropagator}. It is the formal solution to the Neumann series of $G^{(0)}_{f\mathcal{B}}(X_1,X_2)$, \ie 
\begin{align}
    \Delta_{f\mathcal{B}}(X_1,X_2)=&\sum_{n=0}^\infty\Big[G^{(0)}_{f\mathcal{B}}\Big] ^n(X_1,X_2)\\
    =&\mathbb{1}(X_1,X_2)+G^{(0)}_{f\mathcal{B}}(X_1,X_2)+\int\md \Bar{X}_1G^{(0)}_{f\mathcal{B}}(X_1,\Bar{X}_1)\,G^{(0)}_{f\mathcal{B}}(\Bar{X}_1,X_2)+\cdots\label{eq:NeumannSeries}\\
    =&\mathbb{1}(X_1,X_2)+\widetilde{\Delta}_{f\mathcal{B}}(X_1,X_2)\,,\label{eq:TildeDelta}
\end{align}
where in the last line we summarized all terms containing at least one interaction potential into $\widetilde{\Delta}_{f\mathcal{B}}(X_1,X_2)$ and separated them from the identity. Thus, $\widetilde{\Delta}_{f\mathcal{B}}(X_1,X_2)$ contains the true interaction information and, importantly, is subject to an equation analogous to \eqref{eq:Delta_R} given by 
\begin{equation}\label{eq:VolterraTildeDelta}
    \widetilde{\Delta}_{f\mathcal{B}}(X_1,X_2) = G^{(0)}_{f\mathcal{B}}(X_1,X_2) + \int\md\Bar{X}_1G^{(0)}_{f\mathcal{B}}(X_1,\Bar{X}_1)\widetilde{\Delta}_{f\mathcal{B}}(\Bar{X}_1,X_2)\,.
\end{equation}
Its physical meaning is best understood by explicitly computing expectation values. The tree-level expectation value of the one-point density correlation function is given by 
\begin{align}
     \mean{\Psi_f(X_1)}^{(\mathrm{tree})} &= \frac{\delta \mathcal{Z}_0[J_f, J_{\mathcal{B}}]}{\delta J_f(X_1)}\Bigg\rvert_{J_f, J_\mathcal{B}=0}\\
     &=\int\md \Bar{X}_1\Delta_{f\mathcal{B}}(X_1, \Bar{X}_1)\,G_f^{(0)}(\Bar{X}_1)\label{eq:treeleveldensity}\\ 
     &=\mytikz{
     \fdot[X_1]\vertexnew[]\fbprop{f1}{v1}
     }\,.
      \label{eq:Psi_f_diagram}
\end{align}
Iterating the Neumann Series $\eqref{eq:NeumannSeries}$ of the retarded propagator and inserting it into \eqref{eq:treeleveldensity}, we find that the tree-level correlator of the phase space density consists of an infinite sum of interaction chains. A typical chain has the structure
\begin{align}\nonumber
    \int\md\Bar{X}_1\cdots \md \Bar{X}_nG^{(0)}_{f\mathcal{B}}(X_1,\Bar{X}_1)\, G^{(0)}_{f\mathcal{B}}(\Bar{X}_1, \Bar{X}_2)\cdots G^{(0)}_{f\mathcal{B}}(\Bar{X}_{n-1}, \Bar{X}_{n})\,G_f^{(0)}(\Bar{X}_{n})\,.
\end{align}
The contribution to the mean density at $X_1$ thus arises in the following way: A phase space density at $\Bar{X}_n$ described by $G_f^{(0)}(\Bar{X}_{n})$ sources a force acting on the density at $\Bar{X}_{n-1}$ which is thus deflected from its free evolution. This is described by $G^{(0)}_{f\mathcal{B}}(\Bar{X}_{n-1}, \Bar{X}_{n})$. The deflected density then again acts with 
a force on the density at $X_{n-2}$ and so on. The chain ends with a density at $X_1$, contained in $G^{(0)}_{f\mathcal{B}}(X_1, \Bar{X}_{2})$. The causality of the $G^{(0)}_{f\mathcal{B}}$ factors make sure that interactions only affect the future evolution of the density and not its past. Summing over all possible intermediate steps, starting from the free evolution of the density and ending with an infinite chain of interactions, eventually gives the full tree-level density. Thus, the role of $\widetilde{\Delta}_{f\mathcal{B}}(X_1, X_2)$ is to propagate information from a density at $X_2$ to a density at $X_1$ by iteratively deflecting the evolution of one-particle distribution functions. The solution of \eqref{eq:VolterraTildeDelta} then contains infinite orders of the interaction potential. In Fig.\ \ref{fig:tree} we provide a diagrammatic representation of the tree-level evolution of the density for illustration. Importantly, expanding the Neumann series to first order we recover the reducible part of the first-order correction in perturbation theory \eqref{eq:OnePointFirstOrder}.\\ 

\begin{figure}
  \centering
\begingroup%
  \makeatletter%
  \providecommand\color[2][]{%
    \errmessage{(Inkscape) Color is used for the text in Inkscape, but the package 'color.sty' is not loaded}%
    \renewcommand\color[2][]{}%
  }%
  \providecommand\transparent[1]{%
    \errmessage{(Inkscape) Transparency is used (non-zero) for the text in Inkscape, but the package 'transparent.sty' is not loaded}%
    \renewcommand\transparent[1]{}%
  }%
  \providecommand\rotatebox[2]{#2}%
  \newcommand*\fsize{\dimexpr\f@size pt\relax}%
  \newcommand*\lineheight[1]{\fontsize{\fsize}{#1\fsize}\selectfont}%
  \ifx\svgwidth\undefined%
    \setlength{\unitlength}{265.51266888bp}%
    \ifx\svgscale\undefined%
      \relax%
    \else%
      \setlength{\unitlength}{\unitlength * \real{\svgscale}}%
    \fi%
  \else%
    \setlength{\unitlength}{\svgwidth}%
  \fi%
  \global\let\svgwidth\undefined%
  \global\let\svgscale\undefined%
  \makeatother%
  \begin{picture}(1,0.50634351)%
    \lineheight{1}%
    \setlength\tabcolsep{0pt}%
    \put(0,0){\includegraphics[width=\unitlength,page=1]{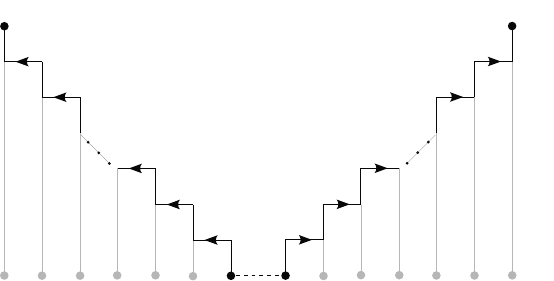}}%
    \put(-0.00055827,0.48488654){\color[rgb]{0,0,0}\makebox(0,0)[lt]{\lineheight{1.25}\smash{\begin{tabular}[t]{l}$X_1$\end{tabular}}}}%
    \put(0.91700377,0.48488654){\color[rgb]{0,0,0}\makebox(0,0)[lt]{\lineheight{1.25}\smash{\begin{tabular}[t]{l}$X_2$\end{tabular}}}}%
  \end{picture}%
\endgroup%

  \caption{An interaction chain which contributes to the tree-level two-point phase space density $\mean{\hat{\mathcal{T}}\Psi_f(X_1)\Psi_f(X_2)}^{(\mathrm{tree})}$ in \eqref{eq:ff_propagator_diagram} is shown here. The dashed line connecting the two phase space points at the base of the diagram represents the connected part of the correlation between those points. phase space points which take part in the interactions are shown in gray.}
  \label{fig:tree_ff}
\end{figure}

A similar calculation reveals for the two-point correlator,
\begin{align}
     \mean{\hat{\mathcal{T}}\Psi_f(X_1)\Psi_f(X_2)}^{(\mathrm{tree})} = \frac{\delta^2\mathcal{Z}[J_f, J_\mathcal{B}]}{\delta J_f(X_1)\delta J_f(X_2)}\Bigg\rvert_{J_f, J_\mathcal{B}=0} = \mytikz{
     \ffpattern[X_1][X_2]\ffprop{f1}{f2} 
     } + \mytikz{\ffdisconnected{X_1}{X_2}
      }\,.
 \end{align}
The connected part of a correlator can, as usual, be extracted by defining the Schwinger functional $\mathcal{W}[J_f, J_\mathcal{B}]$ as
\begin{equation}
    \mathcal{W}[J_f, J_\mathcal{B}] = \ln \Big[\mathcal{Z}[J_f, J_\mathcal{B}]\Big]\,.
\end{equation}
We then find for the connected part of the two-point correlator,
\begin{align}
    \mean{\hat{\mathcal{T}}\Psi_f(X_1)\Psi_f(X_2)}^{(\mathrm{tree})}_c &= \frac{\delta^2\mathcal{W}[J_f, J_\mathcal{B}]}{\delta J_f(X_1)\delta J_f(X_2)}\Bigg\rvert_{J_f, J_\mathcal{B}=0} \\
    &=\Delta_{ff}(X_1, X_2)\\ 
    &= \mytikz{
     \ffpattern[X_1][X_2]\ffprop{f1}{f2} 
     } \,.
     \label{eq:ff_propagator_diagram}
\end{align}
From the definition of the statistical propagator $\Delta_{ff}(X_1, X_2)$ in \eqref{eq:CorrelationPropagator} we see that it describes the connected correlation between two points $X_1$ and $X_2$: Two initially correlated densities, described by $G^{(0)}_{ff}$ are connected to the points $X_1$ and $X_2$ by two causal $\Delta_{f\mathcal{B}}$ propagators, which transport the statistical information of the inner densities as described above. The causal flow is consequently outwards oriented at both ends of $\Delta_{ff}$. A diagrammatic representation of the statistical propagator can be found in Fig.\ \ref{fig:tree_ff}. Again, upon expanding the causal propagators to first order, we recover the reducible part of the last two terms of the microscopic perturbation theory in \eqref{eq:ff_firstOrder}.\\

\begin{figure}
  \centering
\begingroup%
  \makeatletter%
  \providecommand\color[2][]{%
    \errmessage{(Inkscape) Color is used for the text in Inkscape, but the package 'color.sty' is not loaded}%
    \renewcommand\color[2][]{}%
  }%
  \providecommand\transparent[1]{%
    \errmessage{(Inkscape) Transparency is used (non-zero) for the text in Inkscape, but the package 'transparent.sty' is not loaded}%
    \renewcommand\transparent[1]{}%
  }%
  \providecommand\rotatebox[2]{#2}%
  \newcommand*\fsize{\dimexpr\f@size pt\relax}%
  \newcommand*\lineheight[1]{\fontsize{\fsize}{#1\fsize}\selectfont}%
  \ifx\svgwidth\undefined%
    \setlength{\unitlength}{234.32544618bp}%
    \ifx\svgscale\undefined%
      \relax%
    \else%
      \setlength{\unitlength}{\unitlength * \real{\svgscale}}%
    \fi%
  \else%
    \setlength{\unitlength}{\svgwidth}%
  \fi%
  \global\let\svgwidth\undefined%
  \global\let\svgscale\undefined%
  \makeatother%
  \begin{picture}(1,0.5803834)%
    \lineheight{1}%
    \setlength\tabcolsep{0pt}%
    \put(0,0){\includegraphics[width=\unitlength,page=1]{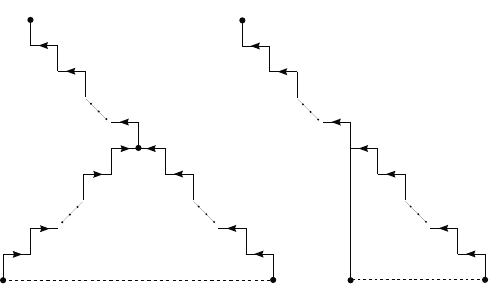}}%
    \put(0.05483973,0.56050501){\color[rgb]{0,0,0}\makebox(0,0)[lt]{\lineheight{1.25}\smash{\begin{tabular}[t]{l}$X_1$\end{tabular}}}}%
    \put(0.48863151,0.56043308){\color[rgb]{0,0,0}\makebox(0,0)[lt]{\lineheight{1.25}\smash{\begin{tabular}[t]{l}$X_1$\end{tabular}}}}%
  \end{picture}%
\endgroup%

  \caption{A diagrammatic representation of the one-loop contributions $\delta\mean{\Psi_f(X_1)}^{(\mathrm{1loop})}$ in \eqref{eq:loop_digrams} is given. The diagrams correspond to the second and first Feynman diagram in \eqref{eq:loop_digrams} respectively. The dashed line connecting the two phase space points at the base of the diagram represents the connected part of the correlation between those points. For clarity, we do not explicitly depict the gray phase space points taking part in the interactions.}
  \label{fig:loops}
\end{figure}

Let us now briefly motivate the one-loop contribution to $\mean{\Psi_f(X_1)}$. It can be computed by including the vertex part of the theory. Since \eqref{eq:Vertexpart} includes an infinite sum of possible vertices, let us restrict ourselves to the contributions coming from $G^{(0)}_{ff\mathcal{B}}$ and $G^{(0)}_{f\mathcal{B}\mathcal{B}}$. The respective correction is then found by computing 
\begin{align}
    \delta\mean{\Psi_f(X_1)}^{(\mathrm{1loop})} \supset& \frac{1}{2!}\int\md \Bar{X}_1\md \Bar{X}_2\md \Bar{X}_3\left[G^{(0)}_{ff\mathcal{B}}(\Bar{X}_1, \Bar{X}_2, \Bar{X}_3)\frac{\delta^4\mathcal{Z}[J_f, J_\mathcal{B}]}{\delta J_f(X_1)\delta J_\mathcal{B}(\Bar{X}_1)\delta J_\mathcal{B}(\Bar{X}_2)\delta J_f(\Bar{X}_3)}\Bigg\rvert_{J_f, J_\mathcal{B}=0}\right.\nonumber\\ &\,\,\,\,\,\,\,\,\,\,\,\,\,\,\,\,\,\,\,\,\left.+ G^{(0)}_{f\mathcal{B}\mathcal{B}}(\Bar{X}_1, \Bar{X}_2, \Bar{X}_3)\frac{\delta^4\mathcal{Z}[J_f, J_\mathcal{B}]}{\delta J_f(X_1)\delta J_\mathcal{B}(\Bar{X}_1)\delta J_f(\Bar{X}_2)\delta J_f(\Bar{X}_3)}\Bigg\rvert_{J_f, J_\mathcal{B}=0}\right]\\
    \supset&\int\md \Bar{X}_1\md \Bar{X}_2\md \Bar{X}_3\left[\Delta_{f\mathcal{B}}(X_1, \Bar{X}_1)G^{(0)}_{ff\mathcal{B}}(\Bar{X}_1, \Bar{X}_2, \Bar{X}_3)\Delta_{\mathcal{B}f}(\Bar{X}_2, \Bar{X}_3)\right.\nonumber\\ &\,\,\,\,\,\,\,\,\,\,\,\,\,\,\,\,\,\,\,\,\left. +\frac{1}{2}\Delta_{f\mathcal{B}}(X_1, \Bar{X}_1)G^{(0)}_{f\mathcal{B}\mathcal{B}}(\Bar{X}_1, \Bar{X}_2, \Bar{X}_3)\Delta_{ff}(\Bar{X}_2, \Bar{X}_3)\right]\\
    =&\mytikz{
     \fdot[X_1]\vertexnew[]\fbprop{f1}{v1}\fbpropCircle{v1}
     } + \frac{1}{2}\mytikz{
     \fdot[X_1]\vertexnew[]\fbprop{f1}{v1}\ffpropCircle{v1}
     }\,,
     \label{eq:loop_digrams}
\end{align}
where we only included the connected contribution to the loop correction. This time, we also have purely internal propagators. The two contributions are schematically depicted in Fig.\ \ref{fig:loops} from which the microscopic interpretation can be deduced. As we can see, in a given diagram the internal $f$ field from a vertex sources the deflections of density statistics due to the $\mathcal{B}$ field of another vertex. The Heaviside function then ensures the correct causal structure of the diagram. Furthermore, we can resum the $G^{(0)}_{f\mathcal{B}}$ contributions by means of \eqref{eq:VolterraTildeDelta}. The effect of more than one interaction acting on a density, or the inclusion of higher density statistics is thus part of the loop-corrections to the density. A more detailed study of the loop corrections, as well as non-perturbative field theoretic methods will be covered in a future work.

\subsection{Tree-level Expectation Values for Operators}
\label{sec:TreeLevel}
As discussed in section \ref{sec:MacroscopicObservables} we are mainly interested in the computation of observables $\mathcal{O}(\vec{q}, t)$ rather then the phase space density itself. Let us therefore shortly discuss how this can be done within the macroscopic field theory.\\

One important feature that we shall exploit is that $G^{(0)}_{f\mathcal{B}}(X_1,X_2)$ is independent of the momentum $\vec{p}_2$ of the response field, which is consequently also the case for $\widetilde{\Delta}_{f\mathcal{B}}$. Thus, we may perform the integration over $\vec{\Bar{p}}_1$ on the right hand side of \eqref{eq:VolterraTildeDelta} by pulling it past $G^{(0)}_{f\mathcal{B}}(X_1,\Bar{X}_1)$ and defining the density projections of $\widetilde{\Delta}_{f\mathcal{B}}$ and $G^{(0)}_{f\mathcal{B}}$ as
\begin{equation}
\begin{aligned}\label{eq:definitionDensity}
    \widetilde{\Delta}_{\rho\mathcal{B}}(\vec{q}_1, t_1, \vec{q}_2, t_2)=&\int\md^3p_1\widetilde{\Delta}_{f\mathcal{B}}({x}_1, t_1, \vec{q}_2, t_2)\,,\\
    G^{(0)}_{\rho\mathcal{B}}(\vec{q}_1, t_1, \vec{q}_2, t_2)=&\int\md^3 p_1G^{(0)}_{f\mathcal{B}}({x}_1, t_1, \vec{q}_2, t_2)\,.
\end{aligned}
\end{equation}
In particular, $\widetilde{\Delta}_{\rho\mathcal{B}}$ itself is now subject to the following self-consistency equation 
\begin{equation}\label{eq:VolterraTildeDeltarhoB}
    \widetilde{\Delta}_{\rho\mathcal{B}}(\vec{q}_1, t_1, \vec{q}_2, t_2)=G^{(0)}_{\rho\mathcal{B}}(\vec{q}_1, t_1, \vec{q}_2, t_2)+\int_{t_2}^{t_1}\md\Bar{t}_1\int\md^3\Bar{q}_1G^{(0)}_{\rho\mathcal{B}}(\vec{q}_1, t_1, \vec{\Bar{q}}_1, \Bar{t}_1)\widetilde{\Delta}_{\rho\mathcal{B}}(\vec{\Bar{q}}_1, \Bar{t}_1, \vec{q}_2, t_2)\,,
\end{equation}
where we have made the causality of $G^{(0)}_{\rho\mathcal{B}}$ explicit through the time integral boundaries. \eqref{eq:VolterraTildeDeltarhoB} is a Volterra-Fredholm integral equation whose solution is, in general, highly non-trivial to find and requires sophisticated numerical techniques. In the next section, however, we present a special case in which an analytical solution can be found. We can now bring \eqref{eq:VolterraTildeDelta} into the form
\begin{equation}\label{eq:DensityToMomentum}
    \widetilde{\Delta}_{f\mathcal{B}}(x_1, t_1, \vec{q}_2, t_2) = \int\md^3\Bar{q}_1\int_{t_2}^{t_1}\md\Bar{t}_1G^{(0)}_{f\mathcal{B}}(x_1, t_1,\vec{\Bar{q}}_1, \Bar{t}_1) \left[\delta_D(\vec{\Bar{q}}_1-\vec{q}_2)\delta_D(\Bar{t}_1-t_2) + \widetilde{\Delta}_{\rho\mathcal{B}}(\Bar{\vec{q}}_1, \Bar{t}_1, \vec{q}_2, t_2)\right]
\end{equation}
where $\widetilde{\Delta}_{\rho\mathcal{B}}$ is a solution to \eqref{eq:VolterraTildeDeltarhoB}. We thus see, that the iteration is not performed with $G^{(0)}_{f\mathcal{B}}$ but rather with $G^{(0)}_{\rho\mathcal{B}}$. The reason for this is that in \eqref{eq:ActionBeforeHST} we artificially included the integration over the momenta in order to keep track of the full phase space information. However, the two particle potential itself only couples to the physical density \eqref{eq:particledensity} defined as the momentum integral over $G_f^{(0)}(X_1)$, 
\begin{equation}
    G_{\rho}^{(0)}(\vec{q}_1, t_1)=\int\md^3p_1G_f^{(0)}(\vec{q}_1, \vec{p}_1, t_1)\,.
\end{equation}
It is therefore sufficient to solve the Volterra-Fredholm equation \eqref{eq:VolterraTildeDeltarhoB} for $\widetilde{\Delta}_{\rho\mathcal{B}}$. Only in the very last step, of the iteration the momentum information is restored in \eqref{eq:DensityToMomentum}. Thus, in order to compute the tree-level or loop contribution, we only need the momentum information of external legs. For instance, if we aim to compute the tree-level expectation value of a one-point operator, 
\begin{equation}
    \mathcal{O}(\vec{q}_1, t_1)=\int\md^3p_1F(\vec{p}_1)f(\vec{q}_1, \vec{p}_1, t_1)\,,
\end{equation}
we find 
\begin{align}\label{eq:OnePointOperatorTreeLevel}
    \mean{\mathcal{O}(\vec{q}_1, t_1)}^{(\mathrm{tree})}=&\int\md^3p_1F(\vec{p}_1)\mean{\Psi_f(\vec{q}_1, \vec{p}_1, t_1)}^{(\mathrm{tree})}\nonumber\\
    =&G_{\mathcal{O}}^{(0)}(\vec{q}_1, t_1)+\int\md\Bar{t}_1\md^3\Bar{q}_1\,\widetilde{\Delta}_{\mathcal{O}\mathcal{B}}(\vec{q}_1, t_1,\vec{\Bar{q}}_1, \Bar{t}_1)\,G_\rho^{(0)}(\vec{\Bar{q}}_1, \Bar{t}_1)\,,
\end{align}
where we defined 
\begin{equation}
\begin{aligned}\label{eq:DensityCumulants}
    G_{\mathcal{O}}^{(0)}(\vec{q}_1, t_1)=&\int\md^3p_1F(\vec{p}_1)G_f^{(0)}(\vec{q}_1, \vec{p}_1, t_1)\,,\\
    \widetilde{\Delta}_{\mathcal{O}\mathcal{B}}(\vec{q}_1, t_1,\vec{\Bar{q}}_1, \Bar{t}_1)=&\int\md^3\Bar{q}_2\int_{\Bar{t}_1}^{t_1}\md\Bar{t}_2G^{(0)}_{\mathcal{O}\mathcal{B}}(\vec{q}_1, t_1,\vec{\Bar{q}}_2, \Bar{t}_2) \left[\delta_D(\vec{\Bar{q}}_1-\vec{q}_2)\delta_D(\Bar{t}_1-t_2) + \widetilde{\Delta}_{\rho\mathcal{B}}(\Bar{\vec{q}}_2, \Bar{t}_2, \Bar{\vec{q}}_1, \Bar{t}_1)\right]\,,\\
    G^{(0)}_{\mathcal{O}\mathcal{B}}(\vec{q}_1, t_1,\vec{\Bar{q}}_1, \Bar{t}_1)=&\int\md^3p_1F(\vec{p}_1)G_{f\mathcal{B}}^{(0)}(\vec{q}_1, \vec{p}_1, t_1, \vec{\Bar{q}}_1, \Bar{t}_1)\,.
\end{aligned}
\end{equation}
The same logic applies to the computation of tree-level two-point correlation functions,
\begin{equation}\label{eq:TwoPointObservable}
    \begin{aligned}
      \mean{\mathcal{O}_1(\vec{q}_1,& t_1)\mathcal{O}_2(\vec{q}_2, t_2)}^{(\mathrm{tree})}_c=\\&G^{(0)}_{\mathcal{O}_1\mathcal{O}_2}(\vec{q}_1, t_1, \vec{q}_2, t_2)+\int_{\ini{t}}^{t_1}\md\Bar{t}_1\int\md^3\Bar{q}_1\widetilde{\Delta}_{\mathcal{O}_1\mathcal{B}}(\vec{q}_1, t_1,\vec{\Bar{q}}_1, \Bar{t}_1)G^{(0)}_{\rho\mathcal{O}_2}(\vec{\Bar{q}}_1, \Bar{t}_1,\vec{q}_2, t_2)\\
      &+\int_{\ini{t}}^{t_2}\md\Bar{t}_2\int\md^3\Bar{q}_2G^{(0)}_{\mathcal{O}_1\rho}(\vec{q}_1, t_1,\vec{\Bar{q}}_2, \Bar{t}_2)\widetilde{\Delta}_{\mathcal{B}\mathcal{O}_2}(\vec{\Bar{q}}_2, \Bar{t}_2,\vec{q}_2, t_2)\\
      &+\int_{\ini{t}}^{t_1}\md\Bar{t}_1\int_{\ini{t}}^{t_2}\md\Bar{t}_2\int\md^3\Bar{q}_1\md^3\Bar{q}_2\,\widetilde{\Delta}_{\mathcal{O}_1\mathcal{B}}(\vec{q}_1, t_1,\vec{\Bar{q}}_1, \Bar{t}_1)G^{(0)}_{\rho\rho}(\vec{\Bar{q}}_1, \Bar{t}_1,\vec{\Bar{q}}_2, \Bar{t}_2)\,\widetilde{\Delta}_{\mathcal{B}\mathcal{O}_2}(\vec{\Bar{q}}_2, \Bar{t}_2,\vec{q}_2, t_2)\,,
\end{aligned}
\end{equation}
where $G^{(0)}_{\mathcal{O}_1\mathcal{O}_2}$, $G^{(0)}_{\mathcal{O}_1\rho}$ and $G^{(0)}_{\rho\rho}$ are defined in the same way as in \eqref{eq:DensityCumulants}. All higher-order correlators and loop corrections can be obtained analogously. In particular, if we are solely interested in information on the physical density $\rho$, we can integrate over all momenta and only need cumulants $G^{(0)}_{\rho\cdots\rho\mathcal{B}\cdots\mathcal{B}}$ of the physical density in all equations. 

\subsection{Tree-level Density and Momentum Correlations for Homogeneous Systems}
\label{sec:HomogeneousSystem}
Last but not least, we illustrate the application of our formalism to the case of a homogeneous system of $N$ particles confined to a finite volume $V$. In the end, we will perform the limit $N,V\rightarrow\infty$ while keeping the mean background density constant, $\Bar{\rho}=N/V=\mathrm{const.}$. Since the system is assumed to be homogeneous at initial time the reduced initial one-particle density is constant in space, \ie
\begin{equation}
    f_1(\vini{q}_1, \vini{p}_1, \ini{t})=\Bar{\rho}\,\varphi(\vini{p}_1)\,,
\end{equation}
where the momentum distribution function $\varphi(\vini{p})$ is normalized,
\begin{equation}
    \int\md^3\ini{p}\varphi(\vini{p})=1\,,
\end{equation}
and describes the momentum dispersion of the individual particles. Due to homogeneity, the reduced initital two-particle density only depends on the relative positions, such that we can generally write,
\begin{align}
    f_2(\vini{q}_1, \vini{p}_1, \vini{q}_2, \vini{p}_2, \ini{t})&=f_2(\vini{q}_1-\vini{q}_2, \vini{p}_1, \vini{p}_2, \ini{t})\,,\\
    g_2(\vini{q}_1-\vini{q}_2, \vini{p}_1, \vini{p}_2, \ini{t})&=f_2(\vini{q}_1-\vini{q}_2, \vini{p}_1, \vini{p}_2, \ini{t})-\Bar{\rho}^2\varphi(\vini{p}_1)\varphi(\vini{p}_2)\,.
\end{align}
We also define the initial radial correlation function $\xi(\vini{q}_1-\vini{q}_2, \ini{t})$ as 
\begin{equation}
    \int\md^3p_1\md^3p_2f_2(\vini{q}_1-\vini{q}_2, \vini{p}_1, \vini{p}_2, \ini{t})=\Bar{\rho}^2\left[1+\xi(\vini{q}_1-\vini{q}_2, \ini{t})\right]\,.
\end{equation}
Our goal will be to derive tree-level expressions for the following density and momentum correlators 
\begin{equation}\label{eq:Example}
    \mean{\rho(\vec{q}_1, t_1)}\,,\,\,\,\mean{\rho(\vec{q}_1, t_1)\rho(\vec{q}_2, t_2)}\,,\,\,\,\mean{\vec{\Pi}(\vec{q}_1, t_1)}\,,\,\,\,\mean{\vec{\Pi}(\vec{q}_1, t_1)\otimes\vec{\Pi}(\vec{q}_2, t_2)}
\end{equation}
in this general setting. Therefore, the required macroscopic phase space density cumulants are $G_f^{(0)}$, $G_{ff}^{(0)}$ and $G_{f\mathcal{B}}^{(0)}$ which can be computed as described in appendix \ref{sec:Cumulants}. For our system they read
\begin{equation}
    \begin{aligned}
    G_f^{(0)}(\vec{p}_1, t_1)=&\,\Bar{\rho}\,\varphi(\vec{p}_1)\,,\\
    G_{f\mathcal{B}}^{(0)}(\vec{R}, \vec{p}_1, t_1, t_2)=&\,\Bar{\rho}\nabla_{\vec{R}}\,v(|\vec{R}-\frac{\vec{p}_1}{m}(t_1-t_2)|,t_2)\cdot\nabla_{\vec{p}_1}\varphi(\vec{p}_1)\Theta(t_1-t_2)\,,\label{eq:homoGrhoB}\\
    G_{ff}^{(0)}(\vec{R}, \vec{p}_1, t_1, \vec{p}_2, t_2)=&\,\delta_D(\vec{R}-\frac{\vec{p}_2}{m}(t_1-t_2))\delta_D(\vec{p}_1-\vec{p_2})\Bar{\rho}\,\varphi(\vec{p}_1)\\&+g_2(\vec{R}-\frac{\vec{p}_1}{m}(t_1-\ini{t})+\frac{\vec{p}_2}{m}(t_2-\ini{t}), \vec{p}_1, \vec{p}_2, \ini{t})\,,
\end{aligned}
\end{equation}
where we made explicit that all cumulants only depend on the difference $\vec{R}=\vec{q}_1-\vec{q}_2$ due to homogeneity. From those we can infer the macroscopic density and momentum cumulants $G_\rho^{(0)}$, $\vec{G}_\Pi^{(0)}$, $G_{\rho\mathcal{B}}^{(0)}$, $\vec{G}_{\Pi\mathcal{B}}^{(0)}$, $G_{\rho\rho}^{(0)}$, $\vec{G}_{\Pi\rho}^{(0)}$, $\vec{G}_{\rho\Pi}^{(0)}$ and $G_{\Pi\otimes\Pi}^{(0)}$. We list their analytical expressions in appendix \ref{sec:CumulantsHomogeneous}. The Volterra-Fredholm integral equation \eqref{eq:VolterraTildeDeltarhoB} for $\widetilde{\Delta}_{\rho\mathcal{B}}$ for the homogeneous system now reads
\begin{align}\label{eq:VolterraFredholmHomogeneous}
    \widetilde{\Delta}_{\rho\mathcal{B}}(\vec{R}, t_1, t_2)=G_{\rho\mathcal{B}}^{(0)}(\vec{R}, t_1, t_2)+\int_{t_2}^{t_1}\md\Bar{t}_1\int\md^3XG_{\rho\mathcal{B}}^{(0)}(\vec{R}-\vec{X}, t_1, \Bar{t}_1)\,\widetilde{\Delta}_{\rho\mathcal{B}}(\vec{X}, \Bar{t}_1, t_2)\,.
\end{align}
As the integral in \eqref{eq:VolterraFredholmHomogeneous} is a convolution in $\vec{X}$, we can Fourier transform it to eliminate the integration over $\vec{X}$,
\begin{equation}\label{eq:Volterra_k}
    \widetilde{\Delta}_{\rho\mathcal{B}}(\vec{k}, t_1, t_2)=G_{\rho\mathcal{B}}^{(0)}(\vec{k}, t_1, t_2)+\int_{t_2}^{t_1}\md\Bar{t}_1 G_{\rho\mathcal{B}}^{(0)}(\vec{k}, t_1, \Bar{t}_1)\,\widetilde{\Delta}_{\rho\mathcal{B}}(\vec{k}, \Bar{t}_1, t_2)\,,
\end{equation}
where $\vec{k}$ is the Fourier conjugate variable to $\vec{R}$. Equation \eqref{eq:Volterra_k} is a simple Volterra integral equation, which can be solved numerically upon time discretization. Due to the causal structure of $G_{\rho\mathcal{B}}^{(0)}$ the resulting matrix equation consists of inverting an upper triangular matrix via forward substitution. If we furthermore restrict to time-translational invariant systems, \ie time independent potentials, we see from \eqref{eq:homoGrhoB} that both, $G_{\rho\mathcal{B}}^{(0)}$ and hence $\widetilde{\Delta}_{\rho\mathcal{B}}$, are functions of time differences $t_1-t_2$ which turns the remaining time integral in \eqref{eq:Volterra_k} into a Laplace-convolution. Hence, it can be solved by means of a Laplace transformation, 
\begin{equation}
    \mathcal{L}\left[\widetilde{\Delta}_{\rho\mathcal{B}}(\vec{k}, t_1- t_2)\right](s) \cdot \left(1-\mathcal{L}\left[G_{\rho\mathcal{B}}^{(0)}(\vec{k}, t_1- t_2)\right](s)\right) = \mathcal{L}\left[G_{\rho\mathcal{B}}^{(0)}(\vec{k}, t_1- t_2)\right](s)\;,
\end{equation}
such that the resulting equation can be solved algebraically for $\widetilde{\Delta}_{\rho\mathcal{B}}$, with the solution 
\begin{equation}\label{eq:LaplaceVolterra}
    \widetilde{\Delta}_{\rho\mathcal{B}}(\vec{k}, t_1- t_2) = \mathcal{L}^{-1}\left[\frac{\mathcal{L}\left[G_{\rho\mathcal{B}}^{(0)}(\vec{k}, t_1- t_2)\right](s)}{1-\mathcal{L}\left[G_{\rho\mathcal{B}}^{(0)}(\vec{k}, t_1- t_2)\right](s)}\right](t_1- t_2)\,,
\end{equation}
where $s$ is the Laplace conjugate to $(t_1-t_2)$.
Independent of how we obtain the solution of $\widetilde{\Delta}_{\rho\mathcal{B}}$, either numerically or analytically, we can now proceed with the computation of the tree-level expectation values \eqref{eq:Example}. We start with the one-point observables $\mean{\rho}$ and $\mean{\vec{\Pi}}$. According to \eqref{eq:OnePointOperatorTreeLevel} and using the explicit expressions listed in \ref{sec:CumulantsHomogeneous} for the density and momentum cumulants we find 
\begin{align}
    \mean{\rho(\vec{q}_1, t_1)}^{(\mathrm{tree})}=&G_\rho^{(0)}(\vec{q}_1, t_1)+\int_{\ini{t}}^{t_1}\md \Bar{t}_1\int\md^3\Bar{q}_1\widetilde{\Delta}_{\rho\mathcal{B}}(\vec{q}_1-\vec{\Bar{q}}_1, t_1, \Bar{t}_1)G_\rho^{(0)}(\vec{\Bar{q}}_1, \Bar{t}_1)\\
    =&\Bar{\rho} + \Bar{\rho}\int_{\ini{t}}^{t_1}\md \Bar{t}_1\int\md^3\Bar{q}_1\widetilde{\Delta}_{\rho\mathcal{B}}(\vec{q}_1-\vec{\Bar{q}}_1, t_1, \Bar{t}_1)\,,
\end{align}
and also 
\begin{align}
    \mean{\vec{\Pi}(\vec{q}_1, t_1)}^{(\mathrm{tree})}=&\vec{G}_\Pi^{(0)}(\vec{q}_1, t_1)+\int_{\ini{t}}^{t_1}\md \Bar{t}_1\int\md^3\Bar{q}_1\vec{G}_{\Pi\mathcal{B}}(\vec{q}_1-\vec{\Bar{q}}_1, t_1, \Bar{t}_1)G_\rho^{(0)}(\vec{\Bar{q}}_1, \Bar{t}_1)\\
    &+\int_{\ini{t}}^{t_1}\md \Bar{t}_1\int_{\ini{t}}^{\Bar{t}_1}\md \Bar{t}_2\int\md^3\Bar{q}_1\md^3\Bar{q}_2\vec{G}_{\Pi\mathcal{B}}(\vec{q}_1-\vec{\Bar{q}}_1, t_1, \Bar{t}_1)\widetilde{\Delta}_{\rho\mathcal{B}}(\vec{\Bar{q}}_1-\vec{\Bar{q}}_2, \Bar{t}_1, \Bar{t}_2)G_\rho^{(0)}(\vec{\Bar{q}}_2, \Bar{t}_2)\nonumber\\
    =&0 +\Bar{\rho}\int_{\ini{t}}^{t_1}\md \Bar{t}_1\int\md^3\Bar{q}_1\vec{G}_{\Pi\mathcal{B}}(\vec{q}_1-\vec{\Bar{q}}_1, t_1, \Bar{t}_1)\\
    &+\Bar{\rho}\int_{\ini{t}}^{t_1}\md \Bar{t}_1\int_{\ini{t}}^{\Bar{t}_1}\md \Bar{t}_2\int\md^3\Bar{q}_1\md^3\Bar{q}_2\vec{G}_{\Pi\mathcal{B}}(\vec{q}_1-\vec{\Bar{q}}_1, t_1, \Bar{t}_1)\widetilde{\Delta}_{\rho\mathcal{B}}(\vec{\Bar{q}}_1-\vec{\Bar{q}}_2, \Bar{t}_1, \Bar{t}_2)\,.\nonumber
\end{align}
Both observables hence formally acquire tree-level corrections to their free evolution. However, as one can easily see from \eqref{eq:homoGrhoB} we find 
\begin{equation}
    \int\md^3R\,G_{f\mathcal{B}}^{(0)}(\vec{R}, \vec{p}_1, t_1, t_2)=0\,,
\end{equation}
due to homogeneity. As a consequence, both tree-level corrections vanish: The mean density stays constant while the mean momentum remains zero. From the physical point of view this was to be expected, since in a homogeneous system with homogeneous potential there is no mean force acting on a given particle. Hence, the system stays consistently homogeneous\footnote{The same can be shown to any loop order, which is covered in future work.}, 
\begin{equation}\label{eq:treelevelmean}
    \mean{\rho(\vec{q}_1, t_1)}^{(\mathrm{tree})}=\Bar{\rho}\,,\,\,\,\,\mean{\vec{\Pi}(\vec{q}_1, t_1)}^{(\mathrm{tree})}=0\,.
\end{equation}
Using \eqref{eq:TwoPointObservable} we find for the connected component of the tree-level density 
\begin{align}
    \mean{\rho(\vec{q}_1, t_1)&\rho(\vec{q}_2, t_2)}^{(\mathrm{tree})}_c=\nonumber\\& G^{(0)}_{\rho\rho}(\vec{q}_1-\vec{q}_2, t_1, t_2)+\int_{\ini{t}}^{t_1}\md\Bar{t}_1\int\md^3X\widetilde{\Delta}_{\rho\mathcal{B}}(\vec{q}_1-\vec{q}_2-\vec{X}, t_1, \Bar{t}_1)G^{(0)}_{\rho\rho}(\vec{X}, \Bar{t}_1, t_2)\nonumber\\
      &+\int_{\ini{t}}^{t_2}\md\Bar{t}_2\int\md^3YG^{(0)}_{\rho\rho}(\vec{Y}, t_1,\Bar{t}_2)\widetilde{\Delta}_{\mathcal{B}\rho}(\vec{q}_1-\vec{q}_2-\vec{Y}, \Bar{t}_2, t_2)\\
      &+\int_{\ini{t}}^{t_1}\md\Bar{t}_1\int_{\ini{t}}^{t_2}\md\Bar{t}_2\int\md^3X\md^3Y\,\widetilde{\Delta}_{\rho\mathcal{B}}(\vec{q}_1-\vec{q}_2-\vec{X}, t_1, \Bar{t}_1)G^{(0)}_{\rho\rho}(\vec{X}, \Bar{t}_1, \Bar{t}_2)\,\widetilde{\Delta}_{\mathcal{B}\rho}(\vec{Y}-\vec{X}, \Bar{t}_2, t_2)\,\nonumber\\
      \equiv&\Bar{\rho}^2\xi(\vec{R}, t_1, t_2)^{(\mathrm{tree})}\,,\nonumber
\end{align}
where in the last line we defined the tree-level unequal-time radial correlation function $\xi(\vec{R}, t_1, t_2)^{(\mathrm{tree})}$. Again, homogeneity ensures that it is only a function of the difference $\vec{R}=\vec{q}_1-\vec{q}_2$. Note that the above equation is again a convolution in the spatial variables, such that the equation is again easier to compute in Fourier space. The same holds true for the two-point momentum expectation value, which results in a similar but significantly longer expression. Omitting all integrals for better readability it reads
\begin{equation}
\begin{aligned}
    \mean{\vec{\Pi}(\vec{q}_1, t_1)&\otimes\vec{\Pi}(\vec{q}_2, t_2)}=G_{\Pi\otimes\Pi}^{(0)}+\vec{G}_{\Pi\mathcal{B}}^{(0)}\otimes  \vec{G}_{\rho\Pi}^{(0)}+\vec{G}_{\Pi\mathcal{B}}^{(0)}\otimes \widetilde{\Delta}_{\rho\mathcal{B}}\cdot \vec{G}_{\rho\Pi}^{(0)}+\vec{G}_{\Pi\rho}^{(0)}\otimes\vec{G}_{\mathcal{B}\Pi}^{(0)}\\&+\vec{G}_{\Pi\rho}^{(0)} \widetilde{\Delta}_{\mathcal{B}\rho}\otimes \vec{G}_{\mathcal{B}\Pi}^{(0)}+\vec{G}_{\Pi\mathcal{B}}^{(0)}\otimes G_{\rho\rho}^{(0)}\cdot \vec{G}_{\mathcal{B}\Pi}+\vec{G}_{\Pi\mathcal{B}}^{(0)}\otimes \widetilde{\Delta}_{\rho\mathcal{B}}\cdot G_{\rho\rho}^{(0)}\cdot \vec{G}_{\mathcal{B}\Pi}\\ & +\vec{G}_{\Pi\mathcal{B}}^{(0)}\otimes G_{\rho\rho}^{(0)}\cdot\widetilde{\Delta}_{\mathcal{B}\rho}\cdot \vec{G}_{\mathcal{B}\Pi}+\vec{G}_{\Pi\mathcal{B}}^{(0)}\otimes\widetilde{\Delta}_{\mathcal{B}\rho}\cdot G_{\rho\rho}^{(0)}\cdot\widetilde{\Delta}_{\mathcal{B}\rho}\cdot \vec{G}_{\mathcal{B}\Pi}\,,
\end{aligned}
\end{equation}
 on a schematic level.

\subsubsection{Low-Temperature Limit}
Before concluding, we present a case in which \eqref{eq:LaplaceVolterra} is exactly solvable. Consider a homogeneous system of particles in which the momentum distribution function $\varphi(\vec{p})$ is given by a Maxwell–Boltzmann distribution, \ie
\begin{equation}
    \varphi(\vec{p})=\left(\frac{m}{2\pi T}\right)^{\frac{3}{2}}\e^{-\frac{m}{2T}\vec{p}^{\,2}}\,,
\end{equation}
with the temperature $T$. From \eqref{eq:AppendixGrB} we find that the exact expression for $G_{\rho\mathcal{B}}^{(0)}(\vec{R}, t_1, t_2)$ is given by 
\begin{equation}
    G_{\rho\mathcal{B}}^{(0)}(\vec{R}, t_1, t_2)=\frac{\Bar{\rho}m^2}{(t_1-t_2)^2}\left(\frac{m}{2\pi T}\right)^{\frac{3}{2}}\int\md^3X\Delta_{\vec{X}}\,v(|\vec{X}|, t_2)\e^{-\frac{m^3}{2T}\left(\frac{\vec{R}-\vec{X}}{t_1-t_2}\right)^2}\,.
\end{equation}
According to the discussion around \eqref{eq:Volterra_k} we Fourier transform $G_{\rho\mathcal{B}}^{(0)}$ and find 
\begin{equation}
    G_{\rho\mathcal{B}}^{(0)}(\vec{k}, t_1, t_2)=-\frac{\Bar{\rho}\vec{k}^{\,2}\widetilde{v}(|\vec{k}|, t_2)(t_1-t_2)}{m}\e^{-\frac{T(t_1-t_2)^2}{m^3}\vec{k}^{\,2}}\,,
\end{equation}
where $\widetilde{v}(|\vec{k}|, t_2)$ is the Fourier transformed two-particle potential which only depends on the absolute value of $\vec{k}$. In the limit of small temperatures we can expand the exponential to lowest order and find 
\begin{equation}
    G_{\rho\mathcal{B}}^{(0)}(\vec{k}, t_1, t_2)\stackrel{T\ll1}{=}-\frac{\Bar{\rho}\vec{k}^{\,2}\widetilde{v}(|\vec{k}|, t_2)(t_1-t_2)}{m}\,.
\end{equation}
If we further assume the potential to be time-independent, we can perform the Laplace transform for $G_{\rho\mathcal{B}}^{(0)}$ \wrt $(t_1-t_2)$ analytically,
\begin{equation}
    \mathcal{L}\left[G_{\rho\mathcal{B}}^{(0)}(\vec{k}, t_1- t_2)\right](s)=-\frac{\Bar{\rho}\vec{k}^{\,2}\widetilde{v}(|\vec{k}|)}{ms^2}\,.
\end{equation}
Using the inverse Laplace transform of the form
\begin{equation}
    \mathcal{L}^{-1}\left[\frac{-A}{s^2+A}\right](t)=-\sqrt{A}\sin\left(\sqrt{A}t\right)\,,
\end{equation}
we get a closed expression for the causal propagator $\widetilde{\Delta}_{\rho\mathcal{B}}(\vec{k}, t_1-t_2)$, given by
\begin{equation}\label{eq:GeneralLowT}
    \widetilde{\Delta}_{\rho\mathcal{B}}(\vec{k}, t_1-t_2)=-\omega(k)\sin\Big(\omega(k)(t_1-t_2)\Big)\,,
\end{equation}
with the $k$-dependent frequency 
\begin{equation}\label{eq:Frequency}
    \omega(k)=\sqrt{\frac{\Bar{\rho}k^2\widetilde{v}(k)}{m}}\,.
\end{equation}
Equation \eqref{eq:GeneralLowT} is the general solution of the causal tree-level propagator in the low-temperature limit of a Boltzmann-Gas. Interestingly, we see that if the potential is repulsive, \ie $\widetilde{v}(k)>0$, $\widetilde{\Delta}_{\rho\mathcal{B}}$ exhibits oscillatory behavior with frequency $\omega(k)$. For instance, consider the Coulomb potential, 
\begin{equation}
    v_C(\vec{q}_1-\vec{q}_2)=\frac{e^2}{4\pi\varepsilon_0}\frac{1}{|\vec{q}_1-\vec{q}_2|}\,,
\end{equation}
whose Fourier transform is
\begin{equation}
    \widetilde{v}_C(k)=\frac{e^2}{\epsilon_0}\frac{1}{k^2}\,.
\end{equation}
Inserting this into \eqref{eq:Frequency} we recover the well-known \textit{plasma frequency} (or \textit{Langmuir} frequency),
\begin{equation}
    \omega(k)=\sqrt{\frac{\Bar{\rho}e^2}{m\epsilon_0}}\equiv\omega_P\,.
\end{equation}
In a homogeneous system where the thermal motion of the particles is negligible, a repulsive inter particle potential, such as the Coulomb potential, will lead to a collective oscillating behaviour
within the system. It arises due to the system's response to a local perturbation as it tries to restore homogeneity.
The charged particles, thus, fall back to their equilibrium position and oscillate around it. This effect is completely covered by the tree-level theory of our approach. \\

On the other hand, if the potential is attractive, $\widetilde{v}(k)<0$, we find 
\begin{equation}
    \omega(k)=i\sqrt{\frac{\Bar{\rho}k^2|\widetilde{v}(k)|}{m}}=i|\omega(k)|\,,
\end{equation}
and thus 
\begin{equation}
    \widetilde{\Delta}_{\rho\mathcal{B}}(\vec{k}, t_1-t_2)=-i|\omega(k)|\sin\Big(i|\omega(k)|(t_1-t_2)\Big)=|\omega(k)|\sinh\Big(|\omega(k)|(t_1-t_2)\Big)\,.
\end{equation}
In this case, the propagator consists of an exponentially growing and an exponentially decaying mode. As an example, consider the attractive Newtonian gravitational potential,
\begin{equation}
    v_G(\vec{q}_1-\vec{q}_2)=-\frac{Gm^2}{|\vec{q}_1-\vec{q}_2|}\,,
\end{equation}
with Fourier transform 
\begin{equation}
    \widetilde{v}_G(k)=-\frac{4\pi Gm^2}{k^2}\,.
\end{equation}
The propagator now reads 
\begin{equation}
    \widetilde{\Delta}_{\rho\mathcal{B}}(t_1-t_2)=\omega_G\sinh\Big(\omega_G(t_1-t_2)\Big)\,,
\end{equation}
with the growth rate 
\begin{equation}
    \omega_G=\sqrt{4\pi Gm\Bar{\rho}}\,.
    \label{eq:growth_rate}
\end{equation}
In a homogeneous system with negligible thermal motion, attractive forces such as the Newtonian gravitational potential lead to exponentially growing structures due to local perturbations. On a physical level a local overdensity in a self-gravitating gas leads to a collapse of the whole surrounding system and thus to an exponentially growing perturbation. 
This result is famously 
known as Jeans instability. The frequency we recovered in \eqref{eq:growth_rate} is related to the characteristic timescale of collapse for a collisionless fluid $\tau$ by $\tau=\frac{1}{\omega_G}$ and is usually referred to as the free-fall collapse time. This same effect is responsible for the growth of cosmic large-scale structures on the largest scales, where the lowest order approximation exhibits ``linear''\footnote{Normally, the gravitational potential in cosmology would be time dependent, due to the expanding background. However, by a convenient choice of the time coordinate this time dependence can be neglected. The resulting structure growth is then linear in the new time variable.} growth of initial perturbations. This effect is also incorporated in the tree-level description of the theory. We can thus easily show that plasma oscillations as well as gravitational collapse in astrophysics and cosmology have the same origin and can both be described already at tree-level. 

\section{Outlook and Conclusion}
In the first part of this paper we have used the path integral representation of the time evolution operator of classical mechanics in order to construct a perturbative approach to the time evolution of a given initial state of the system. We have expanded the full propagator of the theory in a Dyson series and showed that we could express the evolution of the phase space density as a classical analogue of the Lippmann-Schwinger equation. In essence this allows us to consider the (perturbative) solution of the Liouville equation in terms of a scattering process. Our microscopic formalism thus formally resembles many-body quantum mechanics. This resemblance can be well understood through the Koopmann-von Neumann formalism. 

As discussed in Sec.\ \ref{sec:MicroscopicTheory}, the path integral based approach is equivalent to the (iterative) solution of the Liouville equation and as such of the BBGKY-hierarchy. Consequently, it does not provide an improvement over the textbook approach to solving the Liouville equation in terms of an iterative solution of its Green function. In fact, this approach suffers from two major issues:
First, the number of terms that have to be evaluated grows rapidly with increasing order of perturbation theory since the number of possible interactions within a cluster of particles increases quickly if more particles are involved. Just imagine that computing the second order contribution to the two-point correlation $\mean{f(  {x}_1, t_1)f(  {x}_2, t_2)}$ would involve the combinatorics for up to five interacting particles, which results in fifteen terms (see Fig.\ \ref{fig:PT_twoPoint}). As there is no additional consistent truncation criterion, all terms have to be included such that the computation becomes very complex already at low orders of the perturbative expansion. Second, each additional perturbative order introduces a further particle into the cluster, which is also initially correlated with the other particles. Thus the knowledge of increasingly higher-order reduced initial phase space densities is required, which renders the computation impracticable in general. Again, there is no criterion which would consistently truncate the initial correlations that have to be included. 
The main result of this paper is, therefore, the development of a non-perturbative description of the system in terms of macroscopic fields which allows us to overcome both these issues.

Using a slightly modified version of the HST, we have thus constructed an effective field theoretic description in terms of macroscopic fields. The underlying microscopic statistics of the system is encoded in the vertices of this macroscopic field theory. The macroscopic propagator re-sums a particular class of microscopic particle interactions to infinite order. The propagators and vertices of the macroscopic field theory can be directly related to terms appearing in the microscopic perturbative expansion as we discuss in Sec.\ \ref{sec:MicroMacro}. Our macroscopic theory, thus, re-structures microscopic perturbation theory in a convenient way, allowing us to separate effects of the interaction potential from effects of initial correlations. This gives us a much better physical understanding of possible truncations for the macroscopic theory. 

Expansion schemes for the macroscopic theory will typically involve a truncation in terms of vertices. Since the vertices are simply free cumulants of the phase space density, such a truncation automatically induces a truncation on the microscopic level. However, a truncation on the level of vertices of the microscopic theory offers a better way to include (or neglect) certain microscopic effects. For instance, if we only keep pure $f$-cumulants as vertices, we keep only the information on initial correlations of the particles and neglect interactions which would appear in the mixed $f\mathcal{B}$-cumulants. In this way we can decide if we want to include more effects of the interactions between particles or if we deem initial correlation effects more important. In future work we intend to study whether there exists a hierarchy of vertices which can help us in finding a suitable truncation scheme. 

The causal propagator $\Delta_{f\mathcal{B}}$ of our theory is simply given by the Neumann series of $G^{(0)}_{f\mathcal{B}}$ and therefore only contains effects of interactions. It therefore describes how correlations are propagated in time, but does not account for feedback effects, which would appear as self-energy corrections of the statistical propagator such as $\Delta_{ff}$.
In this paper, we have limited our discussion to the tree-level theory. Such self-energy corrections will be the subject of future work where we shall make extensive use of the Feynman diagrams introduced in Sec.\ \ref{sec:Feynman}. 

To round up the discussion on the tree-level theory, we have provided an explicit application of the theory to a spatially homogeneous system in Sec.\ \ref{sec:HomogeneousSystem}. Furthermore, we have specified the momentum distribution to a Maxwell-Boltzmann distribution and computed the causal propagator in the low-temperature limit for a system where the particles interact through a Newtonian gravitational potential and for a system where the particles interact via a Coulomb potential. For the latter, we have recovered the collective oscillating behaviour at the Langmuir frequency related to plasma oscillations. For the gravitational system we have recovered the collective behaviour leading to a gravitational instability with the characteristic time-scale known from Jeans theory.
Through our formalism we could show that both effects are the result of the same collective behaviour.

In a separate paper we will, in fact, apply the formalism presented here to a realistic setting in the field of cosmic large-scale structure formation and show that the analytic results obtained with our tree-level theory indeed reproduce results known from numerical $N$-body simulation and observations of cosmic large-scale structures. 

\section*{Acknowledgements}
We would like to first and foremost thank Robert Lilow for his excellent and most helpful input. We would also very much like to thank Marvin Sipp, Björn Malte Schäfer and Matthias Bartelmann for valuable discussions and their comments on this work.
\paragraph*{Funding information} EK acknowledges funding by the Deutsche Forschungsgemeinschaft (DFG, German Research Foundation) -- $452923686$. TD is funded by the Studienstiftung des deutschen Volkes.

\appendix
\section{Derivation of the Discretized Path Integral}\label{sec:discretization}

In this appendix we describe the derivation of the path integral representation of the classical transition amplitude \eqref{eq:PropagatorPathIntegral} motivated in section \ref{sec:Construction} by the usual time discretization procedure. We follow the derivation of the path integral representation of the Langevin equation (see \eg \cite{Nakazato:1990kk}), which is slightly different from the motivation given in the main text. In order to keep the notation as simple as possible we restrict the construction to a simpler dynamical equation and generalize the result to \eqref{eq:eom}. Let us therefore consider the dynamical equation\footnote{This, of course, corresponds to a one-dimensional Langevin system with vanishing noise. Indeed, classical mechanics can be seen as a Langevin diffusion process where the noise distribution function is set to zero by a Dirac-Delta distribution \cite{Gozzi:1991wi} which makes the evolution is deterministic.},
\begin{equation}\label{eq:DynamicalEquAppendix}
    \dot{x}(t)=F(x(t))\,,
\end{equation}
for some real scalar variable $x(t)$ and force field $F(x(t))$. The initial value $x(\ini{t})$ is randomly distributed by the normalized probability distribution $\varrho(\ini{x}, \ini{t})$. We discretize the time interval between the initial time $\ini{t}\equiv t_0$ and the final time $\fin{t}\equiv t_N$ into $N$ steps of length $\epsilon$ and use the abbreviation $x_k=x(t_k)$. The time discretization now introduces an ambiguity into the evolution equation as it is a priori not clear at which end point of every finite difference time interval the right hand side of \eqref{eq:DynamicalEquAppendix} has to be evaluated. Following \cite{Nakazato:1990kk} we therefore define the discretized version of \eqref{eq:DynamicalEquAppendix} as 
\begin{equation}
    \frac{x_k-x_{k-1}}{\epsilon}=aF(x_{k-1}) + (1-a)F(x_k)\,,
\end{equation}
where the parameter $a\in[0,1]$ has been introduced in order to track this ambiguity in the following derivation. In the context of stochastic dynamics, we have two famous prescriptions, known as the Itô-related interpretation for $a=1$ or the Stratonovich related interpretation for $a=\frac{1}{2}$. For convenience, we introduce 
\begin{equation}\label{eq:Ek}
    \mathcal{E}_{k,k-1}\equiv\frac{x_k-x_{k-1}}{\epsilon}-(aF(x_{k-1}) + (1-a)F(x_k))\,.
\end{equation}
Starting from the initial probability distribution, we define the partition function 
\begin{equation}
    \mathcal{Z}=\int\md\ini{x}\varrho(\ini{x}, \ini{t})\,,
\end{equation}
and update it $N$-times by inserting a ``$1$'' à la Faddeev-Popov, 
\begin{equation}
    \mathcal{Z}=\int\md x_0\underbrace{\left[\prod_{k=1}^N\int\md\mathcal{E}_{k,k-1}\,\delta_D(\mathcal{E}_{k,k-1})\right]}_{=1}\varrho(x_0, t_0)\,.
\end{equation}
We can now perform a change of variables in order to write 
\begin{equation}\label{eq:DeltaEk}
    \mathcal{Z}=\int\md x_0\int\left[\prod_{k=1}^N\md x_k\,\delta_D(\mathcal{E}_{k,k-1})\right]\mathcal{J}\varrho(x_0, t_0)\,,
\end{equation}
where the Jacobian $\mathcal{J}$ is defined as the determinant
\begin{equation}
    \mathcal{J}=\Bigg\rvert\frac{\partial(\mathcal{E}_{1,0}, \cdots, \mathcal{E}_{N, N-1})}{\partial(x_1, \cdots, x_N)}\Bigg\lvert\,.
\end{equation}
Introducing auxiliary fields $\chi_k$ in order to represent the Dirac-Delta distributions in \eqref{eq:DeltaEk} by an exponential function, we find 
\begin{equation}
    \mathcal{Z}=\int\md x_0\int\left[\prod_{k=1}^N\md x_k\, \frac{\md\chi_k}{2\pi}\right]\,\exp\left[\mi \sum_{k=1}^N\chi_k\cdot\mathcal{E}_{k,k-1}\right]\mathcal{J}\,\varrho(x_0, t_0)\,.
\end{equation}
The determinant itself can easily be computed from \eqref{eq:Ek}. It results in the determinant of a lower triangular matrix and is thus the product of the diagonal entries given by
\begin{equation}
    \mathcal{J}=\prod_{k=1}^N\left(\frac{1}{\epsilon}-(1-a)F^\prime(x_k)\right)=\frac{1}{\epsilon^N}\prod_{k=1}^N\left(1-(1-a)\epsilon F^\prime(x_k)\right)\,.
\end{equation}
Importantly, for $a=1$ the determinant is constant, while for $a=\frac{1}{2}$ the determinant depends on the variable $x$. Adopting Itô's convention ($a=1$), we can absorb the $\frac{1}{\epsilon^N}$ factor into the $\chi$ fields and find for the partition function
\begin{equation}
    \mathcal{Z}=\int\md x_0\int\left[\prod_{k=1}^N\md x_k\, \frac{\md\chi_k}{2\pi}\right]\,\exp\left[\mi \epsilon\sum_{k=1}^N\chi_k\cdot\left(\frac{x_k-x_{k-1}}{\epsilon}-F(x_{k-1}) \right)\right]\,\varrho(x_0, t_0)\,,
\end{equation}
where we explicitly included \eqref{eq:Ek}. Note, that there is no divergent $1/\epsilon$ in the above equation. Extracting the integration over $\md x_N=\md \fin{x}$ from the product, we may interpret the remaining integrals as a transition probability, propagating the initial probability distribution from $\ini{t}$ to $\fin{t}$, \ie
\begin{equation}
    \mathcal{Z}=\int\md\fin{x}\varrho(\fin{x}, \fin{t})=\int\md\fin{x}\ini{x}K(\fin{x}, \fin{t}|\ini{x}, \ini{t})\varrho(\ini{x}, \ini{t})\,,
\end{equation}
with 
\begin{align}
    K(\fin{x}, \fin{t}|\ini{x}, \ini{t})=&\int\limits_{\ini{x}}^{\fin{x}}\left[\prod_{k=1}^{N-1}\md x_k\right]\left[\prod_{k=1}^N\frac{\md\chi_k}{2\pi}\right]\,\exp\left[\mi \epsilon\sum_{k=1}^N\chi_k\cdot\left(\frac{x_k-x_{k-1}}{\epsilon}-F(x_{k-1}) \right)\right]\label{eq:discretizedPI}\\
    \overset{N\rightarrow\infty}{\underset{\epsilon\rightarrow0}{\longrightarrow}}&\int\limits_{\ini{x}}^{\fin{x}}\mathcal{D}x(t)\mathcal{D}\chi(t)\exp\left[\mi \int\limits_{\ini{t}}^{\fin{t}}\md t\chi(t)\cdot\left(\dot{x}(t)-F(x(t)) \right)\right]\,,\label{eq:ContinuumPropagator}
\end{align}
where we inserted the definition of the Riemann-integral
\begin{equation}
   \int\limits_{\ini{t}}^{\fin{t}}\md t=\underset{\epsilon\rightarrow0}{\lim}\sum_{k=1}^N\epsilon\,,
\end{equation}
and defined the path integral measure as 
\begin{equation}
    \mathcal{D}x(t)\equiv\underset{N\rightarrow\infty}{\lim}\int\prod_{k=1}^{N-1}\md x_k\,,\,\,\,\,\,\,\mathcal{D}\chi(t)\equiv\underset{N\rightarrow\infty}{\lim}\int\prod_{k=1}^{N}\frac{\md\chi_k}{2\pi}\,.
\end{equation}
Note, that within the transition amplitude, there is no integration over the end points $\ini{x}$ and $\fin{x}$, as they are kept fixed. Replacing the dynamical equation by the actual classical equations of motion \eqref{eq:eom} yields the desired result \eqref{eq:PropagatorPathIntegral}. \\

Note, that there are many more points that should be mentioned, especially concerning the determinant. Since these points clearly exceed our scope, we will briefly mention a few interesting aspects and refer the interested reader to the relevant literature. First note, that the role of the determinant in the context of stochastic quantization has been investigated in \cite{ZINNJUSTIN_1986, Ezewa}. It has been shown that it plays an essential role in the universal supersymmetry arising in such systems and in stochastic systems in general \cite{Ovchinnikov_2016}. Furthermore, in a functional approach the above Jacobian is a functional determinant and can formally be written as an exponential function, \ie
\begin{align}
    \mathcal{J}=\det\left((\partial_t-\nabla_x\,F(x))\delta_D(t-t^\prime)\right)\sim&\exp\left[\mathrm{tr}\ln\left[1-\partial_t^{-1}\nabla_x\,F(x)\right]\right]\\
    =\exp\left[\Theta(0)\int\md t\nabla_xF(x)\right]\,,
\end{align}
where the greens function $\partial_t^{-1}=\Theta(t-t^\prime)$ has been used. Thus, the determinant is indeed dependent on the convention used for the Heaviside function evaluated at zero as mentioned in the main text. Our adopted choice, $\Theta(0)=0$ leads to a constant determinant as in the previously mentioned Itô discretization prescription, while the symmetric Stratonovich convention $\Theta(0)=\frac{1}{2}$ which is mostly used in the filed theoretic approach leads to a determinant which depends on the field variables and thus has to be included. Only in those cases do the Feynman rules in the continuum theory and those derived in the discretized theory coincids \cite{Sato:1976hy}. Note however, that for volume preserving flows (\ie $\nabla_xF(x)=0$) the determinant vanishes as well \cite{Blasone:2004yf}. This is the case in classical mechanics due to the symplectic structure of phase space, \ie $\omega^{ab}\nabla_a\nabla_bH(x)=0$. However, it is nevertheless possible to include the determinant into the classical mechanical path integral formulation. Its role in this context together with the aforementioned emerging supersymmetry has been excessively studied in \cite{Gozzi1988, Gozzi1989, Deotto:2001sy, Abrikosov:2004cf, Gozzi:2010iu}, who showed its connection to the Jacobi fields known from classical mechanics, the Ljapunow exponents in chaotic systems \cite{Gozzi:1993tm} and its crucial role in the cancellation of loop corrections in classical field theories \cite{Cattaruzza:2010wc}.

\subsection{Path Integral with Operators}\label{sec:OperatorInPathint}

In the following we show how the path integral splits, when operators are inserted. As this is a standard textbook method, we keep things short. Again, we restrict our self to the simpler case from the previous section. If we aim to compute the following expectation value,
\begin{equation}
    \mean{\mathcal{O}(t_i)}=\int\limits_{\ini{x}}^{\fin{x}}\mathcal{D}x(t)\mathcal{D}\chi(t)\,\mathcal{O}(x(t_i))\,\exp\left[\mi \int\limits_{\ini{t}}^{\fin{t}}\md t\chi(t)\cdot\left(\dot{x}(t)-F(x(t)) \right)\right]
\end{equation}
we take a step back to the discretized version \eqref{eq:discretizedPI}, making a breakpoint at $t_i$ to arrive at 
\begin{align}
    \mean{\mathcal{O}(t_i)}=&\int\limits_{\ini{x}}^{\fin{x}}\left[\prod_{k=1}^{N-1}\md x_k\right]\left[\prod_{k=1}^N\frac{\md\chi_k}{2\pi}\right]\,\mathcal{O}(x_i)\,\exp\left[\mi \epsilon\sum_{k=1}^N\chi_k\cdot\left(\frac{x_k-x_{k-1}}{\epsilon}-F(x_{k-1}) \right)\right]\\
    =&\int\md x_i\,\int\limits_{x_i}^{\fin{x}}\left[\prod_{k={i+1}}^{N-1}\md x_k\right]\left[\prod_{k=i+1}^N\frac{\md\chi_k}{2\pi}\right]\exp\left[\mi \epsilon\sum_{k=i+1}^N\chi_k\cdot\left(\frac{x_k-x_{k-1}}{\epsilon}-F(x_{k-1}) \right)\right]\times\\&\times\mathcal{O}(x_i)\int\limits_{\ini{x}}^{x_i}\left[\prod_{k=1}^{i-1}\md x_k\right]\left[\prod_{k=1}^i\frac{\md\chi_k}{2\pi}\right]\exp\left[\mi \epsilon\sum_{k=1}^i\chi_k\cdot\left(\frac{x_k-x_{k-1}}{\epsilon}-F(x_{k-1}) \right)\right]\\
    \overset{N\rightarrow\infty}{\underset{\epsilon\rightarrow0}{\longrightarrow}}&\int\md x_iK(\fin{x}, \fin{t}|x_i, t_i)\mathcal{O}(x_i)K(x_i, t_i|\ini{x}, \ini{t})\,.
\end{align}
The same generalizes straight forward to the case of multiple operators as in \eqref{eq:ManyOperatorsinPathintegral}. In particular, if we expand the exponential containing the force field in \eqref{eq:ContinuumPropagator} in a series \eqref{eq:DysonSeries}, we find for the first order,
\begin{align}
    K_1(\fin{x}, \fin{t}|\ini{x}, \ini{t})=&-i\int\limits_{\ini{x}}^{\fin{x}}\mathcal{D}x(t)\mathcal{D}\chi(t)\,\,\exp\left[\mi \int\limits_{\ini{t}}^{\fin{t}}\md t\left(\chi(t)\cdot\dot{x}(t)\right)\right]\int\limits_{\ini{t}}^{\fin{t}}\md t_1\chi(t_1)\cdot F(x(t_1))\,.
\end{align}
We can now repeat similar steps as above in order to find the correct splitting of the propagators. However, care has to be taken, as the inserted observable now also contains a $\chi(t)$ variable. We compute,
\begin{align}
    K_1(\fin{x}, \fin{t}|\ini{x}, \ini{t})=&-i\sum_{l=1}^N\epsilon\int\limits_{\ini{x}}^{\fin{x}}\left[\prod_{k=1}^{N-1}\md x_k\right]\left[\prod_{k=1}^N\frac{\md\chi_k}{2\pi}\right]\,\chi(t_l)\cdot F(x(t_l))\,\e^{\mi \epsilon\sum_{k=1}^N\chi_k\cdot\left(\frac{x_k-x_{k-1}}{\epsilon}) \right)}\\
    =&-i\sum_{l=1}^N\epsilon\int\md x_l\int\md \chi_l\,\int\limits_{x_l}^{\fin{x}}\left[\prod_{k={l+1}}^{N-1}\md x_k\right]\left[\prod_{k=l+1}^N\frac{\md\chi_k}{2\pi}\right]\e^{\mi \epsilon\sum_{k=l+1}^N\chi_k\cdot\frac{x_k-x_{k-1}}{\epsilon}}\times\\&\times\chi_l\cdot F(x_l)\int\limits_{\ini{x}}^{x_l}\left[\prod_{k=1}^{l-1}\md x_k\right]\left[\prod_{k=1}^{l-1}\frac{\md\chi_k}{2\pi}\right]\e^{\mi \epsilon\sum_{k=1}^l\chi_k\cdot\frac{x_k-x_{k-1}}{\epsilon}}\\
    =&-i\sum_{l=1}^N\epsilon\int\md x_l\,\int\limits_{x_l}^{\fin{x}}\left[\prod_{k={l+1}}^{N-1}\md x_k\right]\left[\prod_{k=l+1}^N\frac{\md\chi_k}{2\pi}\right]\e^{\mi \epsilon\sum_{k=l+1}^N\chi_k\cdot\frac{x_k-x_{k-1}}{\epsilon}}\times\\&\times F(x_l)\cdot(-i)\nabla_{x_l}\int\limits_{\ini{x}}^{x_l}\left[\prod_{k=1}^{l-1}\md x_k\right]\left[\prod_{k=1}^{l}\frac{\md\chi_k}{2\pi}\right]\e^{\mi \epsilon\sum_{k=1}^l\chi_k\cdot\frac{x_k-x_{k-1}}{\epsilon}}\\
    \overset{N\rightarrow\infty}{\underset{\epsilon\rightarrow0}{\longrightarrow}}&-\int\limits_{\ini{t}}^{\fin{t}}\md t_1\int\md x_1K_0(\fin{x}, \fin{t}|x_1, t_1)F(x_1)\cdot\nabla_{x_1}K_0(x_1, t_1|\ini{x}, \ini{t})\,.
\end{align}   
In the second step we replaced the $\chi$ field with a gradient \wrt the corresponding $x$ acting on its right. In the above, $K_0$ denotes the propagator \wrt the free theory, \ie with $F=0$. We were thus again able to regroup everything to eventually give a splitted Propagator as above. Replacing $F$ by $-\nabla_{\vec{q}}\,v(\vec{q})$ and $\nabla_x$ by $\nabla_{\vec{p}}$ yields the result used in the main text, \ie the replacement 
\begin{equation}
    \vec{\chi}_p\cdot\nabla_{\vec{q}}\,v(\vec{q})\rightarrow\,\nabla_{\vec{q}}\,v(\vec{q})\cdot\nabla_{\vec{p}}\,,
\end{equation}
within the path integral.

\section{The BBGKY Hierarchy, the Vlasov and the Boltzmann Equation}\label{sec:BBGKY}

In this appendix we present the standard textbook apporach to the time evolution of classical systems. We mainly follow \cite{Balescu}. In order to study the time evolution of expectation values such as in \eqref{eq:standardexpectationvalue} it is thus sufficient to understand the time evolution of the reduced $s$-point densities. These can be obtained from Liouville's equation \eqref{eq:Liouville}
\begin{align}
    \partial_t\varrho_N(\tens{x}, t)=-\tens{\nabla_p}H(\tens{x}, t)\cdot\tens{\nabla_q}\varrho_N(\tens{x}, t)+\tens{\nabla_q}H(\tens{x}, t)\cdot\tens{\nabla_p}\varrho_N(\tens{x}, t)
\end{align}
which, upon insertion of the explicit Hamilton function \eqref{eq:Hamiltonfunction}, takes the form 
\begin{equation}
    \partial_t\varrho_N(\tens{x}, t)=-\sum_{i=1}^N\frac{\vec{p}_i}{m}\cdot\nabla_{\vec{q}_i}\varrho_N(\tens{x}, t)+\sum_{i\neq j=1}^N\nabla_{\vec{q}_i}v(|\vec{q}_i-\vec{q}_j|,t)\cdot\nabla_{\vec{p}_i}\varrho_N(\tens{x}, t)\,.
\end{equation}
The well-known BBGKY hierarchy is then obtained by integrating out the microscopic degrees of freedom as in \eqref{eq:reduceddensity} on both sides in order to obtain evolution equation for the reduced phase space densities, which ultimately yields 
\begin{align}\label{eq:BBGKY}
    \left[\partial_t+\sum_{i=1}^s\frac{\vec{p}_i}{m}\cdot\nabla_{\vec{q}_i}\right]f_s(x_1,\cdots,x_s,t)=&\sum_{i< j=1}^s\nabla_{\vec{q}_i}v(|\vec{q}_i-\vec{q}_j|,t)\cdot(\nabla_{\vec{p}_i}-\nabla_{\vec{p}_j})f_s(x_1,\cdots,x_s,t)\\&+\int\md^6x_{s+1}\sum_{i=1}^s\nabla_{\vec{q}_i}v(|\vec{q}_i-\vec{q}_{s+1}|,t)\cdot\nabla_{\vec{p}_i}f_{s+1}(x_1,\cdots,x_s, x_{s+1},t)\,.\nonumber
\end{align}
The above equation describes the temporal evolution of an $s$-particle cluster. On the left hand side we find the free evolution, while on the right hand side the first term describes interactions among the particles within the $s$-point cluster and the second term describes interactions of any of the $s$ particles with an external one. Thus we obtain a system of equations which relates the evolution of an $f_s$ reduced density to a higher $(s+1)$ reduced density. Most prominently, the equations for $s=1$ and $s=2$ read
\begin{align}
    \left[\partial_t+\frac{\vec{p}_1}{m}\cdot\nabla_{\vec{q}_1}\right]f_1(x_1,t)=&\int\md^6x_{2}\nabla_{\vec{q}_1}v(|\vec{q}_1-\vec{q}_{2}|,t)\cdot\nabla_{\vec{p}_1}f_{2}(x_1,x_2,t)\,,\label{eq:BBGKYk1}\\
    \left[\partial_t+\frac{\vec{p}_1}{m}\cdot\nabla_{\vec{q}_1}+\frac{\vec{p}_2}{m}\cdot\nabla_{\vec{q}_2}\right]f_2(x_1,x_2,t)=&\nabla_{\vec{q}_1}v(|\vec{q}_1-\vec{q}_2|,t)\cdot(\nabla_{\vec{p}_1}-\nabla_{\vec{p}_2})f_2(x_1,x_2,t)\label{eq:BBGKYk2}\\
    +\int\md^6x_{3}\Big[\nabla_{\vec{q}_1}&v(|\vec{q}_1-\vec{q}_{3}|,t)\cdot\nabla_{\vec{p}_1}+\nabla_{\vec{q}_2}v(|\vec{q}_2-\vec{q}_{3}|,t)\cdot\nabla_{\vec{p}_2}\Big]f_{3}(x_1,x_2, x_{3},t)\,.\nonumber
\end{align}
As the system of equations only closes with the inclusion of $f_N$ (there is no $f_{N+1}$ in an $N$-particle ensemble), which is out of scope in almost all cases, one has to find another way in order to truncate?) \eqref{eq:BBGKY}. 
Note that even if at some time $t$ all particles are distributed in a statistically independent way, \ie 
\begin{equation}
    f_s(x_1,\cdots,x_s,t)=\prod_{i=1}^sf_1(x_i, t)\,,
\end{equation}
a statistical dependence among the particles is generated during the evolution due to the presence of inter particle interactions. This is most easily seen in equation \eqref{eq:BBGKYk2} where the first term on the right hand side affects the factorizability of $f_2(x_1, x_2, t)$. Thus, in the very first time step an irreducible $g_2(x_1, x_2, t)$ term is generated by the two-particle potential. In order to study the time evolution of $f_1(x_1, t)$ we can insert \eqref{eq:twopointconnected} into \eqref{eq:BBGKYk1} to obtain the Boltzmann equation,
\begin{align}\label{eq:Boltzmann}
    \left[\partial_t+\frac{\vec{p}_1}{m}\cdot\nabla_{\vec{q}_1}-\nabla_{\vec{q}_1}\Phi(\vec{q}_1, t)\cdot\nabla_{\vec{p}_1}\right]f_1(x_1,t)=&\int\md^6x_{2}\nabla_{\vec{q}_1}v(|\vec{q}_1-\vec{q}_{2}|,t)\cdot\nabla_{\vec{p}_1}g_{2}(x_1,x_2,t)\,,
\end{align}
where we introduced the mean macroscopic potential 
\begin{equation}
    \Phi(\vec{q}_1, t)=\int\md^6x_2v(|\vec{q}_1-\vec{q}_{2}|,t)f_1(x_2,t)\,.
\end{equation}
The last term on the left hand side of \eqref{eq:Boltzmann} involving $\Phi(\vec{q}_1, t)$ is the so-called Vlasov term and the term on the right hand side is usually referred to as the collision term. As we can see, the solution of \eqref{eq:Boltzmann} requires the solution for the evolution equation of $g_2(x_1, x_2, t)$,
 which would require us to solve the BBGKY hierarchy. Since this is not possible, assumptions are made to close the hierarchy. Probably the most famous assumption is the \textit{Stosszahlenansatz} introduced by Boltzmann requiring that $f_2(x_1, x_2, t)=f_1(x_1, t)f_1(x_2, t)$ and thus $g_2(x_1, x_2, t)=0$, effectively neglecting connected inter particle correlations. This ansatz leads to the collision-less Boltzmann equation also known as Vlasov equation, 
\begin{align}\label{eq:CollisionlessBoltzmann}
    \left[\partial_t+\frac{\vec{p}_1}{m}\cdot\nabla_{\vec{q}_1}-\nabla_{\vec{q}_1}\Phi(\vec{q}_1, t)\cdot\nabla_{\vec{p}_1}\right]f_1(x_1,t)=0\,.
\end{align}
The Vlasov equation is formally the Liouville equation for a set of non-interacting particles moving in an ``external'' field which is generated by the particles themselves and depends on their instantaneous distribution. It is thus an inherently non-linear equation. It is easy to see that the Vlasov equation can only describe a non-trivial evolution for spatially inhomogeneous systems, since the Vlasov term vanishes for a homogeneous system. However, even for an inhomogeneous system, the Vlasov term only plays an important role when the interaction potential is long-ranged (compared to the range of variation of the density gradient), which is the case for the Coulomb or the gravitational potential. For a finite range of the interactions or if a system is considered where the mean free path of particles is large, e.g. a dilute gas, the contribution from the Vlasov term becomes negligible. The Vlasov equation is thus only able to describe some of the collective effects in a system, but is insufficient to treat mutual encounters of the particles (or collisions in a broader sense). The latter are not only important for the treatment of homogeneous or dilute systems or systems with a finite interaction range, but necessary to describe the irreversible evolution of a system towards equilibrium. In order to encompass collisions and correlation effects it is, thus, necessary to retain the term on the right hand side of \eqref{eq:Boltzmann}. The goal is then to truncate the BBGKY hierarchy in such a way as to keep the the necessary collisional dynamics. If we neglect three-body correlations, we can write down the evolution equation for $g_2(x_1, x_2, t)$ using \eqref{eq:BBGKYk2},
\begin{align}
     \left[\partial_t+\frac{\vec{p}_1}{m}\cdot\nabla_{\vec{q}_1}+\frac{\vec{p}_2}{m}\cdot\nabla_{\vec{q}_2} \right. 
     &\left. - \nabla_{\vec{q}_1}v(|\vec{q}_1-\vec{q}_2|,t)\cdot(\nabla_{\vec{p}_1}-\nabla_{\vec{p}_2})\right]g_2(x_1,x_2,t)\\
     &=\nabla_{\vec{q}_1}v(|\vec{q}_1-\vec{q}_2|,t)\cdot(\nabla_{\vec{p}_1}-\nabla_{\vec{p}_2})f_1(x_1,t)f_1(x_2,t)\;.
     \label{eq:g_2_evolution}
\end{align}
Using \eqref{eq:Boltzmann} together with \eqref{eq:g_2_evolution} and assuming that the system is homogeneous, the well-known Boltzmann kinetic equation for dilute gases can be derived.\footnote{We reference Balescu, because it is the authors' favourite book on statistical mechanics, but the arguments given here can as well be found in other standard texts on statistical mechanics.} Once the evolution equation \eqref{eq:Boltzmann} for $f_1(x_1,t)$ is solved, expectation values of  macroscopic quantities such as the physical density \eqref{eq:particledensity} or the momentum density \eqref{eq:momentumdensity} can be obtained according to \eqref{eq:expectation_obs} where the expectation value of the Klimontovich phase space is taken over $f_1(x_1,t)$. 
This corresponds to taking the moments of the Boltzmann equation. Taking the zeroth and first moment will then lead to the well-known hydrodynamical balance equations, \ie the continuity, the Euler and the energy-conservation equation.\\

The path integral approach presented in this work is equivalent to the solution of the Liouville equation and therefore to the solution of the BBGKY-hierarchy. It is clear that also in this approach an appropriate truncation scheme has to be chosen in order to solve the evolution of an $N$-particle ensemble of interacting particles. The microscopic perturbation theory discussed in \ref{sec:MicroscopicTheory} is based on an expansion in powers of the interaction potential. If seen from the perspective of the BBGKY-hierarchy, the order of the interaction potential naturally introduces a truncation criterion. For instance, if we wanted to compute the expectation value of the Klimontovich phase space density to first order in interactions, we see from \eqref{eq:BBGKYk1} and \eqref{eq:BBGKYk2} that we only need to insert the free evolution of $f_2(x_1,x_2,t)$ into \eqref{eq:BBGKYk1} since all other terms in \eqref{eq:BBGKYk2} will introduce higher orders of the interaction potential.

\section{Free Macroscopic Generating Functional}\label{sec:MacroscopicZ0}

In this appendix we derive the expression \eqref{eq:FreeMacroGeneratingFunctional} for the free macroscopic generating functional. We start with the path integral expression for $\mathcal{Z}_0[J_f, J_{\mathcal{B}}]$, given by \eqref{eq:FreeMacrosGeneratingFunctionalPathInt} and bring it into a Gaussian form,
\begin{equation}
    \mathcal{Z}_0[J_f, J_{\mathcal{B}}]=\mathcal{N}\int\mathcal{D}\Psi\exp\Big[-\frac{1}{2}\int\md X_1\md X_2\Psi^\top(X_1)\cdot\Delta^{-1}(X_1, X_2)\cdot\Psi(X_2)+\int\md X_1\mathcal{K}^\top(X_1)\cdot\Psi(X_1)\Big]\,,
\end{equation}
where we defined
\begin{align}
    \Psi(X_1) = \begin{pmatrix}
       \Psi_{f}(X_1) \\ \Psi_{\mathcal{B}}(X_1)
    \end{pmatrix}\,,\quad 
    \mathcal{K}(X_1) = \begin{pmatrix}
       J_f(X_1) \\ J_{\mathcal{B}}(X_1)+G^{(0)}_f(X_1)
    \end{pmatrix}\,,\quad 
\end{align}
and the quadratic form 
\begin{equation}
    \Delta^{-1}(X_1,X_2) = \begin{pmatrix}
       0 & \mathbb{1}-G^{(0)}_{\mathcal{B}f} \\
       \mathbb{1}-G^{(0)}_{f\mathcal{B}}  & -G^{(0)}_{ff}
    \end{pmatrix}(X_1,X_2)\,,
\end{equation}
with $G^{(0)}_{\mathcal{B}f}(X_1, X_2)\equiv G^{(0)}_{f\mathcal{B}}(X_2, X_1)$
The solution of the Gaussian functional integral is as usual given by 
\begin{equation}
    \mathcal{Z}_0[J_f, J_{\mathcal{B}}]=\mathcal{N}^\prime\exp\Big[\frac{1}{2}\int\md X_1\md X_2\,\mathcal{K}^\top(X_1)\cdot\Delta(X_1, X_2)\cdot\mathcal{K}(X_2)\Big]\,,
\end{equation}
where $\Delta(X_1, X_2)$ solves the equation
\begin{equation}\label{eq:inverse}
    \int\md \Bar{X}_1\Delta^{-1}(X_1, \Bar{X}_1)\cdot\Delta(\Bar{X}_1, X_2)=\mathbb{1}(X_1, X_2)\,.
\end{equation}
The new normalization constant $\mathcal{N}^\prime$ is chosen such that $\mathcal{Z}_0[0,0]=1$. In order to find the explicit form of $\Delta(X_1, X_2)$ we write 
\begin{equation}
    \Delta(X_1, X_2)=\begin{pmatrix}
        A(X_1, X_2) & B(X_1, X_2) \\ C(X_1, X_2) & D(X_1, X_2)
    \end{pmatrix}\,,
\end{equation}
and read off the four equations from \eqref{eq:inverse} determining the coefficients $A$, $B$, $C$ and $D$,
\begin{align}
    &\int\md \Bar{X}_1\left(\mathbb{1}(X_1, \Bar{X}_1)-G^{(0)}_{\mathcal{B}f}(\Bar{X}_1, X_2)\right)C(\Bar{X}_1, X_2)=\mathbb{1}(X_1, X_2)\\
    &\int\md \Bar{X}_1\left(\mathbb{1}(X_1, \Bar{X}_1)-G^{(0)}_{\mathcal{B}f}(\Bar{X}_1, X_2)\right)D(\Bar{X}_1, X_2)=0\\
    &\int\md \Bar{X}_1\left(\left(\mathbb{1}(X_1, \Bar{X}_1)-G^{(0)}_{f\mathcal{B}}(\Bar{X}_1, X_2)\right)A(\Bar{X}_1, X_2)-G_{ff}^{(0)}(X_1, \Bar{X}_1)C(\Bar{X}_1, X_2)\right)=0\\
    &\int\md \Bar{X}_1\left(\left(\mathbb{1}(X_1, \Bar{X}_1)-G^{(0)}_{f\mathcal{B}}(\Bar{X}_1, X_2)\right)B(\Bar{X}_1, X_2)-G_{ff}^{(0)}(X_1, \Bar{X}_1)D(\Bar{X}_1, X_2)\right)=\mathbb{1}(X_1, X_2)
\end{align}
From this we can conclude 
\begin{align}
    A(X_1, X_2) =&\int\md \Bar{X}_1\md \Bar{X}_2 B(X_1, \Bar{X}_1)G_{ff}^{(0)}(\Bar{X}_1, \Bar{X}_2)C(\Bar{X}_2, X_2)\\
    B(X_1, X_2)=&\left[\mathbb{1}-G^{(0)}_{f\mathcal{B}}\right]^{-1}(X_1, X_2)\\
    C(X_1, X_2)=&\left[\mathbb{1}-G^{(0)}_{\mathcal{B}f}\right]^{-1}(X_1, X_2)\\
    D(X_1, X_2)=&0\,.
\end{align}
Thus the full free generating functional is given by 
\begin{equation}
    \mathcal{Z}_0[J_f, J_\mathcal{B}]=\exp\Bigg[\frac{1}{2}\begin{pmatrix}
        J_f \\ J_\mathcal{B}
    \end{pmatrix}^\top\cdot\begin{pmatrix}
        \Delta_{ff} & \Delta_{f\mathcal{B}}\\ \Delta_{\mathcal{B}f} & 0
    \end{pmatrix}\cdot\begin{pmatrix}
        J_f \\ J_\mathcal{B}
    \end{pmatrix}+J_f\cdot\Delta_{f\mathcal{B}}\cdot G_{f}^{(0)}\Bigg]\,,
\end{equation}
where the integration is implied in the product. Furthermore, we defined the propagators
\begin{align}
    \Delta_{f\mathcal{B}}(X_1, X_2)=&\left[\mathbb{1}-G^{(0)}_{f\mathcal{B}}\right]^{-1}(X_1, X_2)\,,\\
    \Delta_{\mathcal{B}f}(X_1, X_2)=&\Delta_{f\mathcal{B}}(X_2, X_1)\,,\\
    \Delta_{ff}=&\Delta_{f\mathcal{B}}\cdot G_{ff}^{(0)}\cdot  \Delta_{\mathcal{B}f}\,.
\end{align}

\section{Free Cumulants}\label{sec:Cumulants}
\subsection{General Expression for Free $f\mathcal{B}$-Cumulants}

In this appendix we derive general expressions for the free $G_{f\cdots f \mathcal{B}\cdots \mathcal{B}}^{(0)}(X_1,\cdots X_r,X_1^\prime,\cdots X_s^\prime)$ cumulants which constitute the vertices of the macroscopic field theory in section \ref{sec:MacroscopicGeneral}. Our goal will be to express them in terms of a general initial phase space density $\varrho_N(\tini{q},\tini{p},\ini{t})$ and its respective connected reduced phase space densities $g_k(\ini{x}_1,\cdots,\ini{x}_k,\ini{t})$. \\

As described in the main text, the free cumulants $G_{f\cdots f \mathcal{B}\cdots \mathcal{B}}^{(0)}(X_1,\cdots X_r,X_1^\prime,\cdots X_s^\prime)$ are defined as the sum of all connected, time ordered correlation functions containing $r$ phase space densities and $s$ response fields. Thus, we have to compute 
\begin{equation}\label{eq:correlationfunction}
    \mean{\hat{\mathcal{T}}f(  x_1, t_1)\cdots f(  x_r, t_r)\mathcal{B}(  x_1^\prime, t_1^\prime)\cdots \mathcal{B}(  x_s^\prime, t_s^\prime)}^{(0)}\,,
\end{equation}
where $f(x_1, t_1)$ and $\mathcal{B}(x_1, t_1)$ are defined as 
\begin{align}
    f(x_1, t_1)&=\sum_{i=1}^{N}\delta_D(\vec{  {q}}_1-\vec{q}_i(t_1))\delta_D(\vec{  {p}}_1-\vec{p}_i(t_1))\,,\\
    \mathcal{B}(x_1, t_1)&=\mi\sum_{i=1}^{N}\vec{\chi}_{p_i}(t_1)\cdot\nabla_{\vec{q}_i}v(|\vec{q}_i(t_1)-\vec{  {q}}_1|, t_1)\,.
    \label{eq:OperatorsAppendix}
\end{align}
The above correlators are computed as described in the beginning of \ref{sec:MicroscopicTheory} with respect to the free theory. The simplest case of pure density cumulants $G_{f\cdots f}^{(0)}$ has already been discussed in \ref{sec:FreeTheory}. We only have to deduce the connected part form \eqref{eq:freenPoint}. In analogy to \eqref{eq:freelyevolvedfk} we define the freely evolved unequal time irreducible phase space densities $g^{(0)}_k(x_1, t_1, \cdots , x_k, t_k)$ as 
\begin{equation}\label{eq:freelyevolvedgk}
    g_k^{(0)}(x_1, t_1, \cdots, x_k, t_k)=g_2\left( \vec{  {q}}_1-\frac{\vec{  {p}}_1}{m}(t_1-\ini{t}), \vec{  {p}}_1, \cdots, \vec{  {q}}_k-\frac{\vec{  {p}}_k}{m}(t_k-\ini{t}), \vec{  {p}}_k, \ini{t}\right)\,,
\end{equation}
which is simply the initial irreducible $k$-particle phase space density, where each phase space point is again individually shifted by a free trajectory. We thus find for the one-and two point cumulant as described in \ref{sec:FreeTheory},
\begin{align}
    G_{f}^{(0)}(X_1)=&\mean{f(\vec{  {q}}_1, \vec{  {p}}_1, t_1)}^{(0)}\\
    =&f_1^{(0)}(  x_1, t_1)\,,
\end{align}
and 
\begin{align}\label{eq:TwoPointCorrelation}
    G_{ff}^{(0)}(  X_1, X_2)=&\mean{\hat{\mathcal{T}}f(  {x}_1, t_1)f(  {x}_2, t_2)}^{(0)}-\mean{f(  {x}_1, t_1)}^{(0)}\mean{f(  {x}_2, t_2)}^{(0)}\\
    =&\delta_D\left(  {x}_1-  {x}_2(t_1;t_2))\right)f_1^{(0)}(x_1, t_1)+g_2^{(0)}\left(x_1, t_1, x_2, t_2\right)\,.
\end{align}
Interchanging both $f$'s in $G_{ff}^{(0)}$ yields the same term which is accounted for by the ${1}/{2!}$ in \eqref{eq:MacroscopicTheoryGeneralInteractions}. The first term in the last equality comes from the $i=j$ part of the sums belonging to both $f$'s, which yields a one-particle contribution to the two-point density correlation function. It represents the possibility of measuring the correlation by incidentally picking the same particle twice. This explains the appearance of the reduced one-particle correlation function $f_1$ times the Dirac-Delta distribution identifying the phase space position $  {x}_1$ with $  {x}_2(t_1;t_2)$. The second term contains the true two-particle correlation. 
This can be generalised to arbitrary $r$-point density cumulants $G_{f\cdots f}^{(0)}(X_1,\cdots,X_r)$ giving
\begin{equation}
    G_{f\cdots f}^{(0)}(X_1,\cdots,X_r)= \cdots + g_r^{(0)}( x_1, t_1, \cdots, x_r, t_r)\,.
\end{equation}
In the above equation the ellipses is a place holder for all lower order $l$-particle contribution in which at least two indices in the sums are identified. As in the discussion around equation \eqref{eq:TwoPointCorrelation} they correspond to the possibility of randomly picking the same particle multiple times which thus contribute by a lower order irreducible phase space density to the $r$-point cumulant. Every such term will also be accompanied by a Dirac-Delta distribution which identifies the particles at different times as in equation \eqref{eq:TwoPointCorrelation}. Note, that only irreducible phase space densities appear, since we always subtract the connected part, which also affects the lower order contributions. For instance we find for the three point density cumulant $G_{fff}^{(0)}(  X_1,   X_2,  X_3)$ 
\begin{equation}
\begin{aligned}
    G_{fff}^{(0)}(  X_1,   X_2,  X_3)=&\delta_D(  {x}_1-  {x}_{2}(t_1;t_2))\delta_D(  {x}_2-  {x}_{3}(t_2;t_3))f_1^{(0)}\left(  x_1, t_1\right)\\
    &+\delta_D(  {x}_1-  {x}_{2}(t_1;t_2))\,g_2^{(0)}\left(x_1, t_1, x_3, t_3\right)\\
    &+\delta_D(  {x}_2-  {x}_{3}(t_2;t_3))g_2^{(0)}\left(x_1, t_1, x_2, t_2\right)\\
    &+\delta_D(  {x}_1-  {x}_{3}(t_1;t_3))g_2^{(0)}\left( x_1, t_1, x_2, t_2\right)\\
    &+g_3^{(0)}\left(  x_1, t_1, x_2, t_2, x_3, t_3\right)\,.
\end{aligned}
\end{equation}
Having understood the general expression for the pure density cumulants, let us now extend the above discussion to the case of mixed $G_{f\cdots f\mathcal{B}\cdots\mathcal{B}}^{(0)}$ cumulants. First, we note that as has been shown in appendix \ref{sec:discretization}, placing a $\mathcal{B}$ field in front of free propagators amounts to the computation of 
\begin{equation}
    \mathcal{B}(x, t_2)K_0(\tens{x}_2, t_2|\tens{x}_1, t_1)=\sum_{i=1}^N\nabla_{\vec{q}_{2,i}}v(|\vec{q}_{2,i}-\vec{  {q}}\,|, t_2)\cdot\nabla_{\vec{p}_{2,i}}K_0(\tens{x}_2, t_2|\tens{x}_1, t_1)\,,
\end{equation}
since the $\vec{\chi}_{p_i}$ may be replaced by a gradient \wrt the momentum $\vec{p}_i$. Thus we find 
\begin{align}
    G_{\mathcal{B}}^{(0)}(X_1)=\sum_{i=1}^N\int\md\tens{x}_1\md\tini{x}\nabla_{\vec{q}_{1,i}}v(|\vec{q}_{1,i}-\vec{  {q}}_1\,|, t_1)\cdot\nabla_{\vec{p}_{1,i}}K_0(\tens{x}_1, t_1|\tini{x}, \ini{t})\varrho(\tini{x},\ini{t})=0\,,
\end{align}
due to the normalization condition \eqref{eq:NormalizationCondition}. By the same logic a $\mathcal{B}$ field can never be placed on the left, \ie with the latest time in a time ordered expectation value. In particular a pure $\mathcal{B}$ correlator vanishes and consequently 
\begin{equation}
    G_{\mathcal{B}\cdots\mathcal{B}}^{(0)}(X_1, \cdots, X_s) =0\,.
\end{equation}
Thus the simplest mixed correlator is
\begin{align*}
    \mean{\hat{\mathcal{T}}f(x_1, t_1)\mathcal{B}(x_2, t_2)}^{(0)}=\int\md\tens{x}_1\,\md\tens{x}_2\,\md\tini{x}f(x_1, t_1)K_0(\tens{x}_1, t_1|&\tens{x}_2, t_2)\mathcal{B}(x_2, t_2)\times\\&\times K_0(\tens{x}_2|\tini{x}, \ini{t})\varrho_N(\tini{x}, \ini{t})\,,
\end{align*}
which straight forwardly computes to 
\begin{equation}
    \mean{\hat{\mathcal{T}}f(x_1, t_1)\mathcal{B}(x_2, t_2)}^{(0)}=\nabla_{\vec{q}_1}v(\vec{q}_1-\frac{\vec{p}_1}{m}(t_1-t_2), t_2)\cdot\nabla_{\vec{p}_1}f_1^{(0)}(x_1, t_1)\Theta(t_1-t_2)\,, 
\end{equation}
where the Heaviside function ensures that $t_2$ is earlier than $t_1$, \ie the $\mathcal{B}$ field is on the right side of $f$. From $\mean{\mathcal{B}(x_2, t_2)}^{(0)}=0$, we get 
\begin{align}
    G_{f\mathcal{B}}^{(0)}(X_1, X_2)=&\nabla_{\vec{q}_1}v(|\vec{q}_1-\frac{\vec{p}_1}{m}(t_1-t_2)-\vec{q}_2\,|, t_2)\cdot\nabla_{\vec{p}_1}f_1^{(0)}(x_1, t_1)\Theta(t_1-t_2)\\
    =&\hat{\mathcal{L}}_{1\ShortLArrow2}(t_1,t_2;t_2)\,\left[f_1^{(0)}(x_1, t_1)\right]\,,
\end{align}
where we defined the operator 
\begin{equation}
    \hat{\mathcal{L}}_{1\ShortLArrow2}(t_1,t_2;t_3)\equiv\Theta(t_1-t_3)\nabla_{\vec{q}_1}v(|\vec{q}_1-\frac{\vec{p}_1}{m}(t_1-t_2)-\vec{q}_2+\frac{\vec{p}_2}{m}(t_2-t_3)\,|, t_3)\cdot\nabla_{\vec{p}_1}
\end{equation}
describing how the one-particle density distribution at $\vec{q}_1$ at time $t_1$ is influenced by a particle located at $\vec{q}_2$ at time $t_2$ due to an interaction taking place at an earlier time $t_3$. $G_{f\mathcal{B}}^{(0)}$ is thus a one-particle effect in the sense that it physically describes the deflection of the free one-particle density belonging to the single $f$ field of the cumulant. This interpretation generalizes to higher order cumulants. It is straight forward, but tedious, to work out analytic expressions for $G_{ff\mathcal{B}}^{(0)}( X_1,  X_2,   X_3)$ and $G_{f\mathcal{B}\mathcal{B}}^{(0)}(  X_1,  X_2,   X_3)$. In the latter case, there is only one contribution with both $\mathcal{B}$ fields on the right of the single $f$ field. We find in a similar way as above,
\begin{align}\label{eq:GfBB}
    G_{f\mathcal{B}\mathcal{B}}^{(0)}(  X_1,  X_2,   X_3)=&\hat{\mathcal{L}}_{1\ShortLArrow2}(t_1,t_2;t_2)\,\left[\hat{\mathcal{L}}_{1\ShortLArrow3}(t_1,t_3;t_3)\,\left[f_1^{(0)}(x_1, t_1)\right]\right]\\
    =&\hat{\mathcal{L}}_{1\ShortLArrow2}(t_1,t_2;t_2)\,\left[\hat{\mathcal{L}}_{1\ShortLArrow3}(t_1,t_3;t_3)\,\left[G_f^{(0)}(X_1)\right]\right]
\end{align}
where again, both $\mathcal{B}$ fields can be interchanged with each other giving a combinatorial prefactor as above. The causal structure however requires that both times $t_2$ and $t_3$ are earlier than $t_1$. Physically, \eqref{eq:GfBB} describes how the one-particle reduced phase space density at $\vec{q}_1$ at time $t_1$ is affected by two consecutive interactions. First, the freely evolved density is deflected by a particle at $\vec{q}_3$ and then again deflected by a particle at $\vec{q}_2$. The product rule in the evaluation of \eqref{eq:GfBB} signifies that the second interaction now affects either the density or the force which acted beforehand. If we consider all correlators with two $f$ fields and one $\mathcal{B}$ field and substract the disconnected part, we end up with 
\begin{align}
    G_{ff\mathcal{B}}^{(0)}( X_1,  X_2,   X_3)=\left[\hat{\mathcal{L}}_{1\ShortLArrow3}(t_1,t_3;t_3)+\hat{\mathcal{L}}_{2\ShortLArrow3}(t_2,t_3;t_3)\right]G_{ff}^{(0)}(  X_1,   X_2)\,.
\end{align}
Again, interchanging both densities results in a prefactor. However, we now have two contributions to the time ordered product, since the interaction can affect both densities. In analogy to above, $G_{ff\mathcal{B}}^{(0)}$ describes the deformation of the freely evolved two-point phase space density correlation connecting a particle at $x_1$ at time $t_1$ with another particle at $x_2$ at time $t_2$, due to an interaction affecting each of both particles separately. We now have all ingredients to generalise these results in the following way: An arbitrary mixed $G_{f\cdots f \mathcal{B}\cdots \mathcal{B}}^{(0)}(X_1,\cdots X_r,X_1^\prime,\cdots X_s^\prime)$ cumulant consists of a differential operator $\hat{\mathcal{T}}_{r,s^\prime}$ applied to the free pure density cumulant $G_{f\cdots f }^{(0)}(X_1,\cdots X_r)$,
\begin{equation}
    G_{f\cdots f \mathcal{B}\cdots \mathcal{B}}^{(0)}(X_1,\cdots X_r,X_1^\prime,\cdots X_s^\prime)=\hat{\mathcal{T}}_{r,s^\prime}\cdot G_{f\cdots f }^{(0)}(X_1,\cdots X_r)\,.
\end{equation}
The differential operator itself is given by the sum of all possibilities of connecting all the positions $X_1^\prime,\cdots X_s^\prime$ belonging to the response fields $\mathcal{B}$ to the positions $X_1,\cdots X_r$ belonging to the phase space densities $f$ by $\hat{\mathcal{L}}_{x\ShortLArrow x^\prime}(t,t^\prime;t^\prime)$ operators, where $x^\prime$ and $t^\prime$ belong to a response field. In order to avoid the repeated tedious computations it is convenient to summarize the form of $\hat{\mathcal{T}}_{r,s^\prime}$ on a diagrammatic level by the following set of rules:
\begin{enumerate}
    \item Draw one filled circle for every $f$ and one empty circle for every $\mathcal{B}$ appearing in the cumulant and label them with their respective phase space arguments. For a $G_{f\cdots f \mathcal{B}\cdots \mathcal{B}}^{(0)}(X_1,\cdots X_r,X_1^\prime,\cdots X_s^\prime)$ cumulant we draw  
        \begin{equation}
            \mytikz{
            \dotpattern[X_1][X_r][X_1'][X_s']
            }
        \end{equation}
    \item Start exactly one line at each empty circle which can end at any of the filled circles. Multiple lines can end on a filled circle. Note that not all of the filled circles have to have a line ending on them.  
    
    \item A path segment connecting two circles $\mytikz{\fbpattern[X][X^\prime]\straightline{b1}{f1} }$, corresponds to a $\hat{\mathcal{L}}_{x\ShortLArrow x^\prime}(t,t^\prime;t^\prime)$ operator, where the arrow signifies the causal flow due to the Heaviside function inside $\hat{\mathcal{L}}$.
    \item For each diagram all path segments are multiplied. For example,
    \begin{equation}
        \mytikz{
        \fffbbbpattern[X_1][X_2][X_3][X_4][X_5][X_6]\straightline{b1}{f3}\arcline{b3}{f2}\arcline{b2}{f2}
        } = \hat{\mathcal{L}}_{x_2\ShortLArrow x_5}(t_2,t_5;t_5)\hat{\mathcal{L}}_{x_2\ShortLArrow x_6}(t_2,t_6;t_6)\hat{\mathcal{L}}_{x_3\ShortLArrow x_4}(t_3,t_4;t_4)\;.
    \end{equation}
    Note, that if multiple operators end on the same position, the product rule applies. However the order of the operators does not matter. In the above example, the first operator acts also on the second. We could also have switched both giving the same result.
    \item All possible diagrams for given $r$, $s'$ are summed up to obtain $\hat{\mathcal{T}}_{r,s^\prime}$.
    \item If there is no $\mathcal{B}$ field involved, $\hat{\mathcal{T}}_{r,s^\prime}=1$

\end{enumerate}
As an example we give the diagrammatic representations for the following cumulants:
\begin{align}
G_{f\mathcal{B}}^{(0)}(X_1, X_2) = &\left[ \mytikz{
        \fbpattern[X_1][X_2]\straightline{b1}{f1} 
        }\right]\cdot G_{f}^{(0)}(X_1) = \left[\hat{\mathcal{L}}_{x_1\ShortLArrow x_2}(t_1,t_2;t_2)\right]\cdot G_{f}^{(0)}(X_1)\\
    G_{ff\mathcal{B}}^{(0)}(X_1,X_2,X_3) = &\left[\mytikz{\ffbpattern[X_1][X_2][X_3]\arcline{b1}{f1}} + \mytikz{\ffbpattern[X_1][X_2][X_3]\straightline{b1}{f2}}   \right]\cdot G_{ff}^{(0)}(X_1,X_2)\\ = &\left[\hat{\mathcal{L}}_{x_1\ShortLArrow x_3}(t_1,t_3;t_3) + \hat{\mathcal{L}}_{x_2\ShortLArrow x_3}(t_2,t_3;t_3)\right]\cdot G_{ff}^{(0)}(X_1,X_2)\\
    G_{f\mathcal{B}\mathcal{B}}^{(0)}(X_1,X_2,X_3) = &\left[\mytikz{\fbbpattern[X_1][X_2][X_3]\straightline{b1}{f1}\arcline{b2}{f1}} \right]\cdot G_{f}^{(0)}(X_1)\nonumber  \\ = &\left[\hat{\mathcal{L}}_{x_1\ShortLArrow x_2}(t_1,t_2;t_2)\hat{\mathcal{L}}_{x_1\ShortLArrow x_3}(t_1,t_3;t_3)\right]\cdot G_{f}^{(0)}(X_1)\\
    G_{ff\mathcal{B}\mathcal{B}}^{(0)}(X_1,X_2,X_3, X_4) = &\left[\mytikz{\ffbbpattern[X_1][X_2][X_3][X_4]\arcline{b1}{f1}\arcline{b2}{f1}} +\mytikz{\ffbbpattern[X_1][X_2][X_3][X_4]\arcline{b1}{f1}\arcline{b2}{f2}}\right.\\&\left. +\mytikz{\ffbbpattern[X_1][X_2][X_3][X_4]\straightline{b1}{f2}\arcline{b2}{f1}}+\mytikz{\ffbbpattern[X_1][X_2][X_3][X_4]\straightline{b1}{f2}\arcline{b2}{f2}}\right]\cdot G_{ff}^{(0)}(X_1, X_2) \nonumber \\ = &\left[\hat{\mathcal{D}}_{x_1\ShortLArrow x_3}(t_1,t_3)\hat{\mathcal{L}}_{x_1\ShortLArrow x_4}(t_1,t_4;t_4+\hat{\mathcal{L}}_{x_1\ShortLArrow x_3}(t_1,t_3;t_3)\hat{\mathcal{L}}_{x_2\ShortLArrow x_4}(t_2,t_4;t_4)\right.\\&\left.+\hat{\mathcal{L}}_{x_1\ShortLArrow x_4}(t_1,t_4;t_4)\hat{\mathcal{L}}_{x_2\ShortLArrow x_3}(t_2,t_3;t_3)+\hat{\mathcal{L}}_{x_2\ShortLArrow x_3}(t_2,t_3;t_3)\hat{\mathcal{L}}_{x_2\ShortLArrow x_4}(t_2,t_4;t_4)\right]\nonumber\\&\cdot G_{ff}^{(0)}(X_1, X_2)\nonumber
\end{align}
Clearly, the higher order diagrams represent interactions, which deform higher, freely evolved phase space statistics multiple times. The last cumulant for instance describes how the connected two-point statistic can be affected by two interactions.

\subsection{Density and Momentum Cumulants for Homogeneous System}\label{sec:CumulantsHomogeneous}
In this appendix, we list the cumulants for the homogeneous system which contribute to the tree level expectation values of \eqref{eq:Example}. They are computed as described in the previous section.

\begin{align}
        G_\rho^{(0)}(t_1) =&\, \Bar{\rho}\\
        \vec{G}_\Pi^{(0)}(t_1) =& 0\\
        G_{\rho\mathcal{B}}^{(0)}(\vec{R}, t_1, t_2)=&\frac{\Bar{\rho}m^2}{(t_1-t_2)^2}\int\md^3X\,\Delta_{\vec{X}}\,v(|\vec{X}|,t_2)\,\varphi\left(\frac{\vec{R}-\vec{X}}{t_1-t_2}m\right)\,,\label{eq:AppendixGrB}\\
        \vec{G}_{\Pi\mathcal{B}}^{(0)}(\vec{R}, t_1, t_2)=&\frac{\Bar{\rho}m^3}{(t_1-t_2)^3}\int\md^3X\,\Big[(\vec{R}-\vec{X})\Delta_{\vec{X}}\,v(|\vec{X}|,t_2)-\nabla_{\vec{X}}v(|\vec{X}|,t_2)\Big]\,\varphi\left(\frac{\vec{R}-\vec{X}}{t_1-t_2}m\right)\,,\\
        G_{\rho\rho}^{(0)}(\vec{R}, t_1, t_2)=&\frac{\Bar{\rho}m^3}{(t_1-t_2)^3}\varphi\left(\frac{\vec{R}}{t_1-t_2}m\right)\\&+\frac{m^6}{(t_1-\ini{t})^3(t_2-\ini{t})^3}\int\md^3X\md^3Yg_2\left(\vec{Y},\frac{\vec{R}-\vec{X}}{t_1-\ini{t}}m,\frac{\vec{Y}-\vec{X}}{t_2-\ini{t}}m, \ini{t}\right)\nonumber\\
        \vec{G}_{\Pi\rho}^{(0)}(\vec{R}, t_1, t_2)=&\frac{\Bar{\rho}m^4}{(t_1-t_2)^4}\vec{R}\varphi\left(\frac{\vec{R}}{t_1-t_2}m\right)\\&+\frac{m^7}{(t_1-\ini{t})^4(t_2-\ini{t})^3}\int\md^3X\md^3Y(\vec{R}-\vec{X})g_2\left(\vec{Y},\frac{\vec{R}-\vec{X}}{t_1-\ini{t}}m,\frac{\vec{Y}-\vec{X}}{t_2-\ini{t}}m, \ini{t}\right)\nonumber\\
        \vec{G}_{\rho\Pi}^{(0)}(\vec{R}, t_1, t_2)=&\frac{\Bar{\rho}m^4}{(t_1-t_2)^4}\vec{R}\varphi\left(\frac{\vec{R}}{t_1-t_2}m\right)\\&+\frac{m^7}{(t_1-\ini{t})^3(t_2-\ini{t})^4}\int\md^3X\md^3Y(\vec{Y}-\vec{X})g_2\left(\vec{Y},\frac{\vec{R}-\vec{X}}{t_1-\ini{t}}m,\frac{\vec{Y}-\vec{X}}{t_2-\ini{t}}m, \ini{t}\right)\nonumber\\
        G_{\Pi\otimes\Pi}^{(0)}(\vec{R}, t_1, t_2)=&\frac{\Bar{\rho}m^5}{(t_1-t_2)^5}\vec{R}\otimes\vec{R}\varphi\left(\frac{\vec{R}}{t_1-t_2}m\right)\\&+\frac{m^8}{(t_1-\ini{t})^4(t_2-\ini{t})^4}\int\md^3X\md^3Y(\vec{R}-\vec{X})\otimes(\vec{Y}-\vec{X})\times\nonumber\\&\hspace{6cm}\times g_2\left(\vec{Y},\frac{\vec{R}-\vec{X}}{t_1-\ini{t}}m,\frac{\vec{Y}-\vec{X}}{t_2-\ini{t}}m, \ini{t}\right)
        \end{align}

where for $\vec{G}_\Pi^{(0)}$ we made the reasonable assumption that that the mean momentum of the homogeneous system vanishes. For $G_{\rho\mathcal{B}}^{(0)}$ and $\vec{G}_{\Pi\mathcal{B}}^{(0)}$ we performed a partial integration.

\section{Special Case: Field Theory Approach to Self Gravitating System in and out of Equilibrium}\label{sec:SpecialCase}
We will use the explicit example of a self-gravitating system to provide some physical intuition for the newly constructed field theory in Sec.\ \ref{sec:MacroscopicGeneral}. This special case of our theory is a generalisation of \cite{deVega1996}, where the HST has been used to study a self-gravitating gas in terms of a grand canonical equilibrium system.\\

We consider the Hamilton function \eqref{eq:Hamiltonfunction} and assume throughout this subsection an explicit two-particle interaction of the form, 
\begin{equation}\label{eq:Coulomb}
    v(|\vec{q}_i-\vec{q}_j|, t)=\,m_i\,m_j\,v_G(|\vec{q}_i-\vec{q}_j|)\,,\,\,\,\,      v_G(|\vec{q}_i-\vec{q}_j|)=-\frac{G}{|\vec{q}_i-\vec{q}_j|}\,,
\end{equation}
namely the gravitational potential. We introduce the doublet 
\begin{equation}
    \Phi(t_1, \vec{  {q}}_1)=(B_m(t_1, \vec{  {q}}_1),\,\,\rho_m(t_1, \vec{  {q}}_1)  )^\top\,,
\end{equation}
with the particle mass density $\rho_m(t_1, \vec{  {q}}_1)$ and the response field $B_m(t_1, \vec{  {q}}_1)$ defined as 
\begin{align}
    \rho_m(t_1, \vec{  {q}}_1)=&\sum_{i=1}^N m_i\delta_D(\vec{  {q}}_1-\vec{q}_i(t_1))\\
    B_m(t_1, \vec{  {q}}_1)=&-i\sum_{i=1}^Nm_i\vec{\chi}_{p_i}\cdot \nabla_{\vec{  {q}}_1}\delta_D(\vec{  {q}}_1-\vec{q}_i(t_1))\,.
\end{align}
Note that contrary to the response field introduced in Sec.\ \ref{sec:MacroscopicGeneral}, $B_m(t_1, \vec{{q}}_1)$ does not contain the interaction potential.
In analogy to \eqref{eq:ActionBeforeHST}, we bring the interaction part of the action \eqref{eq:interactionaction} into a quadratic form, 
\begin{align}\label{eq:macroscopicInteraction}
    \mi\,\mathcal{S}_I[\tens{x}(t),\tens{\chi}(t)]=&\mi\int\limits_{\ini{t}}^{\infty}\md t\,\tens{\chi}_p(t)\tens{\cdot}\tens{\nabla_q}V(\tens{q}(t))\\
    =&\int\md t_1\md^3  {q}_1\int\md t_2 \md^3  {q}_2\,B_m(t_1,   {q}_1)\,v_G(|  {q}_1-  {q}_2|)\delta_D(t_1-t_2)\,\rho_m(t_2,   {q}_2)\\
    \equiv& \,\frac{1}{2}\int\md Q_1\int\md Q_2 \,\Phi(Q_1)^\top\cdot\sigma(Q_1,Q_2)\cdot\Phi(Q_2)\;.
\end{align}

where in the last line we abbreviated the notation by writing 
\begin{equation}
    \md Q_n\equiv \md t_n \md^3 \vec{  {q}}_n\,,\,\,\,\,\,\,A(t_n, \vec{  {q}}_n)\equiv A(Q_n)\,,
\end{equation}
and defined the interaction matrix 
\begin{equation}
    \sigma(Q_1, Q_2)=\begin{pmatrix}
    0 & \sigma_{ B\rho}(Q_1,Q_2)\\
    \sigma_{ \rho B}(Q_1,Q_2)& 0
    \end{pmatrix}\,,
\end{equation}
with
\begin{equation}
    \sigma_{B\rho}(Q_1,Q_2)=\sigma_{\rho B}(Q_2,Q_1)=v_G(|  \vec{q}_1-  \vec{q}_2|)\delta_D(t_1-t_2)\,.
\end{equation}
Introducing a source doublet 
\begin{equation}
    \mathcal{J}(Q_1)=(J_\rho(Q_1),\,\,  J_B(Q_1)^\top\,,
\end{equation}
we can define the macroscopic generating functional in complete analogy to \eqref{eq:macroscopicGenFuncBeforeTrafo}, 
such that by construction functional derivatives \wrt $J_\rho(Q)$ and $J_B(Q)$ produce integrals over the interaction potential and density and response field correlators,
\begin{align}
    \frac{\delta}{i\delta J_\rho(Q_1)}\mathcal{Z}[J_\rho, J_B]\Big\rvert_{J_\rho, J_B=0}=\int\md \vec{  {q}}^\prime v_G(|\vec{  {q}}_1-\vec{  {q}}^\prime|)\mean{\rho_m(t_1, \vec{  {q}}^\prime)}\label{eq:FuncDerivDensity}\\
    \frac{\delta}{i\delta J_B(Q_1)}\mathcal{Z}[J_\rho, J_B]\Big\rvert_{J_\rho, J_B=0}=\int\md \vec{  {q}}^\prime v_G(|\vec{  {q}}_1-\vec{  {q}}^\prime|)\mean{B_m(t_1, \vec{  {q}}^\prime)}\,,\label{eq:FuncDerivResponse}
\end{align}
which continues analogously for higher order and mixed correlation functions.  In order to solve for the density and response field correlators in \eqref{eq:FuncDerivDensity} and \eqref{eq:FuncDerivResponse}, the gravitational interaction potential \eqref{eq:Coulomb} has to be inverted by means of its Green function, the Laplace operator,
\begin{equation}\label{eq:Greensfunction}
    \frac{1}{4\pi G}\Delta_{  {q}_1}\,v_G(|\vec{  {q}}_1-\vec{  {q}}_2|)=\delta_D(\vec{  {q}}_1-\vec{  {q}}_2)\,.
\end{equation}
This allows us to directly compute correlations of the density and response field as
\begin{align}
    \frac{1}{4\pi G}\Delta_{  {q}_1}\frac{\delta}{i\delta J_\rho(Q_1)}\mathcal{Z}[J_\rho, J_B]\Big\rvert_{J_\rho, J_B=0}=\mean{\rho_m(Q_1)}\label{eq:FuncDerivSolvedDensity}\\
    \frac{1}{4\pi G}\Delta_{  {q}_1}\frac{\delta}{i\delta J_B(Q_1)}\mathcal{Z}[J_\rho, J_B]\Big\rvert_{J_\rho, J_B=0}=\mean{B_m(Q_1)}\,,\label{eq:FuncDerivSolvedResponse}
\end{align}
which can be generalised to higher order correlation functions in a straightforward fashion. 
We now apply the HST and follow the steps \eqref{eq:macroscopicGenFuncAfterTrafo} to \eqref{eq:macroAction} using that the inverse of $\sigma(Q_1,Q_2)$ is given by 
\begin{equation}
    \sigma^{-1}(Q_1, Q_2)=\delta_D(t_1 - t_2)\delta_D(\vec{  {q}}_1-\vec{  {q}}_2)\begin{pmatrix}
    0 & \frac{1}{4\pi G}\Delta_{q_1}\\
    \frac{1}{4\pi G}\Delta_{q_1}& 0
    \end{pmatrix}\,.
\end{equation}
After the HST, the generating functional takes on the form
\begin{align}\label{eq:GeneratingFunctionalMacroscopic}
    \mathcal{Z}[J_\rho, J_B]=\mathcal{N}\int\mathcal{D}\Psi\exp\Bigg[&-\frac{1}{2}\int\md Q_1\md Q_2 \Psi(Q_1)^\top\cdot\sigma^{-1}(Q_1,Q_2)\cdot\Psi(Q_2)\nonumber \\ &+\int\md Q_1\mathcal{J}(Q_1)^\top\cdot\Psi(Q_1)\Bigg]\cdot\mathcal{Z}^{(0)}_{\rho B}\big[\Psi_{B},\Psi_{\rho} \big]\;,
\end{align}
where we defined,
\begin{align}\label{eq:Zmicroscopic}
    \mathcal{Z}^{(0)}_{\rho B}\big[\Psi_{B},\Psi_{\rho} \big]&=\int\md\tfin{x}\int\md\tini{x}\varrho_N(\ini{t},\tini{x})\times \nonumber\\
    &\times\int\limits_{\tini{x}}^{\tfin{x}} \mathcal{D}^\prime\tens{x}(t)\mathcal{D}\tens{\chi}(t)\,\exp\Bigg[i\mathcal{S}_0[\tens{x}(t),\tens{\chi}(t)]+\int\md1\Psi(Q_1)^\top\cdot\Phi(Q_1)\Bigg]\,.
\end{align}
The absence of an interaction term in \eqref{eq:Zmicroscopic} makes it clear that $\mathcal{Z}^{(0)}_{\rho B}\big[\Psi_{B},\Psi_{\rho} \big]$ corresponds to the free $\rho B$-generating functional, whose functional derivatives \wrt $\Psi_{B}$ and $\Psi_{\rho}$ produce free $\rho B$-correlation functions, 
\begin{align}\label{eq:FuncDiffCorr}
    \frac{\delta^{r+s}}{\delta \Psi_{B}(Q_1)\cdots\Psi_{B}(Q_r)\Psi_{\rho}({Q_1}^\prime)\cdots\Psi_{\rho}({Q_s}^\prime) }&\mathcal{Z}^{(0)}_{\rho B}\big[\Psi_{B},\Psi_{\rho} \big]\Big\rvert_{\Psi=0}\nonumber \\
    &=\mean{\rho_m(Q_1)\cdots\rho_m(Q_r)B_m({Q_1}^\prime)\cdots B_m({Q_s}^\prime)}^{(0)}\,.
\end{align}

The physical interpretation of the macroscopic fields can be seen as follows: The expectation value of $\Psi_{\rho}(Q_1)$ can be computed as usual by taking a functional derivative,
\begin{equation}
   \mean{\Psi_{\rho}(Q_1)}=\frac{\delta}{\delta J_\rho(Q_1)}\mathcal{Z}[J_\rho, J_B]\Big\rvert_{J_\rho, J_B=0}\,.
\end{equation}
From \eqref{eq:FuncDerivSolvedDensity} we thus find 
\begin{equation}\label{eq:micromacroidentification}
    \Delta_{  {q}_1}\mean{\Psi_{\rho}(Q_1)}=4\pi G\mean{\rho_m(Q_1)}
\end{equation}
which is the Poisson equation. Thus $\Psi_{\rho}(Q_1)$ corresponds to the macroscopic gravitational field induced by the microscopic 2-particle interactions of the ensemble. The macroscopic correlators are thus correlation functions of the gravitational field itself and we find in general the correspondence
\begin{equation}\label{eq:micromacrocorrespondence}
    \Delta_{  {q}_1}\cdots\Delta_{  {q}_n}\mean{\Psi_{\rho}(Q_1)\cdots\Psi_{\rho}(Q_n)}=(4\pi G)^n\mean{\rho_m(Q_1)\cdots\rho_m(Q_n)}\;.
\end{equation}
Analogous equations can be derived for the macroscopic field $\Psi_B(Q_1)$ which encodes the macroscopic response of the system to fluctuations of the gravitational field. As expected, 
\begin{equation}\label{eq:micromacrocorrespondenceB}
    \Delta_{  {q}_1}\cdots\Delta_{  {q}_n}\mean{\Psi_{B}(Q_1)\cdots\Psi_{B}(Q_n)}=0\;,
\end{equation}
due to $\mean{B_m(Q_1)\cdots B_m(Q_n)}=0$.
Consequently, the HST has inverted the scale on which we describe our system: Instead of microscopic particle dynamics, we now describe the dynamics of a fluctuating macroscopic field whose self-interactions are sourced by the underlying microscopic free statistics.
It is worth noting that in this special case, due to the functional inversion of the interaction potential, we were able to split the generating functional \eqref{eq:GeneratingFunctionalMacroscopic} into a part containing the full interaction potential $\Psi(Q_1)^\top\cdot\sigma^{-1}(Q_1,Q_2)\cdot\Psi(Q_2)$ and a free part \eqref{eq:Zmicroscopic} containing no effects of the interaction potential at all. A very similar procedure has led \cite{deVega1996} to a field theoretic description of a self gravitating gas in equilibrium. In our case however, we are not restricted to the equilibrium case, as we may put the system initially in any desired state. The free 
correlators will then transfer the initial information into the macroscopic field dynamics. Last but not least, we note that the above procedure can be applied to different two-particle potentials, provided, that there exists an appropriate differential operator of which the potential is the Greens function \eqref{eq:Greensfunction}. The physical interpretation of the macroscopic fields will change according to the potential.  

\bibliography{mybib.bib}

\begin{thebibliography}{35}
\providecommand{\natexlab}[1]{#1}
\providecommand{\url}[1]{\texttt{#1}}
\expandafter\ifx\csname urlstyle\endcsname\relax
  \providecommand{\doi}[1]{doi: #1}\else
  \providecommand{\doi}{doi: \begingroup \urlstyle{rm}\Url}\fi

\bibitem[Martin et~al.(1973)Martin, Siggia, and Rose]{MSR1973}
P.~C. Martin, E.~D. Siggia, and H.~A. Rose.
\newblock Statistical dynamics of classical systems.
\newblock \emph{Phys. Rev. A}, 8:\penalty0 423--437, Jul 1973.
\newblock \doi{10.1103/PhysRevA.8.423}.
\newblock URL \url{https://link.aps.org/doi/10.1103/PhysRevA.8.423}.

\bibitem[{Jensen}(1981)]{Jensen1981}
Roderick~V. {Jensen}.
\newblock {Functional integral approach to classical statistical dynamics}.
\newblock \emph{Journal of Statistical Physics}, 25\penalty0 (2):\penalty0
  183--210, June 1981.
\newblock \doi{10.1007/BF01022182}.

\bibitem[Gozzi(1988)]{Gozzi1988}
E.~Gozzi.
\newblock Hidden brs invariance in classical mechanics.
\newblock \emph{Physics Letters B}, 201\penalty0 (4):\penalty0 525–528, 1988.
\newblock ISSN 0370-2693.
\newblock \doi{10.1016/0370-2693(88)90611-9}.
\newblock URL
  \url{https://www.sciencedirect.com/science/article/pii/0370269388906119}.

\bibitem[Gozzi et~al.(1989)Gozzi, Reuter, and Thacker]{Gozzi1989}
E.~Gozzi, M.~Reuter, and W.~D. Thacker.
\newblock Hidden brs invariance in classical mechanics. ii.
\newblock \emph{Phys. Rev. D}, 40:\penalty0 3363--3377, Nov 1989.
\newblock \doi{10.1103/PhysRevD.40.3363}.
\newblock URL \url{https://link.aps.org/doi/10.1103/PhysRevD.40.3363}.

\bibitem[Zinn-Justin(1986)]{ZINNJUSTIN_1986}
J.~Zinn-Justin.
\newblock Renormalization and stochastic quantization.
\newblock \emph{Nuclear Physics B}, 275\penalty0 (1):\penalty0 135--159, 1986.
\newblock ISSN 0550-3213.
\newblock \doi{https://doi.org/10.1016/0550-3213(86)90592-4}.
\newblock URL
  \url{https://www.sciencedirect.com/science/article/pii/0550321386905924}.

\bibitem[Kleinert(2009)]{kleinert2009path}
H.~Kleinert.
\newblock \emph{Path Integrals in Quantum Mechanics, Statistics, Polymer
  Physics, and Financial Markets}.
\newblock EBL-Schweitzer. World Scientific, 2009.
\newblock ISBN 9789814273572.
\newblock URL \url{https://books.google.de/books?id=VJ1qNz5xYzkC}.

\bibitem[{Peliti, L.}(1985)]{Peliti_1985}
{Peliti, L.}
\newblock Path integral approach to birth-death processes on a lattice.
\newblock \emph{J. Phys. France}, 46\penalty0 (9):\penalty0 1469–1483, 1985.
\newblock \doi{10.1051/jphys:019850046090146900}.
\newblock URL \url{https://doi.org/10.1051/jphys:019850046090146900}.

\bibitem[Harsh and Sollich(2023)]{Harsh_2023}
Moshir Harsh and Peter Sollich.
\newblock Accurate dynamics from self-consistent memory in stochastic chemical
  reactions with small copy numbers.
\newblock \emph{Journal of Physics A: Mathematical and Theoretical},
  56\penalty0 (45):\penalty0 455004, oct 2023.
\newblock \doi{10.1088/1751-8121/acfd6a}.
\newblock URL \url{https://dx.doi.org/10.1088/1751-8121/acfd6a}.

\bibitem[Cattaruzza et~al.(2013)Cattaruzza, Gozzi, and Neto]{Cattaruzza_2013}
E.~Cattaruzza, E.~Gozzi, and A.~Francisco Neto.
\newblock Least-action principle and path-integral for classical mechanics.
\newblock \emph{Phys. Rev. D}, 87:\penalty0 067501, Mar 2013.
\newblock \doi{10.1103/PhysRevD.87.067501}.
\newblock URL \url{https://link.aps.org/doi/10.1103/PhysRevD.87.067501}.

\bibitem[Abrikosov et~al.(2005)Abrikosov, Gozzi, and Mauro]{Abrikosov:2004cf}
A.~A. Abrikosov, E.~Gozzi, and D.~Mauro.
\newblock {Geometric dequantization}.
\newblock \emph{Annals Phys.}, 317:\penalty0 24--71, 2005.
\newblock \doi{10.1016/j.aop.2004.12.001}.

\bibitem[Penco and Mauro(2006)]{Penco_2006}
R~Penco and D~Mauro.
\newblock Perturbation theory via feynman diagrams in classical mechanics.
\newblock \emph{European Journal of Physics}, 27\penalty0 (5):\penalty0 1241,
  aug 2006.
\newblock \doi{10.1088/0143-0807/27/5/023}.
\newblock URL \url{https://dx.doi.org/10.1088/0143-0807/27/5/023}.

\bibitem[Das and Mazenko(2012)]{Das_2012}
Shankar~P. Das and Gene~F. Mazenko.
\newblock Field theoretic formulation of kinetic theory: Basic development.
\newblock \emph{JOURNAL OF STATISTICAL PHYSICS}, 149\penalty0 (4):\penalty0
  643–675, NOV 2012.
\newblock ISSN 0022-4715.
\newblock \doi{10.1007/s10955-012-0610-y}.

\bibitem[Kozlikin et~al.(2023)Kozlikin, Lilow, Pauly, Schuckert, Salzinger,
  Bartelmann, and Weidem\"uller]{Kozlikin_2023}
Elena Kozlikin, Robert Lilow, Martin Pauly, Alexander Schuckert, Andre
  Salzinger, Matthias Bartelmann, and Matthias Weidem\"uller.
\newblock {Ultracold plasmas from strongly anti-correlated Rydberg gases in the
  Kinetic Field Theory formalism}.
\newblock \emph{submitted to SciPost Physics}, 2 2023.

\bibitem[Lilow et~al.(2019)Lilow, Fabis, Kozlikin, Viermann, and
  Bartelmann]{Lilow_2019}
Robert Lilow, Felix Fabis, Elena Kozlikin, Celia Viermann, and Matthias
  Bartelmann.
\newblock Resummed kinetic field theory: general formalism and linear structure
  growth from newtonian particle dynamics.
\newblock \emph{Journal of Cosmology and Astroparticle Physics}, 2019\penalty0
  (04):\penalty0 001, apr 2019.
\newblock \doi{10.1088/1475-7516/2019/04/001}.
\newblock URL \url{https://dx.doi.org/10.1088/1475-7516/2019/04/001}.

\bibitem[Pixius et~al.(2022)Pixius, Celik, and Bartelmann]{Pixius_2022}
C.~Pixius, S.~Celik, and M.~Bartelmann.
\newblock Kinetic field theory: perturbation theory beyond first order.
\newblock \emph{Journal of Cosmology and Astroparticle Physics}, 2022\penalty0
  (12):\penalty0 030, dec 2022.
\newblock \doi{10.1088/1475-7516/2022/12/030}.
\newblock URL \url{https://dx.doi.org/10.1088/1475-7516/2022/12/030}.

\bibitem[{Stratonovich}(1957)]{Stratonovich1957}
R.~L. {Stratonovich}.
\newblock {On a Method of Calculating Quantum Distribution Functions}.
\newblock \emph{Soviet Physics Doklady}, 2:\penalty0 416, July 1957.

\bibitem[Hubbard(1959)]{Hubbard1959}
J.~Hubbard.
\newblock Calculation of partition functions.
\newblock \emph{Phys. Rev. Lett.}, 3:\penalty0 77--78, Jul 1959.
\newblock \doi{10.1103/PhysRevLett.3.77}.
\newblock URL \url{https://link.aps.org/doi/10.1103/PhysRevLett.3.77}.

\bibitem[Gozzi and Reuter(1994)]{Gozzi:1993tm}
E.~Gozzi and M.~Reuter.
\newblock {Lyapunov exponents, path integrals and forms}.
\newblock \emph{Chaos Solitons Fractals}, 4:\penalty0 1117--1139, 1994.
\newblock \doi{10.1016/0960-0779(94)90026-4}.

\bibitem[Nakazato et~al.(1990)Nakazato, Okano, Schulke, and
  Yamanaka]{Nakazato:1990kk}
H.~Nakazato, K.~Okano, L.~Schulke, and Y.~Yamanaka.
\newblock {Symmetries in Stochastic Quantization and {Ito-Stratonovich} Related
  Interpretation}.
\newblock \emph{Nucl. Phys. B}, 346:\penalty0 611--631, 1990.
\newblock \doi{10.1016/0550-3213(90)90295-O}.

\bibitem[Ezawa and Klauder(1985)]{Ezewa}
Hiroshi Ezawa and John~R. Klauder.
\newblock {Fermions without Fermions: The Nicolai Map Revisited}.
\newblock \emph{Progress of Theoretical Physics}, 74\penalty0 (4):\penalty0
  904--915, 10 1985.
\newblock ISSN 0033-068X.
\newblock \doi{10.1143/PTP.74.904}.
\newblock URL \url{https://doi.org/10.1143/PTP.74.904}.

\bibitem[Mauro(2002)]{Mauro:2001rm}
D.~Mauro.
\newblock {On Koopman-von Neumann waves}.
\newblock \emph{Int. J. Mod. Phys. A}, 17:\penalty0 1301--1325, 2002.
\newblock \doi{10.1142/S0217751X02009680}.

\bibitem[v.~Neumann(1932)]{vonNeumann}
J.~v.~Neumann.
\newblock Zur operatorenmethode in der klassischen mechanik.
\newblock \emph{Annals of Mathematics}, 33\penalty0 (3):\penalty0 587--642,
  1932.
\newblock ISSN 0003486X.
\newblock URL \url{http://www.jstor.org/stable/1968537}.

\bibitem[{Koopman}(1931)]{Koopman}
B.~O. {Koopman}.
\newblock {Hamiltonian Systems and Transformations in Hilbert Space}.
\newblock \emph{Proceedings of the National Academy of Science}, 17\penalty0
  (5):\penalty0 315--318, May 1931.
\newblock \doi{10.1073/pnas.17.5.315}.

\bibitem[Das et~al.(2015)Das, Panda, and Santos]{Das:2014jia}
Ashok~K. Das, Sudhakar Panda, and J.~R.~L. Santos.
\newblock {A path integral approach to the Langevin equation}.
\newblock \emph{Int. J. Mod. Phys. A}, 30\penalty0 (07):\penalty0 1550028,
  2015.
\newblock \doi{10.1142/S0217751X15500281}.

\bibitem[Bartelmann et~al.(2016)Bartelmann, Fabis, Berg, Kozlikin, Lilow, and
  Viermann]{Bartelmann_2014}
Matthias Bartelmann, Felix Fabis, Daniel Berg, Elena Kozlikin, Robert Lilow,
  and Celia Viermann.
\newblock {A microscopic, non-equilibrium, statistical field theory for cosmic
  structure formation}.
\newblock \emph{New J. Phys.}, 18\penalty0 (4):\penalty0 043020, 2016.
\newblock \doi{10.1088/1367-2630/18/4/043020}.

\bibitem[{Von Roos}(1960)]{vonRoos}
Oldwig {Von Roos}.
\newblock {Formal Solution of Liouville's Equation}.
\newblock \emph{Journal of Mathematical Physics}, 1\penalty0 (2):\penalty0
  107--111, March 1960.
\newblock \doi{10.1063/1.1703639}.

\bibitem[de~Vega et~al.(1996)de~Vega, S\'anchez, and Combes]{deVega1996}
H.~J. de~Vega, N.~S\'anchez, and F.~Combes.
\newblock Fractal dimensions and scaling laws in the interstellar medium: A new
  field theory approach.
\newblock \emph{Phys. Rev. D}, 54:\penalty0 6008--6020, Nov 1996.
\newblock \doi{10.1103/PhysRevD.54.6008}.
\newblock URL \url{https://link.aps.org/doi/10.1103/PhysRevD.54.6008}.

\bibitem[Gozzi(1993)]{Gozzi:1991wi}
E.~Gozzi.
\newblock {Stochastic and nonstochastic supersymmetry}.
\newblock \emph{Prog. Theor. Phys. Suppl.}, 111:\penalty0 115--150, 1993.
\newblock \doi{10.1143/PTPS.111.115}.

\bibitem[Ovchinnikov(2016)]{Ovchinnikov_2016}
Igor Ovchinnikov.
\newblock Introduction to supersymmetric theory of stochastics.
\newblock \emph{Entropy}, 18\penalty0 (4):\penalty0 108, March 2016.
\newblock ISSN 1099-4300.
\newblock \doi{10.3390/e18040108}.
\newblock URL \url{http://dx.doi.org/10.3390/e18040108}.

\bibitem[Sato(1977)]{Sato:1976hy}
Masa-aki Sato.
\newblock {Operator Ordering and Perturbation Expansion in the Path Integration
  Formalism}.
\newblock \emph{Prog. Theor. Phys.}, 58:\penalty0 1262, 1977.
\newblock \doi{10.1143/PTP.58.1262}.

\bibitem[Blasone et~al.(2005)Blasone, Jizba, and Kleinert]{Blasone:2004yf}
Massimo Blasone, Petr Jizba, and Hagen Kleinert.
\newblock {Path integral approach to 't Hooft's derivation of quantum from
  classical physics}.
\newblock \emph{Phys. Rev. A}, 71:\penalty0 052507, 2005.
\newblock \doi{10.1103/PhysRevA.71.052507}.

\bibitem[Deotto et~al.(2001)Deotto, Gozzi, and Mauro]{Deotto:2001sy}
E.~Deotto, E.~Gozzi, and D.~Mauro.
\newblock {Supersymmetry in classical mechanics}.
\newblock 1 2001.

\bibitem[Gozzi and Penco(2011)]{Gozzi:2010iu}
E.~Gozzi and R.~Penco.
\newblock {Three Approaches to Classical Thermal Field Theory}.
\newblock \emph{Annals Phys.}, 326:\penalty0 876--910, 2011.
\newblock \doi{10.1016/j.aop.2010.11.018}.

\bibitem[Cattaruzza et~al.(2011)Cattaruzza, Gozzi, and
  Francisco~Neto]{Cattaruzza:2010wc}
Enrico Cattaruzza, Ennio Gozzi, and Antonio Francisco~Neto.
\newblock {Diagrammar In Classical Scalar Field Theory}.
\newblock \emph{Annals Phys.}, 326:\penalty0 2377--2430, 2011.
\newblock \doi{10.1016/j.aop.2011.05.009}.

\bibitem[{Balescu}(1975)]{Balescu}
R.~{Balescu}.
\newblock {Equilibrium and nonequilibrium statistical mechanics}.
\newblock \emph{NASA STI/Recon Technical Report A}, 76:\penalty0 32809, May
  1975.

\end{thebibliography}
\end{document}